\documentclass{aa}
\usepackage[varg]{txfonts}
\usepackage{graphicx}
\usepackage{natbib}
\bibpunct{(}{)}{;}{a}{}{,}

\newcommand{\one}{\emph{Star~1}}
\newcommand{\two}{\emph{Star~2}}
\newcommand{\three}{\emph{Star~3}}
\newcommand{\four}{\emph{Star~4}}

\newcommand{\Teff}{$T_{\rm eff}$}
\newcommand{\logg}{$\log g$}
\newcommand{\FeH}{[Fe/H]}
\newcommand{\MonH}{[M/H]}

\begin{document}

\title{Comparative Modelling of the Spectra of Cool Giants\thanks{Based on observations obtained at the Bernard Lyot Telescope (TBL, Pic du Midi, France) of the Midi-Pyr{\'e}n{\'e}es Observatory, which is operated by the Institut National des Sciences de l'Univers of the Centre National de la Recherche Scientifique of France.}\fnmsep
\thanks{Tables 6 to 11 are only available in electronic form
at the CDS via anonymous ftp to cdsarc.u-strasbg.fr (130.79.128.5)
or via http://cdsweb.u-strasbg.fr/cgi-bin/qcat?J/A+A/
}}

\author{
T. Lebzelter\inst{1} 
\and U. Heiter\inst{2}
\and C. Abia\inst{3}
\and K. Eriksson\inst{2}
\and M. Ireland\inst{4,20}
\and H. Neilson\inst{5}
\and W. Nowotny\inst{1}
\and J. Maldonado\inst{6}
\and T. Merle\inst{7}
\and R. Peterson\inst{8}
\and B. Plez\inst{9}
\and C.I. Short\inst{10}
\and G.M. Wahlgren\inst{11}
\and C. Worley\inst{7}
\and B. Aringer\inst{12}
\and S. Bladh\inst{2}
\and P. de Laverny\inst{7}
\and A.~Goswami\inst{13}
\and A. Mora\inst{14}
\and R.P. Norris\inst{15}
\and A. Recio-Blanco\inst{7}
\and M. Scholz\inst{16,21}
\and F. Th{\'e}venin\inst{7}
\and T. Tsuji\inst{17}
\and G. Kordopatis\inst{7}
\and B.~Montesinos\inst{18}
\and R.F. Wing\inst{19}
}

\institute{
University of Vienna, T\"urkenschanzstrasse 17, A-1180 Vienna, Austria
\and Department of Physics and Astronomy, Uppsala University, Box 516, SE-75120 Uppsala, Sweden
\and Depto. F\'\i sica Te\'orica y del Cosmos, Universidad de Granada, 18071 Granada, Spain
\and Department of Physics and Astronomy, Macquarie University, NSW 2109, Australia
\and Argelander Institute for Astronomy, University of Bonn, Auf dem Huegel 71, 53121 Bonn, Germany
\and Universidad Aut\'onoma de Madrid, Dpto. F\'isica Te\'orica, M\'odulo 15, Facultad de Ciencias,
Campus de Cantoblanco, E-28049 Madrid, Spain 
\and Universit\'{e} de Nice Sophia Antipolis, CNRS (UMR 6202), Observatoire de la C\^{o}te d'Azur, Cassiop\'{e}e, B.P.4229, 06304 Nice Cedex 04, France 
\and Astrophysical Advances / UCOLick, 607 Marion Pl, Palo Alto, CA  94301  USA 
\and Laboratoire Univers et Particules de Montpellier, Universit\'{e} Montpellier 2, CNRS, F-34095 Montpellier, France 
\and Department of Astronomy \& Physics, Saint Mary's University, 923 Robie Street Halifax, Nova Scotia, Canada B3H 3C3 
\and Goddard Space Flight Center, NASA, Greenbelt MD 20771, USA 
\and INAF-OAPD, Vicolo dell'Osservatorio 5, 35122 Padova, Italy 
\and Indian Institute of Astrophysics, Koramangala, Bangalore, India 
\and ESA-ESAC Gaia SOC. P.O. Box 78, E-28691 Villanueva de la Ca{\~n}ada, Madrid, Spain 
\and Department of Physics, Catholic University of America, 620 Michigan Ave, NE, Washington, DC 20064, USA 
\and Zentrum f\"ur Astronomie der Universit\"at Heidelberg (ZAH), Institut f\"ur
Theoretische Astrophysik, Albert Ueberle-Str. 2, 69120 Heidelberg,
Germany 
\and  Institute of Astronomy, The University of Tokyo, 2-21-1 Osawa, Mitaka, Tokyo, 181-0015 Japan 
\and Centro de Astrobiolog\'ia (INTA-CSIC), LAEFF Campus, European Space Astronomy Center (ESAC),
P.O. Box 78, E-28691 Villanueva de la Ca{\~n}ada, Madrid, Spain 
\and Astronomy Department, Ohio State University, 140 West 18th Avenue,
  Columbus, OH 43210, USA 
\and Australian Astronomical Observatory, PO Box 296, Epping, NSW 1710, Australia 
\and SIfA, School of Physics, University of Sydney, NSW2006, Australia 
}

\date{Received / Accepted}

\abstract{Our ability to extract information from the spectra of stars depends on reliable models of stellar atmospheres and appropriate 
techniques for spectral synthesis. Various model codes and strategies for the analysis of stellar spectra are available today.}
{We aim to compare the results of deriving stellar parameters using different atmosphere models and different analysis strategies. The focus is set on high-resolution spectroscopy of cool giant stars.}
{Spectra representing four cool giant stars were made available to various groups and individuals working in the area of spectral synthesis, asking them to derive stellar parameters from the data provided. The results were discussed at a workshop in Vienna in 2010. Most of the major codes currently used in the astronomical community for analyses of stellar spectra were included in this experiment.}
{We present the results from the different groups, as well as an additional experiment comparing the synthetic spectra produced by various codes for a given set of stellar parameters. Similarities and differences of the results are discussed.}
{Several valid approaches to analyze a given spectrum of a star result in quite a wide range of solutions. The main causes for the 
differences in parameters derived by different groups seem to lie in the physical input data and in the details of the analysis method. 
This clearly shows how far from a definitive abundance analysis we still are.}


\maketitle

\section{Introduction} 
Spectroscopy is the basic tool of modern astrophysics. It is the key for revealing the elemental composition and the physical conditions in the 
spectrum-forming layers of stars.
Interpreting the information contained in the spectra requires knowledge about the physics of the stellar atmosphere, the line formation processes, 
and the atomic and molecular data.  Parameters derived from the analysis of high-resolution spectra via comparison with stellar models have great 
potential but suffer from systematic uncertainties due to insufficient input physics in both the model atmospheres and the spectral synthesis, as well as 
to different fitting approaches.

Several codes to calculate atmospheric models exist today and are used by various groups around the world to analyze
both spectroscopic and photometric data. However, the implementation of the physics, the atomic and molecular data 
used, and the details of the method of deriving stellar parameters from the observed data differ among the
various research groups. Therefore, a comparison of the various codes and their output 
may help us understand uncertainties in the analysis of stellar spectra introduced by these
various components involved in the fitting process. These uncertainties also have
a major impact on the interpretation of photometric data or
the modelling of stellar populations. 

In this paper we present a comparison of a variety of model codes that attempt to analyze the spectra of cool
giants. Red giant stars are quite challenging targets for modelling with their complex atmospheres and
the large number of lines, in particular those of molecular origin. Hence, they provide a good testbed for
exploring the validity of input physics, line data, and modelling approaches. The aim was to test how
comparable the results are when applying different methods.

For this comparison, three of us (U. Heiter, T. Lebzelter, W. Nowotny) designed the following experiment, hereafter referred to as \emph{Experiment~1}: 
colleagues, who are engaged in stellar parameter and abundance determinations for cool giants on a regular basis, received high resolution and 
high signal-to-noise ratio (S/N) spectra of four cool stars and were asked to derive their fundamental stellar parameters, such as the effective 
temperature (\Teff), the surface gravity (\logg), 
and the metallicity ([Fe/H]\footnote{[Fe/H]$\equiv\log\frac{N(\rm Fe)}{N(\rm H)}-\log\frac{N(\rm Fe)}{N(\rm H)}_\odot$. [M/H] is defined accordingly using any atoms heavier than He.}).
In addition, we provided corresponding photometric data in various bands. However, no identifications of the sources were given in order to 
prevent the participants from comparing their findings with data in the literature.  The list of authors of this paper illustrates the high level of interest in this experiment.
The results were compared and discussed during a workshop\footnote{Kindly funded by the ESF within the GREAT network initiative.} 
held at the University of Vienna, August 23-24, 2010.

The choice of cool giants as targets was driven by the motivation of this experiment
within the framework of ESA's upcoming Gaia mission\footnote{http://sci.esa.int/gaia/}. In preparation for the exploitation of
a large amount of spectroscopic and photometric data with the aim of determining accurate stellar parameters,
there is a clear need to identify key areas where model spectra
can and should be improved, and to determine the influence of different methods of analysis on
the results. Giant stars will play an important role within the sample of objects
that will be studied by Gaia. 

At the workshop we agreed to perform a second comparison of our models, hereafter referred to as \emph{Experiment~2}. In this case, each participating group was asked to calculate a high-resolution model spectrum in a pre-defined wavelength range using a given set of stellar parameters.

This article is organized as follows.
In Section~\ref{sect:setup} we provide details on the design of the experiments, and summarize the properties of the two benchmark stars analysed in Experiment~1, $\alpha$~Tau and $\alpha$~Cet.
A detailed description of the modelling approaches used in the experiments is presented in Section~\ref{sect:modelling}.
In Sections~\ref{sect:results} and \ref{sect:discussion} the results of both experiments are presented and discussed. As a by-product we give revised stellar parameters for the two benchmark stars.
The outcome of our experiments forms a basis for future improvements of stellar spectrum modelling. 
This will be an important step towards accurate stellar parameters of giant stars observed by Gaia and Gaia follow-up programmes.

\section{Experiment set-up and benchmark stars} 
\label{sect:setup}
\subsection{Experiment 1 -- stellar parameter determination}
High-resolution and high S/N spectra of four targets were provided for Experiment~1. 
Two of them were of real stars, namely $\alpha$~Tau and $\alpha$~Cet, and covered the visual range of the spectrum between 4900 and 9750\,{\AA}.  They were obtained by one of us (U. Heiter) using the NARVAL spectrograph at the 2m Telescope Bernard Lyot atop Pic du Midi \citep{NARVAL}. The resolution was set to $R$=80\,000.  S/N$>$200 was achieved throughout the whole spectral range for both stars. The data were reduced with the Libre-ESpRIT pipeline \citep{donati97}.  The extracted and calibrated echelle orders were merged by cutting the orders at the centers of the overlap regions.  The spectra were not corrected for telluric features, but a spectrum of a telluric standard star taken in the same night was provided.
Both objects have been studied in detail in the past and have been used as reference targets in several investigations.  In Table~\ref{narvaldata} we summarize the stellar parameters which we assume are the most accurate ones available in the literature.  A more detailed description of the targets is given in Sections~\ref{sect:alftau} and \ref{sect:alfcet}.
Three wavelength ranges were recommended, on which the experiment should focus: $[4900-5400]$, $[6100-6800]$ and $[8400-8900]$ \AA.  Only the second of these regions was judged to be significantly contaminated by telluric lines.

The other two spectra were synthetic ones computed for realistic stellar parameters.  Within the experiment they should allow for a more direct comparison between the models without the uncertainties of stellar parameters and the 
unidentified features we see in observed data. For the artificial data we used two COMARCS
model spectra (for a description of the COMARCS models see Sect.\,\ref{sect:M1}) calculated by W. Nowotny and T. Lebzelter.
To simulate observational effects some gaussian noise was added (S/N=125) and the output was rebinned from $R$=300\,000 to $R$=50\,000. The chosen model parameters are listed in Table~\ref{star34}. The input parameters for \three\ were chosen to reproduce a slightly metal poor K-type giant as might be found in an Large Magellanic Cloud
 cluster. \four\ should resemble a typical  field star on the Asymptotic Giant Branch (AGB).  The corresponding quantities follow the predictions from the stellar evolution models by \citet{marigo08}. The ratio of $^{12}$C to $^{13}$C was solar in both cases.
For the synthetic spectra we chose a wavelength range of $\lambda=$15\,456--15\,674\,{\AA}.  Within this part of the spectrum one finds lines of CO, OH and CN as well as several atomic lines. The wavelength range is almost free of telluric lines \citep[e.g.][]{Arcturusatlas}. The synthetic spectra have a much smaller wavelength coverage than the NARVAL spectra which reflects 
the fact that today's near infrared spectrographs that reach a resolution of $R$=50\,000 (CRIRES, Phoenix) also cover only a comparably small wavelength range at a time. 
Broad band colours of the model stars, listed in Table~\ref{star34} as well, have been calculated from lower resolution spectra ($R$=10\,000) over a wavelength range from 0.45 to 2.6~$\mu$m as described in \citet{2011A&A...529A.129N}.
The participants of the experiment were not informed about the artificiality of these data. 
All four spectra used in Experiment~1 are available online\footnote{ftp://ftp.astro.uu.se/pub/Spectra/ulrike/ComparativeModelling/}.

A specific list of atomic line data was provided, and was
suggested, but not required,  to be used for the analysis.
This line list was extracted on 2010-04-29 
from the Vienna Atomic Line Database (VALD) \citep[Uppsala mirror\footnote{http://www.astro.uu.se/{$<$tilde$>$}vald/php/vald.php};][]{Pisk:95,Kupk:99,2008JPhCS.130a2011H}, and included all lines in the database within the three recommended optical wavelength regions
and the single infrared region.

\begin{table}
\caption{\label{narvaldata} Stellar parameters of $\alpha$~Tau (Section~\ref{sect:alftau}) and $\alpha$~Cet (Section~\ref{sect:alfcet}) based on published data.}
\centering
\begin{tabular}{lcccc}
\hline\hline
                            & $\alpha$ Tau     & Ref. & $\alpha$ Cet      & Ref.\\
                            & \one\            &      & \two\             &\\
\hline
Name                        & Aldebaran        &      & Menkar            & \\
HR                          & 1457             &      & 911               & \\
MK type                     & K5 III           &      & M1.5 IIIa         & \\
$T_{\rm eff}$ (K)           & 3930$\pm$40      & (a)  & 3800$\pm$60       & (a) \\
$T_{\rm eff}$ (K)           & 3920$\pm$130     & (b)  & 3730$\pm$75       & (i) \\
$L$ [$L_\odot$]             & 440$\pm$20       & (c)  & 1870$\pm$130      & (c) \\
log~($g$ (cm\,s$^{-2}$))      & 1.2$\pm$0.1      & (a)  & 0.9$\pm$0.1       & (a) \\
log~($g$ (cm\,s$^{-2}$))      & 1.2$\pm$0.5      & (b)  & 0.7$\pm$0.3       & (i) \\
$M$ [$M_\odot$]             & 1.3$\pm$0.3      & (d)  & 3.0$\pm$0.5       & (d) \\
$[$Fe/H$]$                  & $-$0.22$\pm$0.11 & (e)  & +0.02$\pm$0.03    & (i) \\
\noalign{\smallskip}
\hline
\noalign{\smallskip}
$V-I$                       & 2.17$\pm$0.02    & (f)  & 2.51$\pm$0.02     & (f) \\
$J-K$                       & 0.97$\pm$0.03    & (f)  & 1.08$\pm$0.03     & (f) \\
$V-K$                       & 3.67$\pm$0.03    & (f)  & 4.21$\pm$0.03     & (f) \\
$v_{\rm rad}$ [km~s$^{-1}$] & 54.26$\pm$0.03   & (g)  & $-$26.08$\pm$0.02 & (g) \\
$v$~sin~$i$ [km~s$^{-1}$]   & 5$\pm$1          & (h)  & 3$\pm$2           & (j) \\
\hline
\end{tabular}
\tablefoot{
\tablefoottext{a}{\emph{direct} parameters (see text).}
\tablefoottext{b}{mean of parameters used for spectroscopic \FeH\ (see Table~\ref{tab:paramref}).}
\tablefoottext{c}{from bolometric flux and parallax (see text).}
\tablefoottext{d}{from \Teff\ and $L$, and evolutionary tracks (see text).}
\tablefoottext{e}{see Table~\ref{tab:paramref}.} 
\tablefoottext{f}{\citet{1966CoLPL...4...99J}.}
\tablefoottext{g}{heliocentric radial velocity from \citet{2005A&A...430..165F}.} 
\tablefoottext{h}{\citet{2007AaA...475.1003H}.} 
\tablefoottext{i}{\citet{2008AaA...484L..21M}.}
\tablefoottext{j}{\citet{2008MNRAS.390..377Z}.}
}
\end{table}

\begin{table}
   \caption{\label{star34} Input parameters and model colours for the two artificial stars in Experiment~1.}
\centering
\begin{tabular}{lcc}
\hline\hline
 & \three & \four \\
\hline
$T_{\rm eff}$ (K) & 4257 & 3280\\
$L$ [$L_\odot$] & 319 & 3816\\
log~($g$ (cm\,s$^{-2}$)) & 1.47 & 0.06\\
$M$ [$M_\odot$] & 1.165 & 1.509\\
$[$Fe/H$]$ & $-$0.4 & +0.1\\
C/O & 0.35 & 0.55\\
\noalign{\smallskip}
\hline
\noalign{\smallskip}
$V-I$ & 1.25 & 3.58\\
$J-K$ & 0.82 & 1.23\\
$V-K$ & 2.94 & 6.89\\
\hline
\end{tabular}
\end{table}

\subsection{$\alpha$ Tau}
\label{sect:alftau}
Stellar surface parameters (\Teff\ and \logg) can be determined either from angular diameter measurements in combination with additional data (called \emph{direct} parameters hereafter), or from a model atmosphere analysis of photometric or spectroscopic data.
The \emph{direct} \Teff\ value is obtained from the angular diameter $\theta$ and the bolometric flux $F_{\rm bol}$ according to Eq.~\ref{equ:teff}, where $\sigma$ is the Stefan-Boltzmann constant.
The \emph{direct} \logg\ value is derived from $\theta$, the stellar mass $M$, and the 
parallax $\pi$ according to 
Eq.~\ref{equ:logg}, where $R$ is the linear stellar radius and $G$ is the constant of gravity.

\begin{equation}
   T_{\rm eff}^4 = \frac{F_{\rm bol}}{\sigma(0.5\theta)^2}
   \label{equ:teff}
\end{equation}

\begin{equation}
   g = \frac{GM}{R^2}; \,R=\frac{\theta}{2\pi}
   \label{equ:logg}
\end{equation}

The angular diameter of $\alpha$~Tau was determined recently by \citet{2005A&A...433..305R}, using both lunar occultations and long-baseline interferometry (VLTI-VINCI, 
$K$-band), and taking into account limb darkening.
The integrated absolute flux was measured for $\alpha$~Tau by \citet{1987A&A...188..114D} and \citet{2003AJ....126.2502M}.
The measured \mbox{$\theta=20.58\pm0.03$~mas} and 
\mbox{$F_{\rm bol}=33.57\pm1.35$~nW~m$^{-2}$} result in the direct \Teff\ value given in Table~\ref{narvaldata}.
With the Hipparcos parallax $\pi=48.92\pm0.77$~mas \citep{2007ASSL..350.....V}, this results in $R=45.2\pm0.7~R_\odot$ and in the luminosity $L$ given in Table~\ref{narvaldata}.
We estimate the mass of $\alpha$~Tau using two different sets of stellar evolutionary tracks, those published by the Padova group \citep{2008A&A...484..815B,2009A&A...508..355B} and the Yonsei-Yale (Y2) models \citep{2003ApJS..144..259Y,2004ApJS..155..667D}. For solar metallicity tracks (Padova: $Z$=0.017, $Y$=0.26; Y2: $Z$=0.02, $X$=0.71), the direct \Teff\ and $L$ imply a mass of 1.6$M_\odot$ for both model grids, while the metal-poor tracks corresponding to [Fe/H]=$-$0.3 (Padova: $Z$=0.008, $Y$=0.26; Y2: $Z$=0.01, $X$=0.74) suggest a mass of 1.0 to 1.1~$M_\odot$.
Interpolating between these values results in the mass in Table~\ref{narvaldata}. Thus, we arrive at the \logg\ value for $\alpha$~Tau given in Table~\ref{narvaldata}.

$\alpha$~Tau has been studied with high resolution, high S/N spectra in nine publications since 1980 (according to the PASTEL catalogue, \citealt{Soubiran2010}). 
The stellar parameters derived in these works and the references are given in Table~\ref{tab:paramref}.
Effective temperatures used in the spectroscopic analyses have been derived from various photometric calibrations by most authors. Combining the results given in six publications, the mean published
photometric \Teff\ of $\alpha$~Tau is 3850$\pm$40~K. This is in good agreement with the latest value of \Teff=3880$\pm$40~K determined with the infrared flux method (IRFM),  by \citet{2005ApJ...626..446R}, which is an update of the work by \citet{1999A&AS..139..335A}.
Two publications from 1981 and 2007 derive \Teff\ in a spectroscopic way (excitation equilibrium of iron line abundances) from high-resolution spectra in the optical wavelength range. They arrive at almost the same value, close to 4120~K, which is significantly higher than the photometric one. On the other hand, \citet{2008AaA...484L..21M} obtained a spectroscopic \Teff\ of 3890~K, close to the IRFM value, based on high-resolution infrared spectra centered on 15555~\AA.

The method for determining the surface gravity varies significantly between the publications. Four of the authors 
derive \logg\ in a spectroscopic way in the optical wavelength range (ionization equilibrium of iron line abundances) 
and arrive at a mean value of 1.1$\pm$0.5. Two authors use absolute magnitudes and stellar evolution calculations, 
and cite a higher mean value of 1.6$\pm$0.1. The spectroscopic \logg\ of \citet{2008AaA...484L..21M} determined from 
IR spectra is close to the values obtained from optical spectra. The highest \logg\ value of 1.8 is determined from 
DDO photometry by \citet{1990AJ.....99.1961F}, who cite an error of 0.2.

The metallicity of $\alpha$~Tau is determined in eight publications and found to be below solar (mean value $-0.2\pm0.1$, see Table~\ref{narvaldata}). The results can be divided into two groups, three authors using \Teff$>$3900~K and five authors using \Teff$<$3900~K. The corresponding [Fe/H] values cluster within a few tenths of a dex around $-0.35$ and $-0.15$, respectively.
The mean \Teff\ and \logg\ values used for deriving the mean [Fe/H] are given in Table~\ref{narvaldata}.

The spectral type K5\,III given for $\alpha$~Tau in Table~\ref{narvaldata} was first published in the 
original MKK Atlas \citep{1943QB881.M6.......}
and has been quoted throughout the literature. However, the star's TiO strength, as measured by 
narrow-band classification photometry \citep{2011ASSL..373..145W},
yields a spectral type of K5.7\,III on a scale where type M0.0 immediately follows K5.9. The 
photometric colours also indicate an effective temperature closer to M0.0 than to K5.0.

\begin{table}
\caption{\label{tab:paramref} Results and references for previous spectroscopic works on $\alpha$~Tau.}
\centering
\begin{tabular}{rp{1mm}rp{1mm}rl}
\hline\hline
\Teff & \textit{(a)} & \logg & \textit{(b)} & \FeH & Reference \\
\hline
4140 & sp &  1.0 & sp &  $-$0.33 &  \citet{1981ApJ...248..228L} \\
3830 & ph &  1.2 & sp &  $-$0.14 &  \citet{1983AaA...120...21K} \\
3850 & ph &  1.5 & ev &  \textit{(c)} &  \citet{1985ApJ...294..326S} \\
3800 & ph &  1.8 & ph &  $-$0.17 &  \citet{1990AJ.....99.1961F} \\
3910 & ph &  1.6 & ev &  $-$0.34 &  \citet{1990ApJS...74.1075M} \\
3875 & ph &  0.6 & sp &  $-$0.16 &  \citet{1995AJ....110.2968L} \\
3850 & ph &  0.6 & li &  $-$0.10 &  \citet{1998AaA...338..623M} \\
4100 & sp &  1.7 & sp &  $-$0.36 &  \citet{2007AaA...475.1003H} \\
3890 & sp\tablefootmark{d} &  1.2 & sp\tablefootmark{d} &  $-$0.15 &  \citet{2008AaA...484L..21M} \\
\hline
\end{tabular}
\tablefoot{
\tablefoottext{a}{Method for \Teff\ determination: sp \ldots\ spectroscopic (excitation equilibrium of iron line abundances), ph \ldots\ photometric calibrations.}
\tablefoottext{b}{Method for \logg\ determination: sp \ldots\ spectroscopic (ionization equilibrium of iron line abundances), ph \ldots\ photometric calibrations, ev \ldots\ from absolute magnitudes and stellar evolution calculations, li \ldots\ from literature.}
\tablefoottext{c}{$\alpha$~Tau was used as a reference object for the spectroscopic analysis of 
M giants.}
\tablefoottext{d}{from infrared spectra.}
}
\end{table}

\subsection{$\alpha$ Cet}
\label{sect:alfcet}

The angular diameter of $\alpha$~Cet was determined by \citet{Wittkowski2006b} from long-baseline interferometry, using the same instrumentation as for $\alpha$~Tau, and taking into account limb darkening.
The same authors also determined $F_{\rm bol}$ from integrated absolute flux measurements.
The measured $\theta=12.20\pm0.04$~mas and $F_{\rm bol}=10.3\pm0.7$~nW~m$^{-2}$ result in the direct \Teff\ value given in Table~\ref{narvaldata}.
With the Hipparcos parallax $\pi=13.10\pm0.44$~mas \citep{2007ASSL..350.....V}, this results in $R=100\pm3~R_\odot$ and in the luminosity $L$ given in Table~\ref{narvaldata}.
Using this luminosity and the direct \Teff\ value, we estimate the mass of $\alpha$~Cet (see Table~\ref{narvaldata}) from evolutionary tracks for solar metallicity (see Section~\ref{sect:alftau}). The masses from the two different sets of models agree within 0.1~$M_\odot$ The 2007 Hipparcos parallax is smaller than the ``original'' one by about 10\%, which results in a mass 30\% higher 
than derived by \citet{Wittkowski2006b}.
The mass estimate could be refined in the future through asteroseimology, but in any case the uncertainties for the direct \logg\ value are already much smaller than the spectroscopic uncertainties.
As for $\alpha$~Tau, from these stellar data we can derive an accurate direct \logg\ value for $\alpha$~Cet, given in Table~\ref{narvaldata}. 

For $\alpha$~Cet, there are no previously published high-resolution spectroscopic studies in the optical wavelength range. The \Teff\ value determined with the IRFM value is 3720$\pm$50~K \citep{2005ApJ...626..446R}. The star is included in the infrared spectroscopic study of \citet{2008AaA...484L..21M}, who determine a spectroscopic \Teff\ close to the IRFM value, a spectroscopic \logg = 0.7$\pm$0.3, and solar metallicity (see Table~\ref{narvaldata}). 
The TiO-based spectral type of $\alpha$~Cet, from narrow-band classification photometry 
\citep{2011ASSL..373..145W}, is M1.7\,III, in substantial agreement with the Morgan-Keenan (MK) type shown in 
Table~\ref{narvaldata}.

\subsection{Experiment 2 -- comparison of synthetic spectra for fixed parameters}\label{exp2descr}
For this experiment the set of stellar parameters was predefined in order to be able to compare the output of the various combinations of models and spectral synthesis codes directly. Participants were asked to compute spectra at high resolution ($R \ge$ 300\,000) for $T_{\rm eff} = $3900\,K, \logg = 1.3, and [Fe/H]$ = -0.2$, i.e. a parameter set close to the values corresponding to $\alpha$~Tau. Microturbulence was fixed at 
2.0\,km\,s$^{-1}$, and a mass of 2\,$M_\odot$ was to be assumed. The synthetic spectra covered the three optical
wavelength regions that were recommended for the analysis of \one\ and \two\ in Experiment~1. The same abundance pattern \citep{2009ARA&A..47..481A} was used by all participating groups.

\section{Modelling} 
\label{sect:modelling}
In this section we summarize the main characteristics of the models used for the experiments. In our sample we have two major model `families' -- MARCS (Model Atmospheres in a Radiative Convective Scheme) and 
ATLAS -- where several implementations and individual further developments of the original code were employed, and three alternative models.
For the discussion below we introduce abbreviations for each model and implementation, e.g. `M' for MARCS-based modelling, and `M1' for a specific participating team applying the code.
Where possible we refer to more extensive descriptions of the codes published elsewhere.
Comparisons of different atmospheric models can be found in the original literature describing the codes. For example, \citet{2008A&A...486..951G} write on the comparison between MARCS and ATLAS models:
\begin{quote}
``In view of the fact that these two grids of models are made with two totally independent numerical methods and computer codes, with independent choices of basic data (although Kurucz's extensive lists of atomic line transitions are key data underlying both grids), this overall agreement is both satisfactory and gratifying.''
\end{quote}

\begin{figure}
\begin{center}
   \resizebox{\hsize}{!}{\includegraphics[trim=70pt 40pt 80pt 60pt,clip=true]{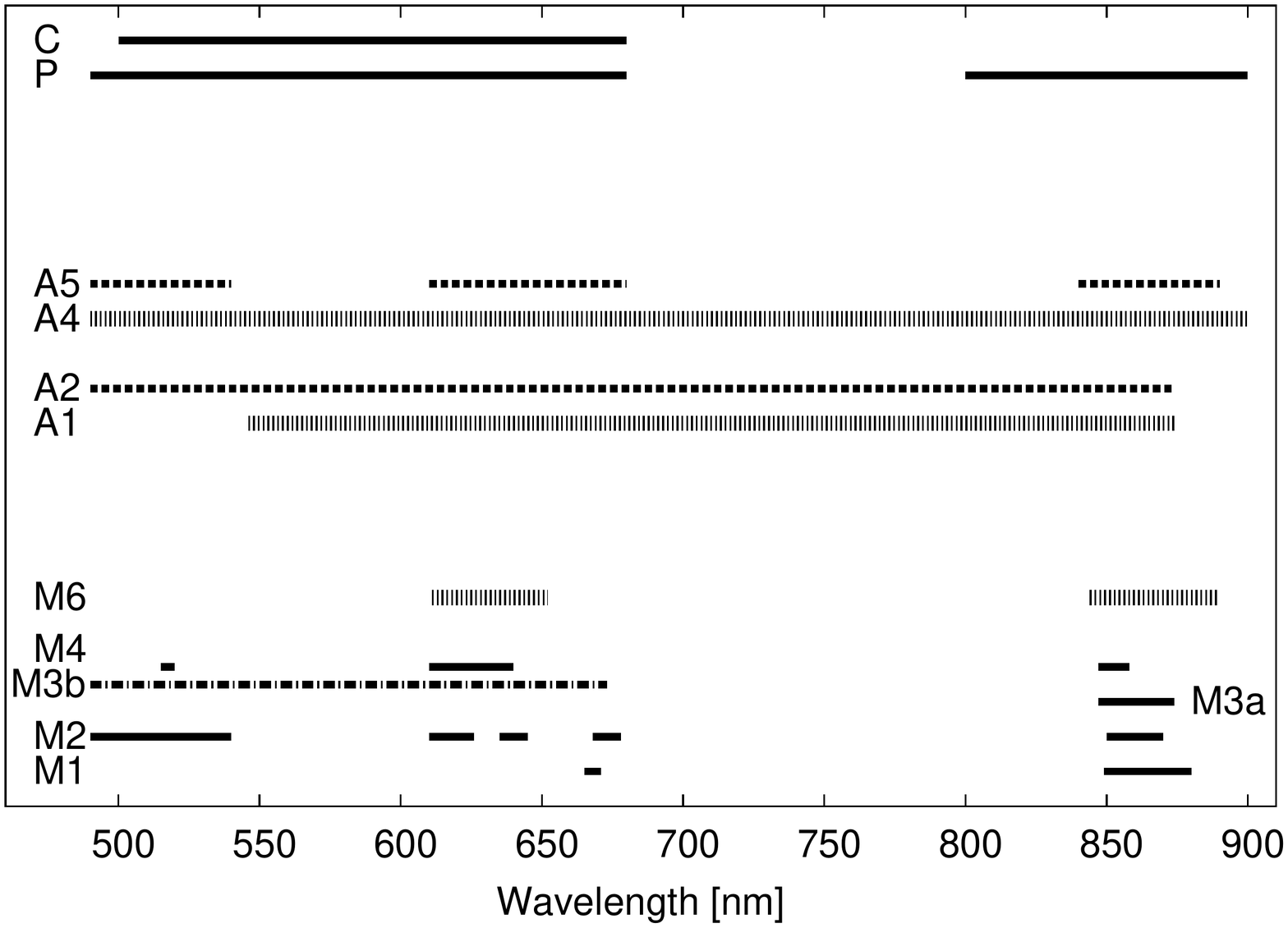}}
\end{center}
\caption{\label{fig:wavelengths} Wavelength intervals used by the eleven different groups analysing the optical spectra. See \S\S~\ref{sect:MARCS} to \ref{sect:T}. For the M3b analysis (\S~\ref{sect:M3}), discontinuous wavelength regions totaling $\sim$435\AA\ were used. The M6, A1, and A4 groups (\S\S~\ref{sect:M6}, \ref{sect:A1}, and \ref{sect:A4}) used spectrum synthesis for individual spectral lines distributed over the indicated regions. The A2 and A5 groups 
(\S\S~\ref{sect:A2}, and \ref{sect:A5}) used equivalent widths for individual spectral lines.}
\end{figure}

Each group worked with different subsections of the available wavelength range. For the optical spectra, Fig.~\ref{fig:wavelengths} shows the wavelength intervals used by the eleven groups who analyzed these spectra.
Most groups analysing the spectra of \one\ and \two\ determined radial velocities, which are given in Tables~\ref{results1a} and \ref{results1b}, and corrected the observed spectra to laboratory wavelengths.
The groups used different approaches to adjust the continuum scale of these spectra, which are described in the individual subsections.

The groups employing MARCS model atmospheres in general used the solar chemical composition from \citet{2007SSRv..130..105G} as a reference (which was used for the calculation of the on-line database of MARCS models). However, M1 (Section~\ref{sect:M1}) adopted the abundances from \citet{1994LNP...428..196G} for C, N, and O, and from \citet{1989GeCoA..53..197A} for all other elements. A3 used only the latter source for their
reference abundance pattern. A4 (Section~\ref{sect:A4}) took
their reference abundance pattern from \citet{1998SSRv...85..161G}. The analysis of A5 (Section~\ref{sect:M5})
refers to the abundances of \citet{2005ASPC..336...25A}.
M5, A1, and A2 (Sections~\ref{sect:M5}, \ref{sect:A1}, and \ref{sect:A2}) used the solar abundances from \citet{2009ARA&A..47..481A}, and P and C (Sections~\ref{sect:P} and \ref{sect:C}) those of \citet{1996ASPC...99..117G}. 
T used abundance values from several sources summarized in Table 4, case (b), of 
\citet{2008A&A...489.1271T}, and in \citet{2002ApJ...575..264T}.

\begin{table}
   \caption{\label{tab:modelgrids} Overview of model grid parameters used by different groups in Experiment~1.}
\centering
\begin{tabular}{llllr}
\hline\hline
ID\tablefootmark{a} & \Teff\ & \logg\  & \FeH\ \tablefootmark{b} \\
 & (K)\tablefootmark{b} & (cm\,s$^{-2}$)\tablefootmark{b} & \\
\hline
M1  & 3600 to 4000\tablefootmark{c} /  50 & 0.5, 1.0, 1.5 & 0.0, $-$0.3, $-$0.5 \\
M2  & 3500 to 4100\tablefootmark{d} / 100 & 0.5 to 2.0 / 0.5 & $-$0.5 to +0.5 / 0.25 \\
M3\tablefootmark{e}  & 2500 to 8000 & 0.0 to 5.0 & $-$5.0 to +1.0 \\
M4  & 3800 to 4300 / 50 & 0.0 to 3.0 / 0.5 & 0.0, $-$0.25 \\
M5  & 2500 to 8000 / 250\tablefootmark{f} & $-$1 to 5 / 0.5 & $-$5 to 1 / \tablefootmark{g} \\
M6  & 3750 to 4250 / 200 & 0.0 to 2.0 / 1.0 & -1.0 to 0.5 / 0.1 \\
A1  & 3500 to 13000 / 250 & 0.0 to 5.0 / 0.5 & $-$1.5 to +0.5 / 0.5 \\
A2  & 3500 to 6000 / 250 & 0.0 to 5.0 / 0.5 & $-$1.5 to +0.5 / 0.5 \\
A3  & 3000 to 8000 / 100 & 0.0 to 3.0 / 0.1 & 0.0\tablefootmark{j}\\
A4  & 3500 to 6250 / 250 & 0.0 to 5.0 / 0.5  & +0.5, +0.2, +0.0; \\
 & & & $-$0.5 to $-$4.5 / 0.5\tablefootmark{h} \\
A5  & 3500 to 6000 / 250 & 0.0 to 5.0 / 0.5 & 0.0, -1.0 \\
P   & 3500 to 4500 / 125 & 1.0 to 2.5 / 0.5 & 0.0, $-$0.5 \\
T & 2800 to 4000 / 100 & $-$0.52 to 1.34 / \tablefootmark{g} & 0.0\\
C   & 2600 to 4000 / 200 & $-$0.5, 0.0 & $-$0.5, 0.0 \\
\hline
\end{tabular}
\tablefoot{
\tablefoottext{a}{For the group IDs see the subsections of Section~\ref{sect:modelling}.}
\tablefoottext{b}{A range of parameters is given in the format: \emph{minimum value} to \emph{maximum value} / \emph{step size}. Comma-separated values indicate a discrete set of parameter values.} 
\tablefoottext{c}{\Teff\ range chosen on the basis of an educated guess from the strength of the TiO features in the spectra, 81 models.}
\tablefoottext{d}{\Teff\ range centered on \Teff\ from photometric calibrations.}
\tablefoottext{e}{One value of [$\alpha$/Fe] and 2905 models for M3a, five values of [$\alpha$/Fe] 
and 16\,783 models for M3b.}
\tablefoottext{f}{Step size is 100\,K between 2500 and 4000\,K. In addition, a small set of C-rich models was used.}
\tablefoottext{g}{Variable step size.}
\tablefoottext{h}{\FeH\ $-$1.5 to 0.2 / 0.1 for a subgrid with \Teff\ 3000 to 4000\,K.}
\tablefoottext{j}{One model with a solar abundance pattern was calculated, and one model
with $\alpha$ elements enhanced by +0.4.}
}
\end{table}

Some of the groups determined stellar parameters from the photometric colours provided together with the spectra for each star. These parameters, and the corresponding photometric calibrations, are summarized in Tables~\ref{results1a} to \ref{results2b}.
Table~\ref{tab:modelgrids} summarizes the range of parameters of the model grids used by several of the groups for the determination of the best-fit spectrum.

Sections~\ref{sect:MARCS} to \ref{sect:T} contain detailed accounts of the analysis work of the participating groups. 
In these sections, \one\ and \two\ refer to $\alpha$~Tau and $\alpha$~Cet, respectively.
Each modelling description also includes a brief discussion of the individual fitting results.
The impatient reader may at this point skip to Section~\ref{sect:results}, where we start with an overview of the main aspects of each analysis, and the detailed descriptions may serve for later reference. A general discussion comparing the results from the various groups is provided in Section~\ref{sect:discussion}.

\subsection{MARCS model atmospheres (M)}
\label{sect:MARCS}
\subsubsection{M1}
\label{sect:M1}
The Padova-Vienna team included B. Aringer, T. Lebzelter, and W. Nowotny. This analysis used the COMARCS hydrostatic atmosphere models, which are a further development of the original MARCS code of \citet{gustafsson75} and \citet{jorgensen92}.
Opacities are treated using opacity tables created in advance with the opacity generation code \emph{coma08} \citep{aringer09}. Compared to the MARCS models of \citet{2008A&A...486..951G}, the opacities are computed with different interpolation schemes.
The temperature and, accordingly, the pressure stratification of the models are derived under the assumption of a spherical configuration in hydrostatic and local thermodynamic equilibrium (LTE). A detailed description of the spectral synthesis is given in \citet{2010A&A...514A..35N}. Sources for spectral line data are given in \citet{2009A&A...494..403L}.

Recent applications to the fitting of observed high-resolution spectra can be found in \citet{lebzelter08} and \citet{lederer09}. Here the same approach of finding the best fit was used, which will be briefly sketched in the following. Note that only the two spectra in the visual range, \one\ and \two\, were fitted. The 
same codes were used to calculate the two artificial sample spectra, \three\ and \four\ , 
and the corresponding broad band photometry. 

The M1 analysis started by calculating a grid of model spectra (see Table~\ref{tab:modelgrids}) at $R$=300\,000, rebinned to $R$=80\,000 afterwards.
Broad band model colours were determined, too, following descriptions in \citet{2011A&A...529A.129N}. Microturbulence and macroturbulence were set to fixed values (see Tables~\ref{results1a} and \ref{results1b}).  The value for the macroturbulence was checked with the typical line widths in the observed spectrum, and was found to be reasonably chosen. 

The first step was to identify features in the optical range that are primarily sensitive to changes of a single 
stellar parameter. The starting point for finding the best fit was the calcium triplet which constrained 
the value for \logg.  This estimate was validated by checking a few obviously \logg\ sensitive lines between 8500 and 
8800\,{\AA}. Because the strength of these lines is also dependent on the metallicity, it was decided to determine 
the best fitting \logg\ value for all three [Fe/H] values of the grid independently. Next, the temperature was fixed 
based on the strength of the TiO band heads at 6652, 6681, and 6698\,{\AA}.  The temperature 
was derived for each of the 
three metallicity / \logg\ pairs from the previous step.  A semiempirical approach combining $\chi^{2}$ analysis with 
a check of the fit by eye was applied.  
In cases where the best fit seemed to fall between two points of the model grid, the final fit parameters were
determined by interpolation between these grid points.
For \one\, this resulted in the best fitting parameter combinations ($T_{\rm eff}$, \logg, [Fe/H]) = 
(3900\,K, 1.25, 0.0), (3800\,K, 0.75, $-$0.3), and (3750\,K, 0.5, $-0.5$). 
For \two, the corresponding values are (3800\,K, 1.0, 0.0), (3700\,K, 0.5, $-$0.3), and (3675\,K, 0.5, $-$0.5).

The best combination of metallicity, temperature and \logg\ should then be determined by testing the overall fit of the atomic lines and molecular bands at shorter wavelengths.  However, no clear decision could be made in that respect, either for \one\ or for \two.  It turned out that the fit of the spectrum below 6000\,{\AA} became rather poor with several missing lines.  Furthermore, several lines came out too strong from the model, and it is suspected that incorrect line blends are responsible for this problem. Tentatively, an [Fe/H] value below solar seems to be more appropriate for both stars.

To further constrain the metallicity of the star, synthetic colours for the chosen best fit combinations were compared with the observed values. For \one, the three given colours (see Table~\ref{narvaldata}) are best represented by the solution for [Fe/H] = $-0.3$. For \two, the solution for [Fe/H] = $-0.5$ fits best, 
and the solar abundance model can be excluded.
These solutions are given in Tables~\ref{results1a} and \ref{results1b} (see \S 4 below). The error bars are derived from the different metallicities considered.

\subsubsection{M2} 
\label{sect:M2}
In the M2 analysis, \one\ and \two\ were studied by B. Plez.
Spectra were computed using the TURBOSPECTRUM code \citep{Alvarez1998}. Model atmospheres were extracted from the MARCS database \citep{2008A&A...486..951G}. Refinement in the grid of models was obtained by interpolation using the routines by \citet{Masseron2006}, provided on the MARCS site\footnote{http://marcs.astro.uu.se}.
Molecular line lists are described in \citet{2008A&A...486..951G}, and atomic lines were extracted from the VALD database. As in the MARCS calculations, collisional line broadening is treated following \citet{Bark:00}, with particular broadening coefficients for many lines.

The comparison between observations and calculations was done using a mean square differences computation. 
As a first step, the photometry provided with the data was used to derive a first estimate 
of \Teff. Colour  -- \Teff\ calibrations based on MARCS models were used \citep{1998A&A...333..231B,2011ASPC..445...71V}.
Spectra were then computed for spherical models of one solar mass,
with \Teff\ $\pm$200~K around these values, gravities typical for giants, a range of metallicities,
and fixed values for the microturbulence parameter (see Table~\ref{tab:modelgrids}).
The differences between observed ($o_i$) and calculated ($c_i$) spectra with $N$ wavelength points were characterized by $\sigma=\sum_{i=1}^{N}\frac{(o_i-c_i)^2}{N}$ computed for wide portions of the spectra (see Fig.\,\ref{fig:wavelengths}).

This computation 
quickly showed that models with different parameter combinations (e.g. a lower \Teff\ and a lower metallicity) could give similarly small values of $\sigma$. Also, different spectral regions would lead to best fits for different values of the stellar parameters. The reason is that the spectra are dominated by numerous atomic and molecular lines, for which the line position and strength data might not be sufficiently accurate. The result is then that many small differences add up to a relatively large $\sigma$ that becomes quite insensitive to local improvements on a few ``good'' lines. Another problem is caused by the normalization of the observed spectra, which leaves residual slopes relative to the calculations. This is especially problematic for spectra of cooler objects with many molecular bands. Robust methods indeed use only the information from selected spectral ranges, where line data are carefully calibrated, and where spectra can be renormalized easily.  This selection was not done owing to 
the limitations of present experiment.

Values of $\sigma$ were also computed for a number of macroturbulence parameters, but they tended to decrease 
systematically with increasing macroturbulence. This is again explained by the presence of numerous lines, many of which were not well fitted. 
It proved more efficient to use values derived from an eye-inspection of the spectra (see Tables~\ref{results1a} and \ref{results1b}).
The best set of parameters (\Teff, \logg, metallicity) was derived for the various spectral regions 
(Fig.\,\ref{fig:wavelengths}) from the model giving the smallest $\sigma$, and for the few models giving similarly small values. Later, an inspection by eye assigned higher weight to regions that gave a better global fit (and a correspondingly smaller $\sigma$).
The results for \one\ and \two\ are given in Tables~\ref{results1a} and \ref{results1b} and are discussed below.

\paragraph{\one\ and \two:}
Photometry suggested the \Teff\ values given in Tables~\ref{results1a} and \ref{results1b}. The values were consistent for $J-K$ and $V-K$ in the case of \two, but not for \one.
The preferred calibration for cool stars is \Teff--$(V-K)$, which is quite insensitive to metallicity, and is strongly dependent on \Teff. It is, however, also sensitive to the C/O ratio, especially when C/O$>$0.9; this is the reason for using a combination of $V-K$ and $J-K$ to determine S-type star parameters \citep{2011ASPC..445...71V}.

For \one, the best $\sigma$-values for the first four spectral windows (Fig.~\ref{fig:wavelengths}) were 0.12, 0.047, 0.055, and 0.055. The number of models with a $\sigma$ within 10\% of the smallest were 35, 5, 13, and 7. This procedure yielded an estimate of 
\Teff\ of about 3800~K and \logg\ around 1.5, but there were other models with \Teff=3700~K to 3900~K, 
and \logg\ from 0.5 to 2.0, which also gave small values of $\sigma$ . The metallicity was not well constrained (solar, or maybe sub-solar). Attempts were made for the \ion{Ca}{II} lines (8500--8700~\AA) in the Gaia Radial Velocity Spectrometer (RVS) range, leading to a best fit for \Teff=3850~K, \logg=1.25, and solar metallicity. The individual Ca abundance was also varied, once the other model parameters were fixed, and was found to be solar to within 0.1~dex. The fit of the 5000--5200~{\AA} region (\ion{Mg}{I} and MgH lines) gave consistent stellar parameters, and a solar Mg abundance.
For \two, a similar approach resulted in the parameters given in Table~\ref{results1b}. This gave a best fit in the 6400~\AA\ region but not in the \ion{Ca}{II} IR triplet region, unless Ca is slightly overabundant (+0.2~dex).

\paragraph{\three\ and \four:}
Photometry gave the \Teff\ values given in Tables~\ref{results2a} and \ref{results2b}.
No further work was done on these stars.
The \Teff\ values were consistent for $J-K$ and $V-K$ in the case of \three, but not for \four.
A consistent \Teff=3400~K could have been obtained from both colours, if the star is s-element rich (about +2~dex relative to the Sun). As this toy-star had a solar composition, this points to a difference between MARCS and COMARCS model spectra, presumably due to differences in the opacities.

\subsubsection{M3}
\label{sect:M3}
The M3 team consisted of C.~Worley, P.~de Laverny, A.~Recio-Blanco and G.~Kordopatis.
In this case, two separate spectral analyses were carried out for each of \one\ and \two\ using pre-existing procedures that have been established for two separate research projects. Both projects consist of automated pipelines that feed the spectra into the stellar classification algorithm MATISSE (MaTrix Inversion for Spectrum SynthEsis). Thus, the two stars were analyzed in a completely blind way using automated procedures. The spectra were degraded to much lower resolutions 
(R$\approx$6500 for M3a and R$\approx$15000 for M3b) 
and no photometry was used to obtain prior estimates of the stellar parameters. Only the observed spectra were used in the search across a wide range of stellar parameters (dwarf to giant, metal-rich to metal-poor FGKM stars). The entire process, during which the spectra undergo wavelength selection, cosmic ray cleaning, radial velocity determination and correction, normalisation, and analysis in MATISSE, lasts only a few minutes for each pipeline. A key feature is the iteration between the normalisation procedure and the stellar parameter determination in MATISSE, whereby synthetic spectra generated for the preceding set of stellar parameters are used to normalise the observed spectrum for the next parameter determination. In this manner there is convergence to the final set of stellar parameters, the corresponding synthetic spectra, and the final normalised observed spectra.

The MATISSE algorithm, initially developed to be used in the analysis of the Gaia RVS spectra, is based on a local multi-linear regression method \citep{Recio-Blanco2006,Bijaoui2008}. A stellar parameter $\theta$ (in this case: \Teff, \logg, or [M/H]) of a star is determined by the projection of the observed spectrum, $O(\lambda)$, onto a vector function $B_{\theta}(\lambda)$. The $B_{\theta}(\lambda)$ vector is an optimal linear combination of spectra, $S_i(\lambda)$, in a grid of theoretical spectra. The product $\hat{\theta} = \frac{1}{N} \sum\limits_{\lambda = 1,N} B_{\theta}(\lambda) \ O(\lambda)$ is calculated.

In the training phase of MATISSE the $B_{\theta}(\lambda)$ vectors are created from the grid of synthetic spectra, with $B_{\theta}(\lambda) = \sum \ \alpha_{i} \ S_{i}(\lambda)$, and $\alpha_{i}$ being the weight associated with the spectrum $S_{i}$ giving the maximum correlation between $\hat{\theta}_i$ and $\theta_i$ in the training grid.
The sensitivity of a wavelength region to the particular stellar parameter $\theta$ is reflected in the corresponding $B_{\theta}(\lambda)$ vector.
The synthetic spectra cover the entire optical domain and have been built using the spectral line formation code TURBOSPECTRUM (Plez, private communication; \citealt{Alvarez1998}) and the MARCS stellar atmosphere models \citep{2008A&A...486..951G}.
We decided not to perform any convolution (spectral resolution and instrumental profile, stellar
rotation, macroturbulence,...) for these spectra in order to keep an easy and fast adaptation to the 
properties of any spectrograph and/or astrophysical application \citep[see also][]{2012A&A...544A.126D}.
At the comparably low spectral resolution the M3 analyses were performed, these broadening parameters were
of negligible relevance.

This grid covers the range of parameters given in Table~\ref{tab:modelgrids}.
The atomic line lists were taken from VALD (August 2009) and for M3a were calibrated to the Sun and Arcturus. The molecular line lists were provided by B.~Plez and included CH, OH, MgH plus
several isotopic compositions, SiH, CaH, FeH, C$_{2}$, CN, TiO, VO, and ZrO.
The grid is presented in more detail in de Laverny et al. (2012, to be submitted).
For each of the two research projects there was a training phase during which the $B_{\theta} (\lambda)$ functions for each spectral set-up (wavelength regions and resolution) were generated from the initial synthetic spectra grid.

\paragraph{FLAMES/GIRAFFE MATISSE Analysis:}
The first analysis method used (M3a) has been developed for the analysis of $\sim$700 galactic disk stars observed using FLAMES/GIRAFFE 
\citep{Kordopatis2011a,Kordopatis2011b}. The wavelength domain and resolution of this project correspond to that of the Gaia RVS (8470--8740 \AA\ and R$\sim$6500), of which a key spectral feature is the \ion{Ca}{II} Infra-Red Triplet (LR8 setup of GIRAFFE). For this analysis some small features were masked in order to eliminate sky lines and the cores 
of some strong lines which were not well reproduced by the synthetic spectra. Also the line list was calibrated to the Sun and Arcturus. For this analysis standard galactic $\alpha$ element enhancements were assumed for the stellar atmospheric models; this is the reason for the lower number of nodes in the synthetic spectra grid, compared to M3b (see below).
For a full description of the FLAMES/GIRAFFE analysis including the adaption of the line lists 
and the error analysis, see \citet{Kordopatis2011a,Kordopatis2011b}.

\paragraph{AMBRE-FEROS MATISSE Analysis:}
The second approach (M3b) has been designed for the analysis of the archived spectra of the 
FEROS spectrograph (AMBRE Project\footnote{performed 
under a contract between the Observatoire de la C\^{o}te d'Azur (OCA) and the 
European Southern Observatory (ESO).}); see \citet{Worley2011} for details. 

The grid of synthetic spectra for the AMBRE Project includes the full range of parameters, 
and the spectra cover the entire optical domain \citep{2012A&A...544A.126D}. 
Due to the large wavelength range no calibration of the line list was carried out. 
Comprehensive comparisons were made between the AMBRE-FEROS stellar parameters and corresponding stars in the S$^4$N library, the PAram{\`e}tres STELlaires 
(PASTEL) database \citep{Soubiran2010} and the dwarf stellar sample in \citet{Bensby2003}. For the combined sample of 178 stars from S$^4$N and 
PASTEL, the dispersions in $T_{\rm eff}$, $\log g$, and [M/H], are 150~K, 0.25~dex, and 0.12~dex respectively. For the 66 stars in \citet{Bensby2003} the dispersions in $T_{\rm eff}$, $\log g$, [M/H], and [$\alpha$/Fe] are 86~K, 0.17~dex, 0.06~dex and 0.04~dex respectively. Hence there is excellent agreement between AMBRE-FEROS and these high quality, non-automated, spectroscopic analyses.

For the AMBRE-FEROS analysis of \one\ and \two\ (M3b), as the spectra did not encompass the full AMBRE-FEROS wavelength range, only the regions between 4900--6730\AA\ were extracted ($\sim$435\AA\ in total). The spectra were then convolved to R$\sim$15\,000.  High resolution Hinkle atlases of the Sun and Arcturus \citep{2007assp.book.....W,2000vnia.book.....H} were analyzed concurrently with \one\ and \two. The parameters that were determined are \Teff~=~5783$\pm$80~K, $\log g$~=~4.30$\pm$0.15, [M/H]~=~0.01$\pm$0.10, [$\alpha$/Fe]~=~0.00$\pm$0.10 for the Sun, and \Teff~=~4306$\pm$80~K, $\log g$~=~1.80$\pm$0.15, [M/H]~=~$-$0.66$\pm$0.10, [$\alpha$/Fe]~=~0.24$\pm$0.10 for Arcturus, in excellent agreement with accepted values. 

Results for \one\ and \two\ for both approaches are listed in Tables~\ref{results1a} and \ref{results1b}.
Both analyses confirm that these stars are cool giants. For \one\ and \two\ both the M3a and M3b analyses show very good agreement in $T_{\rm eff}$ and [Fe/H] in 
comparison to the stellar parameters quoted in Table~\ref{narvaldata}. For \one\ there is an absolute difference in gravity of $\Delta \log g = 0.60$ between M3a and M3b. However both values differ by only 0.3~dex from the adopted value stated in Table~\ref{narvaldata}.

For the M3a analysis the small spectral domain of 8470--8740\AA\ at this fairly low spectral resolution is known to contain rather poor (and few) spectral signatures that are sensitive to gravity, leading to degeneracies in the derivation of the stellar parameters \citep[see][]{Kordopatis2011a}. 
Gravity problems for the \ion{Ca}{II} IR triplet arise specifically for cool dwarfs (lack of signatures) and subgiants
(\Teff-$\log g$ deceneracy). The problem is much less important for the giant stars studied in this
paper.

For the M3b analysis it was determined that the normalisation of the magnesium 
triplet at $\sim$5170\AA\ was driving the issues with the gravity determination. At these cool temperatures this is a difficult region to normalise and, as the MgIb triplet is a dominant contributor to the gravity determination, the less than ideal normalisation of this region at each normalisation/MATISSE iteration drove the automated process to less precise gravity values, in particular for \two. In the full AMBRE-FEROS analysis a three times wider wavelength interval is used which includes 
significant, and better distributed, contributors to the gravity. This reduces 
the impact of the MgIb triplet at these cool temperatures.

\subsubsection{M4} 
\label{sect:M4}
The M4 analysis was done for \one\ by K. Eriksson.
A grid of model atmospheres was calculated for the parameters given in Table~\ref{tab:modelgrids}.
All models were spherical with one solar mass.
For more details on the MARCS model atmospheres see \citet{2008A&A...486..951G}.
For each of these model atmospheres synthetic spectra were calculated, for the three wavelength regions indicated in Fig.~\ref{fig:wavelengths} with a wavelength step of 0.02~\AA\ using the line formation code BSYN. 
The spectra were then downgraded to a resolution of 80\,000.

For the atomic lines the VALD data base was used. For computing the atmosphere models, absorption data for transitions of 18 molecules were included (see Table~2 of \citealt{2008A&A...486..951G}). The molecular species included in the synthetic spectrum calculations are listed in Table \ref{tab:M4}.

\begin{table}[htbp] 
   \caption{\label{tab:M4} Molecular species used in the M4 synthetic spectrum calculations.}
   \centering
   \begin{tabular}{@{} l c c c @{}} 
      \hline\hline
      Species & 4900--5400~\AA & 6100--6800~\AA & 8400--8900~\AA  \\
      \hline
      C$_2$ & $\bullet$ &  &  \\
      CN       & $\bullet$ & $\bullet$ & $\bullet$ \\
      MgH   & $\bullet$ & $\bullet$ &  \\
      SiH    & $\bullet$  &  &  \\
      TiO    & $\bullet$ & $\bullet$   &  $\bullet$ \\
      CaH  &        & $\bullet$ &  \\
      ZrO   &        &  $\bullet$ & \\
      FeH  &       &          &  $\bullet$ \\
       \hline
   \end{tabular}
\end{table}

The flux was adjusted by a continuum scale factor determined from 100~\AA-wide regions for the different parts of the spectrum.
The synthetic spectra were then cross-correlated with the given observed spectrum for \one\ for a number of wavelength regions and a $\chi^2$-value was calculated ($\chi^2\equiv\sum_{\Delta\lambda}\frac{(m-o)^2}{o}$, where $\Delta\lambda$ is the wavelength interval, $m$ is the model spectrum, and $o$ is the observed spectrum).
The minimum $\chi^2$-value should then give the best model atmosphere for that wavelength region if the included line lists are complete and accurate enough.
The results for some representative wavelength regions are as follows.

\begin{figure*}[htbp]
   \centering
   \resizebox{\hsize}{!}{\includegraphics[angle=180]{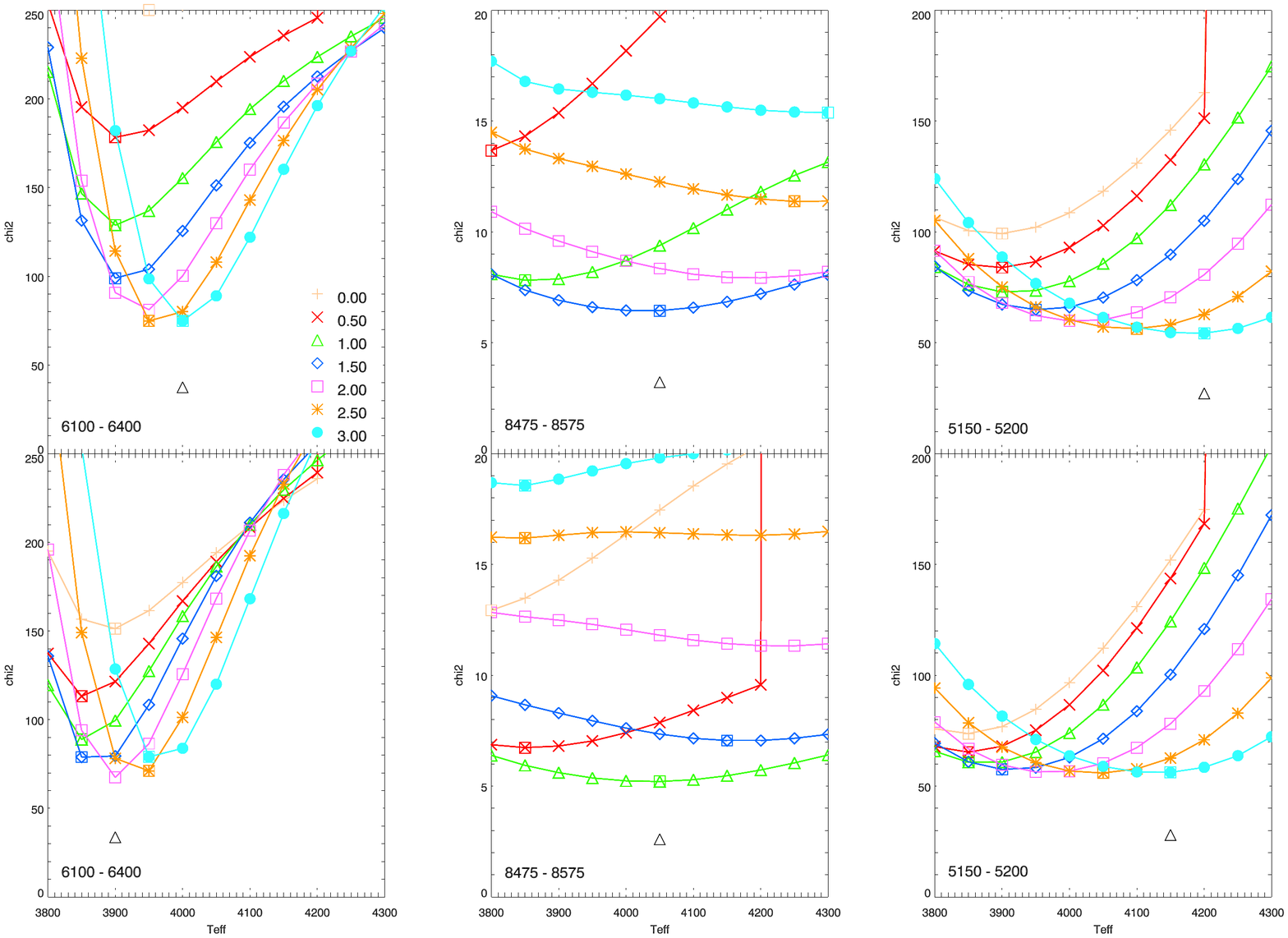}} 
   \caption{M4 analysis of \one: $\chi^2$ values as a function of \Teff\ for different \logg\ values (see legend) and for three different wavelength regions (left, middle, right).  Top row: solar abundances, bottom row: \FeH = $-$0.25.}
   \label{fig:M4chi2}
\end{figure*}

\paragraph{6100--6800 \AA:}
The region 6400--6700 \AA\ was used to set the best flux scaling of the observed spectrum. The scale factor depended to some degree on the model atmosphere used, but did not vary by more than a few percentage units.
With this scaling factor the wavelength region 6100--6400 \AA\ was investigated. The result for solar abundances can be seen in the top left panel of Fig.~\ref{fig:M4chi2}, where the $\chi^2$-values are presented as a function of \Teff\ with different curves for different \logg\ values. The best fit was obtained for an effective temperature of 4000~K or slightly less and for a \logg\ of 3.0 or less.
The same procedure for model atmospheres with a metal abundance of \FeH = $-$0.25 yielded the result shown in Fig.~\ref{fig:M4chi2} (bottom left). Note that the best fit is shifted to lower \Teff\ values and lower \logg\ values by 100~K and 1~dex, respectively.

\begin{figure}[htbp]
   \centering
   \resizebox{\hsize}{!}{\includegraphics[viewport=70 50 770 554]{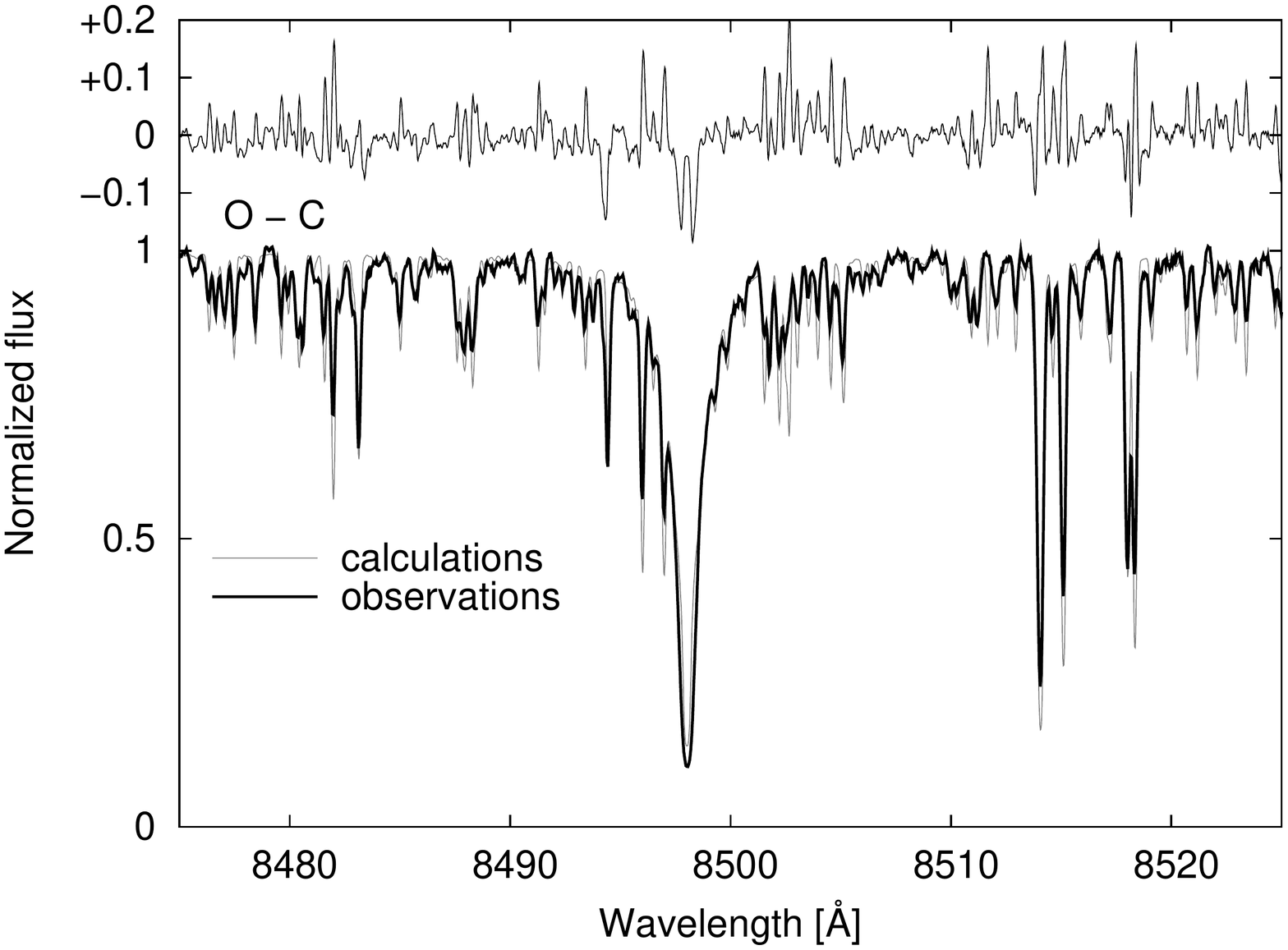}} 
   \caption{M4 analysis of \one: The best fit for the 8475--8575~\AA\ region, showing a \ion{Ca}{II} IR triplet line, for solar abundances (\Teff=4050~K, \logg=1.5). Bottom: Scaled flux, observed and calculated; top: observed minus calculated flux.}
   \label{fig:M4spec1}
\end{figure}

\paragraph{8400--8900 \AA:}
In this wavelength region, very interesting from a Gaia point of view, we find the \ion{Ca}{II} infrared triplet, which is strong in cool giant stars. The $\chi^2$ plots for the 8475--8575~\AA\ wavelength interval for solar abundances and for \FeH = $-$0.25 can be found in Fig.~\ref{fig:M4chi2} (top and bottom center, respectively). The best fit for solar abundances is shown in Fig.~\ref{fig:M4spec1}.
This region is not very sensitive to the effective temperature, and quite sensitive to the surface gravity. Using a lower metal abundance resulted in a lower surface gravity.

\begin{figure}[htbp]
   \centering
   \resizebox{\hsize}{!}{\includegraphics[viewport=70 50 770 554]{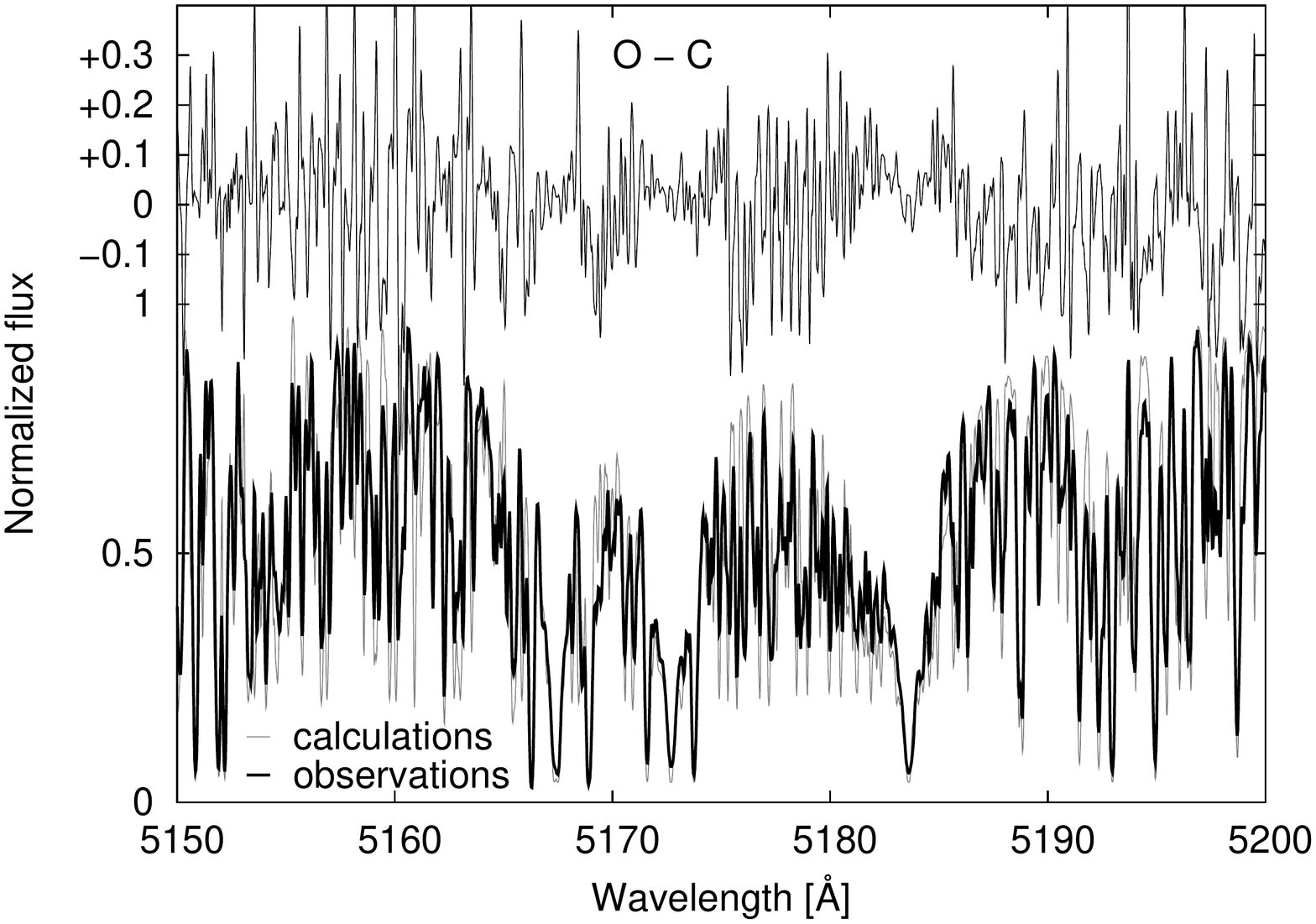}} 
   \caption{M4 analysis of \one: The best fit for the 5150--5200~\AA\ region, showing the \ion{Mg}{I}b triplet lines, for sub-solar abundances (\FeH = $-$0.25; \Teff=4150~K, \logg=3.0). Bottom: Scaled flux, observed and calculated; top: observed minus calculated flux.}
   \label{fig:M4spec2}
\end{figure}

\paragraph{4900--5400 \AA:}
This wavelength region has a much weaker temperature- and gravity sensitivity, as 
can be seen in Fig.~\ref{fig:M4chi2} (right) for the interval 5150--5200~\AA\ containing the \ion{Mg}{I}b triplet lines.
The ``best'' fit for the \FeH = $-$0.25 models is shown in Fig.~\ref{fig:M4spec2}.

In conclusion, within the M4 analysis, the 6100--6400~\AA\ region, with its 
temperature-sensitive TiO bands, was best suited for estimating the effective temperature for \one.
The 8500~\AA\ region was best suited for estimating the surface gravity.
The solutions for the two adopted metallicity values are given in Table~\ref{results1a}.
It should be noted that the M4 analysis did not result in a unique set of best-fit parameters: the temperature-sensitive region gave a higher \logg\ than the gravity-sensitive region, and vice versa.
The blue-green region (5150--5200~\AA) gave both a higher \Teff\ and a higher \logg\ than the other two regions.
Most of the other tested wavelength intervals behaved similarly to the blue-green region, i.e. showing very broad minima at rather high \Teff\ and \logg\ in the $\chi^2$ plots. These were four 50--100~\AA\ wide intervals between 4900 and 5400~\AA, and a similar number of intervals in the region 8400--8900~\AA, which did not contain any of the \ion{Ca}{II} triplet lines.

An analysis of individual (e.g. \ion{Fe}{I}) lines to derive an estimate of [Fe/H] was not attempted.  Neither was a $\chi^{2}$ analysis in the metallicity dimension (although one can note that these values are on average somewhat smaller for the $-$0.25 models in Fig.\,\ref{fig:M4chi2}).
A further conclusion is that this procedure is to some degree dependent on the scale factors, i.e. it is important to have a good fit to the stellar continuum (which is hard to do, due to the effect of molecular bands).

\subsubsection{M5}
\label{sect:M5}
This part of the experiment was led by C. Abia.
The M5 analysis focused on the coolest stars of the experiment, \three\ and \four, because 
the M5 team members are more familiar with the spectral region to be analyzed in these stars (the $H$-band) where, furthermore, telluric lines are absent.

The original atomic line list was taken from the VALD database (version 2009). For the molecules, C$_2$ lines are from \citet{Wahlin05}, CO lines 
come from \citet{Goorvitch94}, an the CN and CH lines were assembled from the best 
available data as described in \citet{Hill02} and \citet{Cayrel04}.  The molecular lists also include lines of  OH, TiO, 
CaH, SiH, FeH  and  H$_2$O taken from the HITRAN database \citep{Rothman05}.
First, the atomic and molecular lines were calibrated by obtaining astrophysical $gf$ values using the high resolution Solar \citep{Livingston91} and Arcturus \citep{Arcturusatlas} spectra in the 1.5$\mu$m region. In both analyses MARCS atmosphere models \citep{2008A&A...486..951G} were used, with parameters \Teff/\logg/\FeH=5777/4.4/0.0 and 4300/1.5/$-$0.5 for the Sun and Arcturus, respectively. The reference abundances were those by \citet{2009ARA&A..47..481A} for the Sun, and those derived by \citet{Peterson93} for Arcturus. 
The 1.5$\mu$m region is dominated by CO, OH, CN and C$_2$ absorptions (in order of decreasing
importance). The main features were calibrated on the Arcturus spectrum. Synthetic spectra in LTE were computed using the TURBOSPECTRUM V9.02 code described in \citet{Alvarez1998}.
The theoretical spectrum was convolved with a Gaussian function with a FWHM of $\sim$300~m{\AA}.
The stellar parameters were estimated from the available photometry using the recent photometric calibrations by \citet{Worthey11} (see Tables~\ref{results2a} and \ref{results2b}).

\paragraph{\three:}
A spherical (1~M$_\odot$) model atmosphere with parameters determined from the photometric colours was taken from the grid of \citet{2008A&A...486..951G}.
The fit (by eye) with this set of parameters to the observed spectrum was quite good in the full spectral range.
The simultaneous fit to some OH, CO and CN lines allowed an estimate of the CNO abundances, namely C/N/O$=8.37/7.90/8.70$, i.e. C/O$=0.46$. A rough estimate of the metallicity from fits to some metallic lines in the region was compatible with [Fe/H] $\approx 0.0\pm 0.2$. Finally, according to the line list used, 
some CO lines are sensitive to variations in the $^{12}$C/$^{13}$C ratio. Several fits unambiguously resulted in a carbon isotopic ratio lower than the solar value (89). However, due to the weakness of the $^{13}$CO lines only a lower limit was derived, $^{12}$C/$^{13}$C$>25$.

\paragraph{\four:}
With a model atmosphere with parameters determined from the photometric colours it was impossible to fit the observed spectrum. In particular, the predicted intensity of the CO lines was lower than observed. In this case, the stellar parameters were estimated through a $\chi^2$ test comparing observed and theoretical spectra computed for several choices of \Teff/\logg/\FeH\ ($\chi^2(i)\equiv\frac{(c(i) - o(i))^2}{c(i)+o(i)}$, where $c$ is the computed spectrum, and $o$ is the observed spectrum).

The best fit was obtained with a spherical 2~M$_\odot$ MARCS model (K. Eriksson, private 
communication) of parameters \Teff/\logg/\FeH = 3500/0.0/0.0 ($\chi^2<1.5$ over 20--30 {\AA} regions). With these stellar parameters the corresponding photometric colours according to \citet{Worthey11} are instead $(V-I)$ = 2.49, $(J-K)$ = 1.18, and $(V-K)$ = 5.15.  In the next step, the CNO abundances were estimated in a similar way to \three, resulting in C/N/O = $8.52/7.88/8.68$, i.e. C/O$ = 0.69$, and a lower limit of 35 for the $^{12}$C/$^{13}$C. The estimated metallicity 
was also compatible with [Fe/H]$ = 0.0$. The microturbulence parameter was set to a fixed value (see Table~\ref{results2b}). Note that the C/O ratio in the model atmosphere used (close to 1) was not consistent with the C/O ratio derived. The fact that this M-type star has some carbon enrichment may explain why a theoretical spectrum computed with the stellar parameters giving the best fit to the photometric colours does not fit the observed spectrum. In cool carbon enhanced stars, the actual C/O ratio is a critical parameter which determines the structure of the atmosphere affecting also the photometric colours. 
The results of the M5 analysis for both stars are summarized in Table~\ref{results2b}.

\subsubsection{M6}
\label{sect:M6}
The M6 team, consisting of T.~Merle and F.~Th{\'e}venin, did not try to reproduce molecular lines and 
chosen to work with only \one\, 
which has less affected spectra.  
For the M6 analysis, the photometry information was used to get starting values for \Teff, \logg, and [Fe/H]. With the proposed approximate colours and 
using the theoretical calibrations by \citet{2000AJ....119.1448H}, the \Teff\ and \logg\ values given in Table~\ref{results1a} were obtained.
For K giant stars, a good approximation is to take a mass of 1~M$_\odot$.
The spectroscopic analysis started with the selection of \ion{Fe}{I} and \ion{Fe}{II} lines. Care was taken to remove strong lines, multiplets with 
no dominant log $gf$ ($>$ 10\%) and polluted or unclear line shapes in the observed spectrum. 21 \ion{Fe}{I} lines were chosen in the near-IR range. 
Unfortunately, all \ion{Fe}{II} lines in the near-IR part were too weak 
to use for a fit.  Then, six \ion{Fe}{II} lines were selected 
in the $[6100-6800]$\AA\ range.  For each line, radiative damping was estimated using the Uns\"old formula; hydrogen elastic collisional damping 
came from ABO theory \citep{1995MNRAS.276..859A,1997MNRAS.290..102B,1998MNRAS.296.1057B} when available (e.g. the six \ion{Fe}{II} lines), otherwise, 
classical Uns\"old theory was used with an enhancement factor of 1.5 (e.g. the 21 \ion{Fe}{I} lines).
The line list is given in Table~\ref{tab:Merle}.

\begin{table} 
\center
\caption{Iron line selection for the M6 stellar parameter determination (Sect.\,\ref{sect:M6}). 
For each line, the wavelength $\lambda$, the lower level excitation potential $E_{\rm low}$, the $gf$ value, and the source for the $gf$ value are given.}
\label{tab:Merle}\begin{tabular}{llrl}
\hline\hline
 $\lambda$ & $E_{\rm low}$ & log($gf$) & Source$^a$ \\
 (nm) & (eV) & & \\
\hline
\ion{Fe}{I} &     &             & \\
843.957 & 4.549 &    $-0.587$  & K07 \\
847.174 & 4.956 &    $-1.037$  & K07 \\
852.667 & 4.913 &    $-0.760$  & BWL \\
852.785 & 5.020 &    $-1.625$  & K07 \\
857.180 & 5.010 &    $-1.414$  & K07 \\
859.295 & 4.956 &    $-1.066$  & K07 \\
859.883 & 4.386 &    $-1.089$  & BWL \\ 
860.105 & 5.112 &    $-1.577$  & K07 \\ 
861.394 & 4.988 &    $-1.247$  & K07 \\ 
863.241 & 4.103 &    $-2.341$  & K07 \\ 
869.945 & 4.955 &    $-0.380$  & BWL \\ 
872.914 & 3.415 &    $-2.871$  & K07 \\ 
874.743 & 3.018 &    $-3.176$  & K07 \\ 
879.052 & 4.988 &    $-0.586$  & BWL \\ 
879.807 & 4.985 &    $-1.895$  & K07 \\ 
881.451 & 5.067 &    $-1.793$  & K07 \\ 
881.689 & 4.988 &    $-2.203$  & K07 \\ 
883.402 & 4.218 &    $-2.558$  & K07 \\ 
884.674 & 5.010 &    $-0.777$  & K07 \\ 
889.140 & 5.334 &    $-1.108$  & K07 \\ 
\hline
\ion{Fe}{II} &     &  &\\
611.332 &  3.221 &   $-4.230$ & RU \\
614.774 &  3.889 &   $-2.827$ & RU \\
641.692 &  3.892 &   $-2.877$ & RU \\
643.268 & 10.930 &   $-1.236$ & RU \\
645.638 &  3.903 &   $-2.185$ & RU \\
651.608 &  2.891 &   $-3.432$ & RU \\
\hline
\end{tabular}
\tablefoot{
\tablefoottext{a}{K07 ... \citet{KuruczWeb}, BWL ... \citet{BWL}, RU ... \citet{RU}}
}
\end{table}

The final values of the stellar parameters for the M6 analysis were constrained by the method of ionization equilibria between \ion{Fe}{I} and 
\ion{Fe}{II} and excitation equilibria for \ion{Fe}{I}. For this analysis, the MOOG2009 code \citep{1973PhDT.......180S} with spherical MARCS model 
atmospheres \citep{2008A&A...486..951G} and atomic line lists from VALD were used. The inconsistency between spherical atmosphere and plane parallel 
radiative transfer is negligible \citep{2006A&A...452.1039H}. The chemical composition was from \citet{2007SSRv..130..105G}. For each model atmosphere 
the abundance was determined for each line by fitting the observed spectrum while varying the iron 
abundance in steps of 0.1~dex around the abundance value
 adopted for the model atmosphere. The continuum level was determined using a local scaling factor.
Then, [\ion{Fe}{I}/H] was plotted as a function of the lower level excitation potential $E_i$ of the lines, and using linear regression, the model 
with the smallest slope was selected.
For the ionization equilibrium, the mean [\ion{Fe}{I}/H] and [\ion{Fe}{II}/H] was compared for each model, and the model with the smallest abundance 
difference was selected.
The best model atmosphere had parameters as given in Table~\ref{results1a}. The uncertainties come from the large steps in the stellar parameters (200\,K for \Teff\ , 1 for \logg\ , and 0.1 dex in 
metallicity)
taken to perform the analysis. With these large steps, the same atmospheric parameters were obtained both from the slope 
of [Fe/H] vs $E_i$ and from the ionization equilibrium.

The radial velocity was determined using 21 weak and relatively unblended lines in the $[6100-6800]$ and $[8400-8900]$ \AA\ domains. The lines were synthesized using the MOOG2009 LTE code and shifted in steps of 0.5~km\,s$^{-1}$ for all the lines selected.

\subsection{ATLAS model atmospheres (A)}
\subsubsection{A1}
\label{sect:A1}
This working group consisted of G. Wahlgren and R. Norris.
Using synthetic photometric colours from \citet{2005AaA...442..281K}, initial estimates of temperature, 
gravity, and metallicity were made. With a least sum squares approach, the colours of the provided stars were compared to the colours in Ku{\v c}inskas' table, and the parameters of the best matching synthetic colours were interpolated to the given stars (see Tables~\ref{results1a} to \ref{results2b}).

Using these estimates of stellar parameters, appropriate ATLAS9 model atmospheres were selected from the grid of \citet{2003IAUS..210P.A20C}, and synthetic spectra were created with SYNTHE \citep{1993KurCD..18.....K}. ATLAS is an LTE, plane-parallel model atmosphere code. Molecules from the Kurucz suite of line lists \citep[][hydrides, CN, CO, TiO, SiO, H$_2$O] {1995KurCD..23.....K} as well as other molecules were included in the statistical equilibrium calculations. The use of 
the gravities and metallicities obtained from broadband colours produced synthetic spectra which did not match the workshop spectra (at lines known to be sensitive to gravity and metallicity). Therefore, specific spectral lines as indicators of gravity and metallicity were used.
The A1 analysis was continued for the K-type stars \one\ and \three, because the contributing authors were not convinced that they had sufficient molecular opacities for late M-type stars.

Atomic line data in this analysis come from Kurucz (approximately 2008). 
Tables~\ref{tab:Wahlgren1} and \ref{tab:Wahlgren2} list the lines used in the abundance determination for \one\ and \three, respectively. Lines from neutral species only were used.  

\paragraph{\one:}
Synthetic colours suggested a cool, metal deficient giant or supergiant. Comparison to UVES spectra showed close matches with both HR~7971 (K3~II/III) and HR~611, which has ben classified as 
M0.8~III from narrow-band TiO/CN photometry \citep{Wing78}. 
The infrared \ion{Ca}{ii} triplet (8498, 8542, 8662~{\AA}) and the \ion{Fe}{i} 
calibration lines at 8327, 8468, 8514, 
and 8689~{\AA} 
noted in \citet{1945ApJ...101..265K} were used to determine \logg\ of 1.5 in 
a temperature range of 3900 -- 4000~K. Deficiencies in iron (see Table~\ref{results1a}) and the $\alpha$ elements were of particular note ([Si/H]=$-0.45\pm0.03$, [Ca/H]=$-0.90\pm0.15$, [Ti/H]=$-0.38\pm0.03$).
The lines used in these measurements are listed in Table~\ref{tab:Wahlgren1}.

\begin{table} 
\center
\caption{Lines used in the A1 analysis to determine abundances for \one. 
For each line, the atomic number $Z$, the wavelength $\lambda$, the $gf$ value, 
the lower level excitation potential $E_{\rm low}$, and the line abundance are given. \label{tab:Wahlgren1}}
\begin{tabular}{llllrl}
\hline\hline
{$Z$} & {$\lambda$} & {log($gf$)} & {Source\tablefootmark{a}} & \multicolumn{1}{l}{$E_{\rm low}$} & {Abundance}\\
 & {(nm)} &  &  & \multicolumn{1}{l}{[cm$^{-1}$]} & {(log(N$_{H}$)=12.00)}\\
\hline
14 & 568.448 & $-$1.650 & GARZ & 39955.053 &  6.95\\
14 & 569.043 & $-$1.870 & GARZ & 39760.285 &  7.10\\
14 & 570.110 & $-$2.050 & GARZ & 39760.285 &  7.08\\
14 & 577.215 & $-$1.750 & GARZ & 40991.884 &  7.04\\
14 & 579.307 & $-$2.060 & GARZ & 39760.285 &  7.15\\ 
20 & 558.875 & $+$0.210 & NBS  & 20371.000 &  4.83\\
20 & 559.011 & $-$0.710 & NBS  & 20335.360 &  5.23\\
20 & 559.446 & $-$0.050 & NBS  & 20349.260 &  4.81\\
20 & 560.128 & $-$0.690 & NBS  & 20371.000 &  5.13\\
20 & 615.602 & $-$2.200 & NBS  & 20335.360 &  5.88\\
20 & 645.560 & $-$1.350 & NBS  & 20349.260 &  5.48\\
20 & 817.329 & $-$0.546 & K88  & 39349.080 &  5.90\\
20 & 820.375 & $-$0.848 & K88  & 36575.119 &  5.90\\
20 & 829.218 & $-$0.391 & K88  & 39340.080 &  5.80\\
22 & 546.047 & $-$2.880 & SK   & 386.874   &  4.55\\
22 & 556.271 & $-$2.870 & MFW  & 7255.369  &  4.55\\
22 & 802.484 & $-$1.140 & MFW  & 15156.787 &  4.45\\
22 & 806.825 & $-$1.280 & MFW  & 15108.121 &  4.65\\
22 & 841.229 & $-$1.483 & MFW  & 18482.860 &  4.55\\
22 & 867.527 & $-$1.669 & MFW  & 8602.340  &  4.50\\
22 & 868.299 & $-$1.941 & MFW  & 8492.421  &  4.50\\
22 & 869.233 & $-$2.295 & MFW  & 8436.618  &  4.70\\
22 & 873.471 & $-$2.384 & MFW  & 8492.421  &  4.70\\
26 & 552.554 & $-$1.330 & FMW  & 34121.580 &  6.70\\
26 & 552.890 & $-$2.020 & FMW  & 36079.366 &  7.30\\
26 & 552.916 & $-$2.730 & FMW  & 29371.811 &  7.10\\
26 & 553.275 & $-$2.150 & FMW  & 28819.946 &  6.70\\
26 & 554.394 & $-$1.140 & FMW  & 34017.098 &  6.70\\
26 & 555.798 & $-$1.280 & FMW  & 36079.366 &  6.65\\
26 & 556.021 & $-$1.190 & FMW  & 35767.561 &  6.75\\
26 & 829.351 & $-$2.126 & K88  & 38678.032 &  6.90\\
26 & 832.705 & $-$1.525 & FMW  & 17726.981 &  6.70\\
\hline
\end{tabular}
\tablefoot{
\tablefoottext{a}{ References as listed in \citet{1995KurCD..23.....K}: http://kurucz.harvard.edu/LINELISTS/LINES/gfall.ref}
}
\end{table}

Of the $\alpha$ elements, the abundance of calcium resulting from the A1 analysis is particularly low. For the six lines below 8000~\AA, which have a lower excitation energy close to 20\,000~cm$^{-1}$, weaker $gf$-values correlated to higher abundances. For example, the 5589~{\AA} and 5601~{\AA} lines have the same lower excitation energy but have 
different $gf$-values. Of these, 5601~{\AA} with a log($gf$)=$-$0.690~dex suggests an abundance of 5.13 whereas 5589~\AA with log($gf$)=0.210~dex suggests an abundance of 4.83.
This trend does not exist for the three calcium lines above 8000~{\AA} which were included in the abundance determination. Although each of these lines has a more energetic lower excitation energy, each suggests a higher abundance than the shorter 
wavelength lines with similar $gf$-values. These three lines use \citet{KuruczTrans} calculated $gf$-values as opposed to the NBS \citep{1969atp..book.....W} values used for the shorter wavelength lines.
The \citet{KuruczWeb} calculated values for the four lines below 6000~{\AA} have larger $gf$-values than NBS reports. Presuming that this correlation between calculated and experimental values would continue for longer wavelength lines, this indicates that for the three lines over 8000~\AA, the reported $gf$-value is an upper bound.
If the correlation between higher abundance and low $gf$-value in the shorter wavelength values is not entirely the result of poor atomic data; it is likely the result of non-LTE effects, to which calcium lines are particularly sensitive (for a non-LTE analysis of a different set of calcium lines see Section~\ref{sect:nlte}).
The results of the A1 analysis for \one\ are given in Table~\ref{results1a}.

\paragraph{\three:}
Synthetic colours suggested a metal deficient giant of effective temperature 4300--4400~K. Comparison of the workshop spectrum with synthetic spectra produced with parameters obtained from broadband colours showed that while the \logg\ and temperature indicated 
by synthetic colours fit, the metallicity was closer to the solar value than the synthetic colours suggested.
The A1 team determined abundances of carbon, nitrogen, and oxygen for \three\ by fitting molecular lines. There are strong OH and CN lines present, as well as a CO bandhead, in the wavelength region 
of the experiment. Unblended OH lines were used as indicators of the oxygen abundance and CN and CO lines as indicators of carbon and nitrogen abundances. 
Several iron lines, listed in Table~\ref{tab:Wahlgren2}, served for the determination of the iron abundance.
Despite hints from synthetic colours that the star was metal deficient, carbon, nitrogen, oxygen, and iron were all enhanced. 
Lines of some other elements, though not analyzed, suggested similarly enhanced abundances. The star was found to be oxygen rich.
The results of the A1 analysis for \three\ are summarized in Table~\ref{results2a}.

\begin{table} 
\center
\caption{\ion{Fe}{I} lines used in the A1 analysis to determine iron abundances for \three.
For each line, the wavelength $\lambda$,  the $gf$ value, the lower level excitation potential $E_{\rm low}$, 
and the line abundance are given. \label{tab:Wahlgren2}}
\begin{tabular}{lllll}
\hline\hline
{$\lambda$} & {log($gf$)} & {Source\tablefootmark{a}} & {$E_{\rm low}$} & {Abundance}\\
{(nm)} &  &  & {[cm$^{-1}$]} & {[log(N$_{H}$)=12.00]}\\
\hline
1549.034 & $-$4.574 & O    & 17726.987  &  7.85\\
1558.826 & $+$0.323 & K94  & 51359.489  &  7.70\\
1561.115 & $-$3.822 & K94  & 27523.001  &  7.65\\
1564.851 & $-$0.714 & K94  & 43763.977  &  7.90\\
1565.283 & $-$0.476 & K94  & 50377.905  &  7.80\\
\hline
\end{tabular}
\tablefoot{
\tablefoottext{a}{References as listed in 
\citet{1995KurCD..23.....K}: http://kurucz.harvard.edu/LINELISTS/LINES/gfall.ref}
}
\end{table}

\subsubsection{A2} 
\label{sect:A2}
The A2 team (consisting of J. Maldonado, A. Mora, and B. Montesinos) decided to work only with the optical region of the spectrum, and therefore, only \one\ and \two\ were analyzed.

To determine the radial velocity, the spectra of the target stars were cross-correlated against spectra of several radial velocity standards \citep{2010A&A...521A..12M}.  Cross-correlation was performed with the IRAF\footnote{IRAF is distributed by the National Optical Astronomy Observatory, which is operated by the Association of Universities for Research in Astronomy, Inc., under contract with the National Science Foundation.} task {\it fxcor}. Spectral ranges with prominent telluric lines were excluded from the cross-correlation.

A grid of synthetic spectra was calculated for the parameter range given in Table~\ref{tab:modelgrids}.
To compute the spectra, team A2 used ATLAS9 model atmospheres and the SYNTHE code \citep{1993KurCD..13.....K,1993KurCD..18.....K}, adapted to work under the Linux platform by \citet{2004MSAIS...5...93S} and \citet{2005MSAIS...8...61S}\footnote{http://wwwuser.oat.ts.astro.it/atmos/Download.html}. 
Line data are from the Kurucz web site (2005 version).
The following molecules are included: C$_{2}$, CH, CN, CO, H$_{2}$, MgH, NH, OH, SiH, SiO.
Atomic line data are listed in Table~\ref{A2_line_list}.  The new opacitiy distribution functions from \cite{2003IAUS..210P.A20C} were used. A mixing length parameter of 1.25 was used.
All spectra were computed with a resolution of 300\,000,

To obtain the ``best parameters'' of the target stars, A2 compared the equivalent widths (EWs) of a sample of spectral lines measured in the target stars with the EWs of the same lines measured in each synthetic spectrum. To compile a list of ``well behaved'' lines, the A2 team
calculated synthetic spectra for two stars with accurately known stellar parameters (namely, the Sun, G2V, and Procyon, F5IV) 
and compared them with high-resolution observed spectra. 
Relatively 
isolated lines, with a reasonable stretch 
of flat continuum around the limits, were selected to estimate the EWs with high accuracy.
Since in all cases the synthetic lines fit the observed spectrum, it can be expected that the atomic parameters are fairly reliable. The list of 
lines used is given in Table~\ref{A2_line_list}. 
The abundance ratios of the individual elements relative to each other were kept
fixed to the solar ratios.
Final parameters were obtained by using a reduced $\chi^{2}$ fitting method. Lines with large EWs ($>$100~m\AA) were not used. The results are listed in Tables~\ref{results1a} and \ref{results1b}.

The chosen method clearly can be further improved to obtain more accurate parameters. 
The first problem identified during the A2 analysis was the line selection. Although an
attempt was made 
to compute a list of well behaved lines, it is clear that such a list depends on the spectral type of the target stars. The selection was started using as reference the Sun and Procyon, but some well behaved lines in the Sun were not well behaved in Procyon and vice versa. In addition, since the program 
stars of this experiment were cooler than the chosen reference stars, the selection had to be reviewed several times in order to avoid blended lines or lines not present in the program stars.

\begin{table*} 
\centering
\caption{Lines used by the A2 participants (Sect.\,\ref{sect:A2}). 
For each line, the species code, the wavelength $\lambda$ and the $gf$ value are given (taken from the line lists of Kurucz). Lines included in the final fit have been marked with the symbol $^{\star}$.
The species code follows Kurucz's notation, e.g. for iron lines (atomic
number 26): 26.00 = \ion{Fe}{I}, 26.01=\ion{Fe}{II}, etc. References as listed in \citet{1995KurCD..23.....K}\tablefootmark{a}. Furthermore, the table gives the measured equivalent widths of the
lines in the two benchmark stars.}
\label{A2_line_list}
\begin{tabular}{lcccrr|lcccrr}
\hline\hline\noalign{\smallskip}
Ion   & $\lambda$ & $\log$(gf) & Source & EW Star 1 & EW star 2 &  Ion   & $\lambda$ & $\log$(gf) & Source & EW star 1 & EW star 2\\
      &   (nm)   &            &        & (mA) & (mA) &         &   (nm)   &            &    & (mA) & (mA)    \\
\hline\noalign{\smallskip}
26.00$^{\star}$  &  490.514 &  -2.050 &  FMW &  73.63 & 30.33 & 26.00           &  536.162 &  -1.430 & FMW & 269.68 & 225.63\\
26.01            &  492.393 &  -1.320 &  FMW &  -- & -- & 26.00$^{\star}$ &  537.371 &  -0.860 & FMW & 82.82 & 97.37\\
26.00            &  492.477 &  -2.220 &  FMW &  107.96 & 183.82 & 26.00           &  537.958 &  -1.480 & FMW & 102.03 & 113.47\\
24.00            &  493.634 &  -0.340 &  MFW &  110.10 & 82.99 & 26.00           &  538.634 &  -1.770 & FMW & -- & --\\
26.00$^{\star}$  &  496.258 &  -1.290 &  FMW &  52.201 & -- & 26.00$^{\star}$ &  538.948 &  -0.410 & FMW & 20.50 & 85.97\\
28.00            &  501.094 &  -0.870 &  FMW &  -- & -- & 28.00           &  539.233 &  -1.320 & FMW & -- & 4.94\\
26.00            &  504.422 &  -2.150 &  FMW & 169.70 & 181.76 & 26.00            &  612.025 &  -5.950 & FMW & 112.24 & 113.17\\
26.00            &  504.982 &  -1.420 &  FMW & -- & -- & 26.00$^{\star}$  &  615.938 &  -1.970 & FMW & 26.72 & 40.58\\
26.00$^{\star}$  &  505.465 &  -2.140 &  FMW & 25.21 & 25.79 & 26.00           &  622.674 &  -2.220 & FMW & 23.82 & 6.31\\
22.00$^{\star}$  &  506.406 &  -0.270 &  MFW & 2.45 & 2.50 & 26.01$^{\star}$ &  636.946 &  -4.253 & K88 & -- & --\\
26.00$^{\star}$  &  508.334 &  -2.958 &  FMW & 1.16 & 11.24 & 26.00$^{\star}$  &  639.254 &  -4.030 & FMW & 64.58 & 67.31\\
26.00            &  509.078 &  -0.400 &  FMW & -- & 4.39 & 21.01            &  660.460 &  -1.480 & MFW & 142.65 & 127.39\\
28.00$^{\star}$  &  509.441 &  -1.080 &  FMW & 23.30 & 22.79 & 26.00$^{\star}$  &  672.536 &  -2.300 & FMW & 61.93 & 8.42\\
26.00$^{\star}$  &  512.735 &  -3.307 &  FMW & 72.89 & 74.44 & 26.00$^{\star}$  &  848.198 &  -1.647 & K94 & 39.32 & 30.20\\
26.00$^{\star}$  &  514.174 &  -2.150 &  FMW & 73.17 & 91.99 & 14.00$^{\star}$  &  850.222 &  -1.260 & KP & 14.75 & 1.58\\
26.00$^{\star}$  &  514.373 &  -3.790 &  FMW & 27.32 & 28.61 & 26.00            &  851.510 &  -2.073 & O & 152.76 &  132.81\\
22.00$^{\star}$  &  514.547 &  -0.574 &  MFW & -- & -- & 26.00            &  858.225 &  -2.133 & O  & 146.53 & 141.05 \\
26.00$^{\star}$  &  515.191 &  -3.322 &  FMW & 83.65 & 104.74 & 26.00$^{\star}$  &  859.295 &  -1.083 & K94 & 83.14 & 81.46\\
26.00$^{\star}$  &  515.906 &  -0.820 &  FMW & 74.80 & 80.40 & 14.00$^{\star}$  &  859.596 &  -1.040 & KP & 42.03 & 38.56\\
26.00            &  518.006 &  -1.260 &  FMW & -- & -- & 14.00$^{\star}$  &  859.706 &  -1.370 & KP & 60.52 & 47.79\\
26.00$^{\star}$  &  518.791 &  -1.260 &  FMW & 56.57 & 37.12 & 26.00$^{\star}$  &  859.882 &  -1.088 & O & 71.36 & 59.76 \\
26.00            &  519.495 &  -2.090 &  FMW & 73.31 & 113.68 & 26.00$^{\star}$  &  860.707 &  -1.463 & K94 & 48.49 & 49.34\\
26.00$^{\star}$  &  519.547 &   0.018 &  K94 & 85.83 & 21.61 & 26.00            &  861.180 &  -1.900 & FMW & 217.20 & 226.04\\
22.01            &  521.154 &  -1.356 &  K88 & -- & -- & 26.00            &  862.160 &  -2.321 & O & 128.62 & 135.64 \\
22.00$^{\star}$  &  521.970 &  -2.292 &  MFW & 174.09 & 173.95 & 26.00$^{\star}$  &  867.474 &  -1.850 & FMW & 46.99 & 25.25\\
26.00            &  522.985 &  -0.241 &  K94 & -- & 26.94 & 14.00            &  868.635 &  -1.200 & KP & -- & 4.67\\
26.01            &  523.463 &  -2.050 &  FMW & -- & -- & 26.00            &  868.862 &  -1.212 & FMW & 351.04 & 340.18\\
26.00            &  524.250 &  -0.840 &  FMW & 114.71 & 172.90 & 26.00$^{\star}$  &  869.870 &  -3.433 & K94 & 86.81 & 87.78\\
26.01$^{\star}$  &  525.693 &  -4.250 &  K88 & 57.75 & -- & 26.00$^{\star}$  &  869.945 &  -0.380 & O & 77.67 & 69.31\\
26.00            &  526.331 &  -0.970 &  FMW & 129.66 & 135.07 & 26.00$^{\star}$  &  871.039 &  -0.555 & K94 & 87.34 & 86.76\\
24.00$^{\star}$  &  528.718 &  -0.907 &  MFW & 61.55 & 19.62 & 14.00$^{\star}$  &  872.801 &  -0.610 & KP & 24.91 & 22.03\\
24.00            &  529.670 &  -1.400 &  MFW & 214.13 & 203.87 & 26.00$^{\star}$  &  872.914 &  -2.951 & K94 & 91.38 & 96.13\\
27.00$^{\star}$  &  535.205 &   0.060 &  FMW & 62.30 &  50.28 &               &          &         &  & & \\
\noalign{\smallskip}\hline\noalign{\smallskip}
\end{tabular}
\tablefoot{
\tablefoottext{a}{http://kurucz.harvard.edu/LINELISTS/LINES/gfall.ref}
}
\end{table*}

This is related to the second problem, namely how to measure EWs.  Although EWs can be measured ``by hand'' (for example using the IRAF task {\it splot}), this is not feasible even for this experiment with only two target stars, since more than six hundred synthetic spectra need to be analyzed. The A2 team developed a code which performs an integration around the center of each line using a fixed width.  This could definitely 
be improved. Ideally, the code should be ``intelligent enough'' to decide on the width of the integration band according to the line profile.  An alternative option, on which the A2 team is still working, is to compute EWs using model atmospheres 
and an abundance computation program such as WIDTH9 \citep{2005MSAIS...8...44C}.

\subsubsection{A3} 
\label{sect:A3}
For A3, H. Neilson attempted to determine the effective temperature, gravity, iron abundance, and carbon-to-oxygen ratio for \three\ and \four\ based on the broad-band colours and spectra provided. Model stellar atmospheres and synthetic spectra were computed using Fortran 90/95 versions of the ATLAS code \citep{Lester2008}. The new versions of the code can compute model stellar atmospheres assuming either plane-parallel or spherically symmetric geometry and either opacity distribution functions or opacity sampling.
\cite{Lester2008} demonstrated that model stellar atmospheres computed with this code predict temperature structures consistent with plane-parallel ATLAS9 and ATLAS12 models  as well as spherically symmetric PHOENIX \citep{Hauschildt1999} and MARCS \citep{2008A&A...486..951G} models.
Furthermore, \cite{Neilson2008} showed that spherical model atmospheres predict intensity distributions that fit interferometric observations of red giant stars \citep{Wittkowski2004, Wittkowski2006b, Wittkowski2006a} with center-to-limb intensity profiles from model atmospheres, and 
they determined stellar parameters consistent with results using ATLAS9 and PHOENIX models.

For the A3 analysis, models were computed assuming plane-parallel 
geometry and using the opacity distribution functions to minimize computing time. 
Derivation of stellar parameters for the two stars was done in the following manner.  First, \emph{synthetic colours} were computed from a grid of model atmospheres spanning a range in \Teff\ of 3000 to 8000~K, and in \logg\ of 0 to 3 with solar metallicity. 
Comparing the synthetic colours to the given colours, values for \Teff\ and $\log g$ were estimated.
Next, a new grid of stellar model atmospheres and \emph{synthetic spectra} for a range of \Teff\ and $\log g$ about the preliminary estimates was computed such that $\Delta$\Teff\ = $\pm 400$~K and $\Delta \log g = \pm 1$, while also varying the iron abundance. 
Line data were taken from Kurucz database\footnote{http://kurucz.harvard.edu}.
For each synthetic spectrum, a $\chi^2$-fit was computed and a new best-fit \Teff, \logg, and [Fe/H] was determined. Using these values, a new grid was computed, varying \Teff, \logg, and the silicon abundance. A new value for \Teff\, $\log g$, and the silicon abundance was found and the process was repeated for oxygen, carbon and calcium.

The best-fit stellar parameters for \three\ and \four\ resulting from the A3 analysis can be found in Tables~\ref{results2a}
and \ref{results2b}.
The results for \four\ are consistent with the parameters given in Table~\ref{star34}. However, the results for \three\ do not agree.  This disagreement is due to the method which was employed.  This method did not use any specific absorption lines to constrain the gravity or abundance.  Instead, the parameters were determined using a blind $\chi^2$-fit. Furthermore, possible degeneracies between $\log g$ and various abundances were ignored.  For instance, a synthetic spectrum for a model with \Teff\ = 4200~K, $\log g = 2.25$, [Fe/H]$ = 0.6$ and C/O$ = 0.15$ had a fit to the \three\ spectrum with a $\chi^2$ value that was $\lesssim 10\%$ different than the $\chi^2$ value for the best-fit model.

\subsubsection{A4} 
\label{sect:A4}
A4 (R. Peterson) analyzed the optical spectrum of \one, as part of an ongoing analysis of standard stars spanning a wide range of temperature, gravity, and metallicity.  For these analyses, stellar parameters and abundances were derived by matching each stellar spectral observation to theoretical spectra calculated with an updated version of the \citet{1993KurCD..18.....K} SYNTHE program and the static, one-dimensional stellar atmosphere models selected from the grid of \citet{2003IAUS..210P.A20C}. A4 interpolated an appropriate model for each star, and used as 
input a list of molecular and atomic line transitions with species, wavelengths, energy levels, $gf$-values, and damping constants. 

\begin{figure*}
\centering
   \resizebox{\hsize}{!}{\includegraphics[scale=0.6]{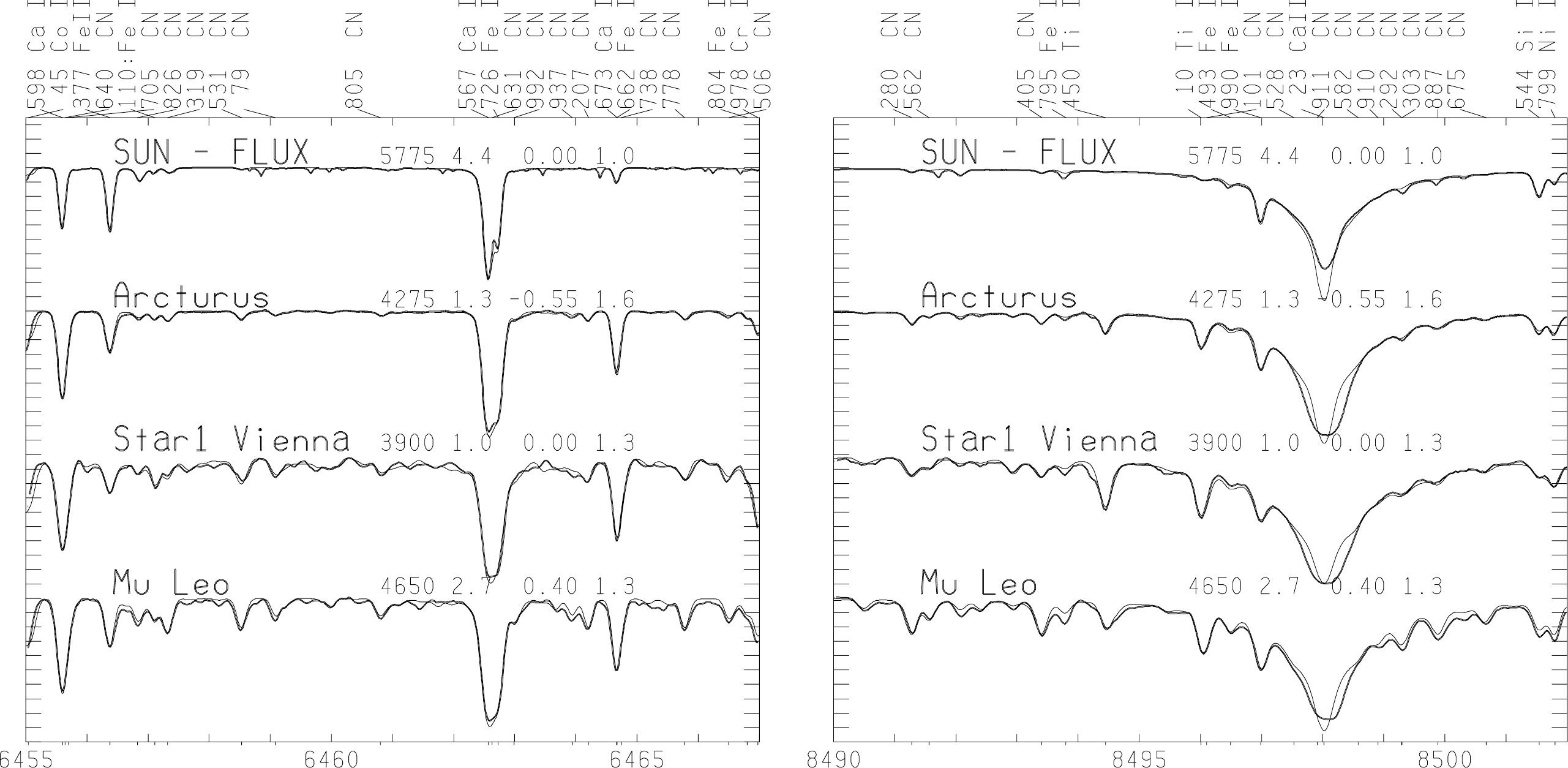}} 
   Wavelength [{\AA}]
\caption{A4 analysis: The plot compares observed spectra (heavy lines) to calculated spectra (light lines) for \one\ and three standard stars. The two panels display two separate wavelength regions, as indicated in \AA\ at the bottom of each panel.  The stars are displaced vertically for clarity.  Each y-axis tick denotes 10\% of the full height of the normalized spectrum.  The star name appears in bold above the continuum of each plot, followed by the model stellar parameters adopted for the calculation: effective temperature \Teff, surface gravity \logg, logarithmic iron-to-hydrogen ratio [Fe/H] with respect to that of the Sun, and microturbulence $v_{\rm mic}$. Above the upper boundary of each panel appear identifications of the strongest lines in the calculation for $\mu$~Leo: the three digits after the decimal point of the wavelength in \AA\ of the line is followed by a colon for a ``missing'' line, and then the species identification.}
\label{Peterson_fig1}
\end{figure*}

The Kurucz \emph{gfhy}\footnote{http://kurucz.harvard.edu/LINELISTS/GFHYPER100/} lists of atomic
lines with known energy levels (``laboratory'' lines) were modified by comparing calculations to echelle spectra of 
standard stars.  Moving from weak-lined to stronger-lined stars, first each spectrum was calculated, then the $gf$-
values were adjusted individually for atomic lines and as a function of band and energy for molecular lines, and, 
finally, a guess was made on the identifications of ``missing'' lines, those appearing in the spectra but not in the 
laboratory line list. This process was iterated until a match was
achieved in each case. 
\citet{2008STScINews...4.1P} shows an example of the fits achieved in the near-UV for turnoff stars, from solar to extremely low metallicities. Fig.\,\ref{Peterson_fig1} shows optical spectra for stronger-lined stars.

Stellar parameters were derived from the spectra, and not from colours.  The effective temperature \Teff\ was constrained by demanding that the same abundance emerge from low- and high-excitation lines of the same species (usually \ion{Fe}{I}), and by fitting the Balmer line wings in stars of \Teff\ $\sim$ 5000\,K or hotter.  The gravity was inferred from the wings of other strong lines; comparing \ion{Fe}{I} and \ion{Fe}{II} abundances provided a check.  Demanding no trend in abundance with line strength set $v_{{\rm mic}}$.  The iron abundance and other elemental abundance ratios stemmed from matching relatively unblended weak lines.  The resulting uncertainties are typically 0.1 -- 0.2\,dex in [X/Fe] for element X.  \citet{2001ApJ...559..372P} provide details.

A4 first compared \one\ against the metal-poor K1.5\,III giant Arcturus ($\alpha$~Boo, HR~5340, HD~124897) and the super-metal-rich K2,III giant $\mu$~Leo (HR~3905, HD~85503), two well-observed stars. This quickly established \one\ to be cooler than any giant with solar metallicity higher than one-third solar that A4 has analyzed before.  Moreover, throughout the red region 
the star exhibited a multitude of absorption lines not seen in either of the other two K giants. The majority proved to be TiO lines. Consequently, the TiO line list of \citet{1998FaDi..109..321S}, downloaded from the Kurucz website\footnote{http://kurucz.harvard.edu/molecules/TiO/tioschwenke.idasc-gz} was added. 

Several iterations were required to fit reasonably well the observed spectrum for \one.
The first iterations refined \Teff, and simultaneously [Fe/H], then \logg\ and $v_{\rm mic}$. Ultimately the best fit was obtained for a model with the parameters given in Table~\ref{results1a}. The fit does deteriorate below 6000\AA, where ``missing'' lines remain significant.

With one exception, there was no need to alter the relative abundance of any element with respect to that of iron in \one\ from the values adopted for the Sun.  For nitrogen in \one, however, the abundance was lowered to [N/Fe] = $-$0.1~dex to match the multitude of CN lines in the red. That oxygen in \one\ is solar, [O/Fe] = 0, was confirmed from three separate diagnostics: the [\ion{O}{I}] lines at 6300.3\,\AA\ and 6363.8\,\AA, the high-excitation \ion{O}{I} triplet at 7771.9\,\AA, 7774.2\,\AA, and 7775.4\,\AA, and a fit of the red TiO lines. For the Sun the older, higher oxygen abundance log(O/H) = $-$3.07 was adopted.

For Arcturus and $\mu$ Leo, relative nitrogen abundances were increased by 0.1~dex.  Relative abundances were also changed for other elements, notably sodium and aluminum, the light elements Ca, Mg, Si, and Ti, and elements beyond the iron peak, whose proportions are all known to vary among old stars \citep{2008ARA&A..46..241S}.

The degree to which the A4 calculations match the observed spectra of \one\ and other standards is illustrated in Fig.~\ref{Peterson_fig1}, which shows a comparison in two wavelength regions of the observed spectrum to the calculated spectrum for four stars: the Sun, Arcturus, \one, and $\mu$ Leo. Both wavelength regions are relatively free of TiO absorption in \one, but its inclusion is nonetheless important to better define the continuum. Both have many CN features, which must be closely approximated to define both continuum and blends. 

Although the match is not perfect, it is largely satisfactory in all four stars. Whenever a line is observed to be significantly too strong in one star, it is usually also significantly too strong in the others -- indicating either a missing or wrong identification, or an erroneous $gf$-value.  The \ion{Ca}{II} core mismatch is due to the chromospheric contribution to this line. The \logg\ values are supported by the agreement between \ion{Ca}{I} and \ion{Ca}{II} lines and by \ion{Fe}{II} lines. Temperatures are supported by the wide range of lower excitation potentials spanned by lines of \ion{Si}{I}, \ion{Ti}{I}, and \ion{Fe}{I}. 

\subsubsection{A5}
\label{sect:A5}
The A5 analysis was performed by A. Goswami.
The stellar atmospheric parameters (\Teff, \logg, and [Fe/H]), for \one\ were determined by an LTE analysis of the 
equivalent widths of atomic lines using a recent version of MOOG \citep{1973PhDT.......180S}.  
The 59 cleanest lines of \ion{Fe}{I} and two lines of \ion{Fe}{II} within 
the three wavelength ranges recommended for Experiment~1 were used in the analysis (see Tab.~\ref{tab:goswami}). 
The Fe lines covered a range in excitation potential (1.0 -- 5.0~eV) and equivalent widths (10 -- 165~m\AA). The excitation potentials and oscillator strengths of the lines were from various sources listed in the atomic spectral line database from CD-ROM 23 of R. L. Kurucz\footnote{http://www.cfa.harvard.edu/amp/ampdata/kurucz23/}.  Model atmospheres were selected from the Kurucz grid of model atmospheres available at the Kurucz web site\footnote{http://cfaku5.cfa.harvard.edu/}. Kurucz models both with and without convective overshooting (see \S~\ref{sect:convection}) were employed, the latter from those that are computed with better opacities and abundances and labelled with the suffix ``odfnew''. It was found that the derived temperatures from both the options agree well within the error limits.

\begin{table}
\center
\caption{Lines analysed by A5 (Sect.\,\ref{sect:A5}). For each line, 
the wavelength, the $gf$ value, the lower level excitation potential $E_{\rm low}$, the equivalent width (EW), and the source of the line data according to
 \citet{1995KurCD..23.....K}\tablefootmark{a} are given.}
\label{tab:goswami}
\begin{tabular}{lcccc}
\hline\hline
$\lambda$ &   $E_{\rm low}$ & log($gf$) & EW   & Reference \\
(nm)          &   (eV) & & (mA) & \\
\hline
\ion{Fe}{i} & & & & \\
  491.0325    &     4.191  &  -0.459   &   143.3   &   K88\\
  491.6662    &     3.929  &  -2.960   &    26.8   &   FMW\\
  491.8020    &     4.231  &  -1.360   &    68.9   &   FMW\\
  492.7417    &     3.573  &  -1.990   &    78.1   &   FMW\\
  494.5636    &     4.209  &  -1.510   &    82.5   &   FMW\\
  495.2639    &     4.209  &  -1.665   &    73.6   &   K88\\
  501.0300    &     2.559  &  -4.577   &    11.0   &   K88\\
  505.8496    &     3.641  &  -2.830   &    45.1   &   FMW\\
  513.7395    &     4.177  &  -0.400   &   116.9   &   FMW\\
  523.6205    &     4.186  &  -1.720   &    51.3   &   FMW\\
  526.2608    &     4.320  &  -2.280   &    30.4   &   FMW\\
  530.0412    &     4.593  &  -1.750   &    19.1   &   FMW\\
  531.5065    &     4.371  &  -1.550   &    70.2   &   FMW\\
  532.2041    &     2.279  &  -3.030   &   156.2   &   FMW\\
  532.6799    &     4.415  &  -2.100   &    31.5   &   FMW\\
  536.4858    &     4.445  &  +0.230   &   136.7   &   FMW\\
  538.5579    &     3.694  &  -2.970   &    32.9   &   FMW\\
  538.6959    &     3.642  &  -2.624   &    84.9   &   K88\\
  615.1617    &     2.176  &  -3.299   &   152.8   &   FMW\\
  616.5361    &     4.143  &  -1.55    &    82.5   &   FMW\\
  621.3429    &     2.223  &  -2.66    &   160.5   &   FMW\\
  625.3829    &     4.733  &  -1.660   &    54.6   &   FMW\\
  627.1276    &     3.332  &  -2.950   &    65.7   &   FMW\\
  630.1498    &     3.653  &  -0.745   &   143.1   &   K88\\
  630.2494    &     3.687  &  -1.203   &   112.1   &   K88\\
  630.7856    &     3.642  &  -3.535   &    33.6   &   K88\\
  631.5809    &     4.076  &  -1.710   &    72.3   &   FMW\\
  632.2690    &     2.588  &  -2.426   &   151.1   &   FMW\\
  633.6823    &     3.687  &  -1.050   &   145.7   &   FMW\\
  640.8016    &     3.687  &  -1.048   &   161.0   &   K88\\
  641.1647    &     3.653  &  -0.820   &   148.9   &   FMW\\
  648.1869    &     2.279  &  -2.984   &   151.1   &   FMW\\
  651.8365    &     2.830  &  -2.750   &   131.2   &   FMW\\
  660.9110    &     2.559  &  -2.692   &   158.4   &   FMW\\
  660.9676    &     0.990  &  -4.936   &   153.7   &   K88\\
  663.3746    &     4.559  &  -0.780   &    81.4   &   FMW\\
  664.6932    &     2.609  &  -3.990   &    81.6   &   FMW\\
  664.8079    &     1.011  &  -5.275   &   118.3   &   FMW\\
  666.3437    &     2.424  &  -2.479   &   161.5   &   FMW\\
  666.7417    &     2.453  &  -4.400   &    56.7   &   FMW\\
  670.3568    &     2.759  &  -3.160   &   100.2   &   FMW\\
  670.5101    &     4.607  &  -1.496   &    63.1   &   K88\\
  671.0316    &     1.484  &  -4.880   &   125.8   &   FMW\\
  672.5353    &     4.103  &  -2.30    &    30.8   &   FMW\\
  673.3151    &     4.638  &  -1.580   &    57.4   &   FMW\\
  673.9520    &     1.557  &  -4.950   &    97.0   &   FMW\\
  843.9563    &     4.549  &  -0.698   &   132.7   &   K88\\
  847.1739    &     4.956  &  -0.863   &    67.7   &   K88\\
  848.1982    &     4.187  &  -1.631   &    60.1   &   K88\\
  852.6667    &     4.913  &  -0.513   &    86.4   &   K88\\
  858.2257    &     2.991  &  -1.993   &   156.3   &   K88\\
  859.8825    &     4.387  &  -1.428   &    75.8   &   K88\\
  863.2412    &     4.104  &  -1.958   &    47.1   &   K88\\
  878.4434    &     4.955  &  -1.393   &    52.8   &   K88\\
  879.3338    &     4.608  &  -0.219   &   127.6   &   K88\\
  880.4623    &     2.279  &  -3.234   &   162.0   &   FMW\\
  886.6920    &     4.549  &  -0.065   &   157.9   &   K88\\
  887.8248    &     2.991  &  -3.600   &    82.1   &   K88\\
\ion{Fe}{ii} & & & & \\
  528.4109    &     2.891  &  -3.190   &    56.4   &   FMW\\
  645.6383    &     3.903  &  -2.075   &    45.2   &   K88\\
\hline
\end{tabular}
\tablefoot{
\tablefoottext{a}{http://kurucz.harvard.edu/LINELISTS/LINES/gfall.ref}
}
\end{table}

The microturbulence $v_{\rm mic}$ was estimated at a given effective temperature by demanding that there should be no dependence of the derived \ion{Fe}{I} abundance upon the equivalent widths of \ion{Fe}{I} lines (see Tables~\ref{results1a} and \ref{results1b}). The effective temperature was obtained adopting the derived value of the microturbulence parameter by the method of excitation balance, forcing the slope of the abundances from the \ion{Fe}{I} lines versus excitation potential to be near zero. Using the \ion{Fe}{I}/\ion{Fe}{II} ionisation equilibrium, the surface gravity of the star was obtained. The results are listed in Table~\ref{results1a}.

\subsection{Phoenix model atmospheres (P)}
\label{sect:P}
A model fit using Version 15 of PHOENIX \citep{hauschildt_b99} was provided by C.I.~Short.
A grid of spherical LTE atmospheric models was computed with a parameter range as given in Table~\ref{tab:modelgrids}.
The formal numerical precision for the fitting in $T_{\rm eff}$ was about $\pm30$~K.
The mass was held fixed at 1\,M$_\odot$ to determine radii for each value of $\log g$.
Convection was treated in the mixing length approximation with a mixing length parameter, $l$, of 1.0 pressure scale height. Atomic line data were taken from the ATLAS9 list \citep{1993KurCD..13.....K}.
Most molecular line data are from \citet{Rothman05} and \citet{1993KurCD..13.....K} except for the line lists
of CO \citep{Goorvitch94} and TiO \citep{1998FaDi..109..321S}.

For the purpose of Experiment~1 the approach of investigating the global fit to two wavelength regions (``visible'' and ``near-IR'', see Fig.~\ref{fig:wavelengths}) was adopted in the P analysis.
The radial velocity correction was determined by fitting the Doppler shift, $\Delta\lambda$, 
to relatively unblended weak spectral lines at the blue and red ends of the observed spectra.
Synthetic spectra were computed with a spectral resolution, $R$, of 300\,000. 
Assuming that giants are slowly enough rotating that vsini would be minor compared to the effect of microturbulence,
no rotational broadening was taken into account.
An initial rectification was performed by dividing the synthetic spectra by unblanketed spectra for each model; the spectra 
were then re-rectified to the observed spectra piece-wise with a single-point calibration in each of the two fitting regions.  
The visible region is affected by significant telluric contamination.  In principle, every feature in the entire spectrum contains information about the stellar parameters, provided they can be modelled correctly, and this experiment can be seen as a test of how well a model ``blindly'' fitted to the broad spectrum with equal weight on all spectral lines would recover the stellar parameters of a known star.  

In the near-IR region the strongest diagnostics are the \ion{Ca}{II} IR triplet lines.  However, it was found that PHOENIX consistently produces profiles for these lines that are too bright throughout the damping wings for any realistic stellar parameters, even with line damping parameters tuned to match the solar line profiles.  A comparison of unblanketed spectra among the workshop participants revealed that PHOENIX predicts a larger ``background'' continuum flux throughout the near-IR 
band than the other codes, and a possibility to be investigated that is consistent with both of these results is that PHOENIX under-estimates the {\it continuous} opacity in this region.

\begin{figure}
\resizebox{\hsize}{!}{\includegraphics[trim=40 40 0 0,clip]{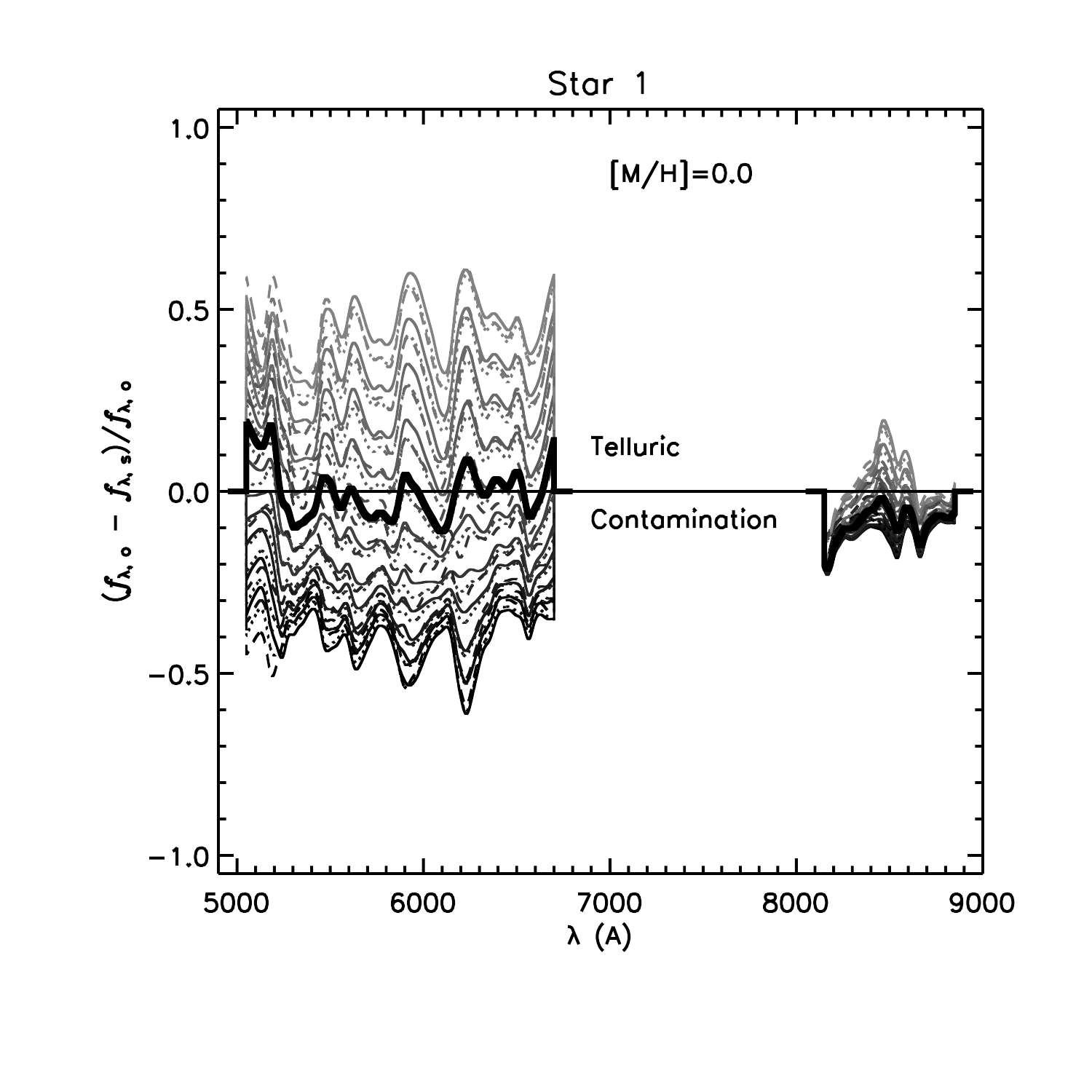}} 
\caption{P analysis: Relative flux difference, $(f_{\lambda, {\rm o}} - f_{\lambda, {\rm S}})/f_{\lambda, {\rm o}}$, 
smoothed to a resolution corresponding to $\Delta\lambda=50$ \AA, for a subset
of the grid with [M/H]=0.0 around the best-fit parameters for Star 1 are shown in
gray-scale. The best-fit model (Teff/log g/[M/H] = 3940/2.0/0.0) is overplotted
with a thicker line style.} 
\label{shortf1}
\end{figure}

\begin{figure}
\resizebox{\hsize}{!}{\includegraphics[trim=40 40 0 0,clip]{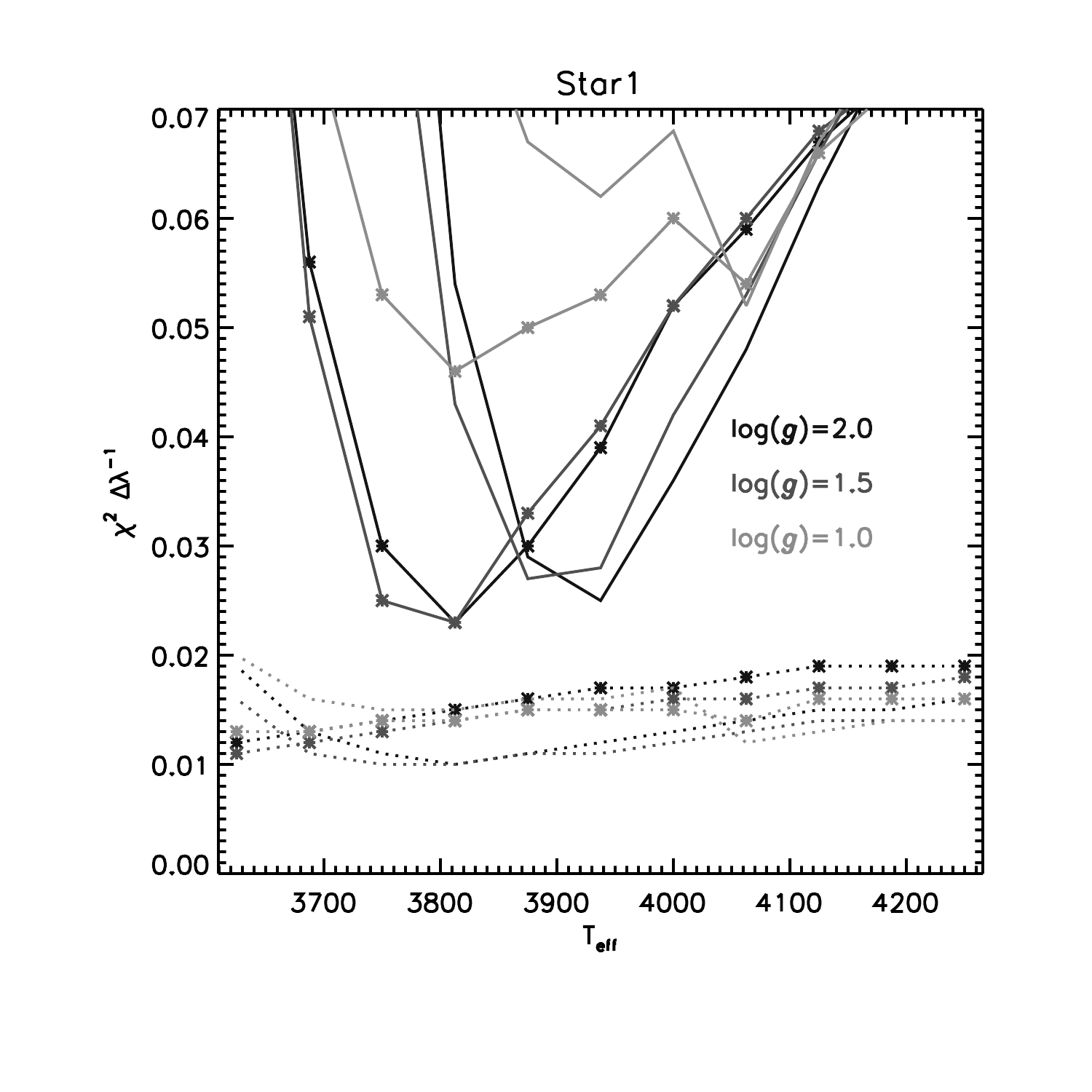}} 
\caption{P analysis: Values of $\chi^{\rm 2} (\Delta\lambda)^{\rm -1}$ for \one.  Solid lines: Fit to visible region; Dotted lines: Fit to near-IR region.  
No point symbol: \MonH = 0.0; Asterisks: \MonH = $-$0.5.  Dark to light gray lines: \logg = 2.0 to 1.0. } 
\label{shortf2}
\end{figure}

\begin{figure}
\resizebox{\hsize}{!}{\includegraphics[trim=40 40 0 0,clip]{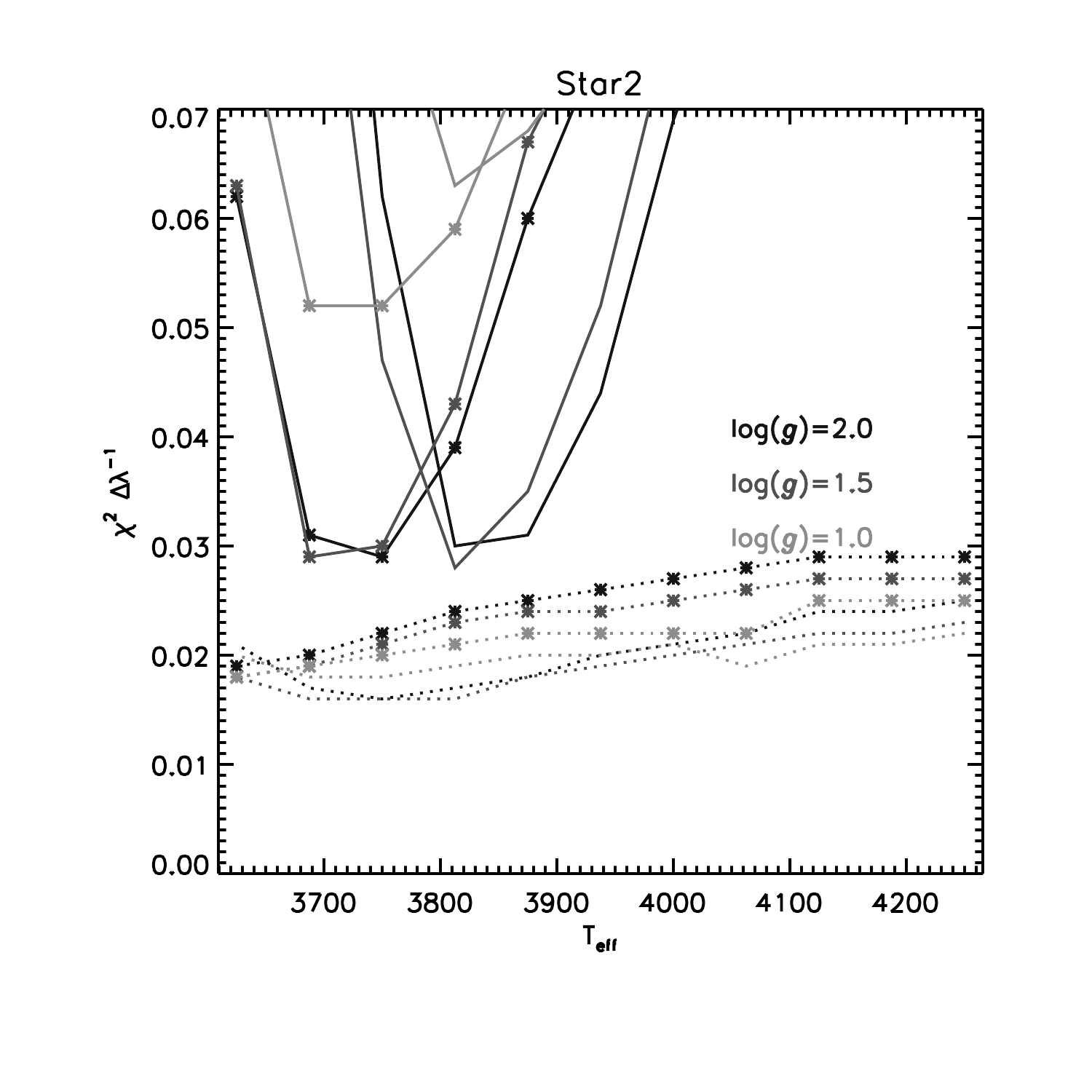}} 
\caption{P analysis: Same as Fig.~\ref{shortf2} except for \two.} 
\label{shortf3}
\end{figure}

Figure~\ref{shortf1} shows the relative flux residuals, $(f_\lambda,{\rm o} - f_\lambda,{\rm s})/f_\lambda,{ \rm o}$, where $f_\lambda,{\rm o}$ and $f_\lambda,{\rm s}$ are observed and synthetic flux, respectively, for \one\ and models of \MonH = 0.0.  Figures~\ref{shortf2} and \ref{shortf3} show the values of $\chi^{\rm 2}$ per interpolated $\lambda$ point for \one\ and \two, respectively, and all models.  

\paragraph{\one}
For the visible fitting region, the $\chi^{\rm 2}$ values showed distinct global minima as functions of \Teff\ (minimum $\chi^{\rm 2} (\Delta\lambda)^{\rm -1}\approx0.025$), but picked out best-fit values of \logg\ and \MonH\ less clearly. 
Models of \MonH=0.0 yielded best-fit \Teff\ values of 3940 and 3875 ($\pm 30$~K, formally) for \logg\ values of 2.0 and 1.5, respectively.  For \MonH=$-$0.5 these \logg\ values yielded best fit \Teff\ values of 3800~K.  Models of \logg=1.0 provided a significantly worse fit. There was marginal evidence that models of \MonH = $-$0.5 might provide a better fit than those of \MonH = 0.0.  For the near-IR 
region the $\chi^{\rm 2}$ values showed a much more shallow minimum around 3800 K (minimum $\chi^{\rm 2} (\Delta\lambda)^{\rm -1}\approx 0.01$) for \MonH = 0.0 and \logg\ values of 2.0 and 1.5, and did not show a minimum at all for \MonH = $-$0.5.

\paragraph{\two}
The fits yielded a similar result as that for \one, but with all the $\chi^{\rm 2} (\Delta\lambda)^{\rm -1}$ minima shifted downward in \Teff\ by about 100~K for the visible band, and 50~K for the near-IR 
band.  For the visible band the best fit value for \logg\ was either 1.5 or 2.0, and the best fit values for \Teff\ were 3800~K (\MonH = 0.0) or 3700 to 3750 K (\MonH = $-$0.5).  Again, only models of \MonH = 0.0 showed a minimum for the IR band, with a best fit \Teff\ value being indicated by a very shallow minimum of $\sim$3750~K.  Models of \MonH\ of 0.0 and $-$0.5 provided the same quality of fit, so the nominal best fit value was found to be $-0.25 \pm 0.25$.  

In conclusion, the near IR band needs attention in the P modelling to cultivate it as a more effective diagnostic of stellar parameters for M giants.  The lack of distinct minima in the fitting statistic may be caused by the poor quality of the fit to the broad \ion{Ca}{II} IR triplet lines, as discussed above.  Therefore, the best-fit parameters for the P analysis given in Tables~\ref{results1a} and \ref{results1b}
are based entirely on the fit to the visible region.  

\subsection{CODEX model atmospheres (C)}
\label{sect:C}
  The authors contributing the C analysis were M. Ireland and M. Scholz.
  The Cool Opacity-sampling Dynamic EXtended (CODEX) models as 
  published in \citet{Ireland08} are based on a combination of three separate codes -- 
  a grey dynamical atmosphere code, a temperature iteration code and a 
  spectrum computation code. In addition, the equation of state is pre-tabulated 
  based on the 1970s code from T. Tsuji, including updates from \citet{Sharp90} 
  and \citet{Jeong03}. The CODEX models are designed to predict  the observables in the
  atmospheres of Mira variables, and the main algorithms involved are described in 
  \citet{SchmidBurgk75} and \citet{SchmidBurgk84}.
  For the static models computed for this paper, the dynamical code was not 
  used and the pressure stratification was iterated simultaneously with the 
  temperature.
  
An important property of these models relevant to the test stars in this paper is that only atomic lines of neutral atoms were included, so that the code was not applicable and has not been used for effective temperatures above 4000\,K. Furthermore, there were numerical instabilities 
that are not fully understood, which means that the spherical code could only be used in significantly extended configurations.  In practice, this implied a restriction to \logg\ values less than approximately $-$0.5 to 0. Within these constraints, the model structures produced by the CODEX models agree with MARCS structures within 50\,K.
The H2O lines come from \citet{1997JChPh.106.4618P} and those of TiO
from \citet{1998FaDi..109..321S}. Other diatomic molecular (CO, OH,CN, SiO, MgH)
and neutral metal atomic lines come from the input to the ATLAS12
models \citep{1994KurCD..19.....K}. VO was not taken into account here.
  
The full grid of models used to fit the stars in Experiment~1 had the parameters given in Table~\ref{tab:modelgrids}, and a fixed extension of 100 times solar.  The atmospheric extension in this context was defined as $(T_{\rm eff}/5770)(R/{\rm R}_\odot)(M/{\rm M}_\odot)^{-1}$.  
A spectral fit was obtained at fixed gravity by minimising the mean square difference in between the model and the sample spectrum after interpolating bi-linearly in metallicity and effective temperature.  Additional fit parameters were macroturbulence (or almost equivalently spectrograph resolution/stellar rotation velocity), radial velocity, continuum flux and continuum slope.
A single continuous spectral window was used for the analysis of \two\ (see Fig.~\ref{fig:wavelengths}).
  
The only stars for which reasonable spectral fits could be obtained were \two\ and \four, as they were the only stars with parameters that were covered by the model grid. Uncertainties in the fitted parameters were not estimated in the C analysis, but fits were noticeably poorer by eye at parameters different by 0.2~dex in metallicity or 50~K in effective temperature, and 
so these can be taken as uncertainties. Gravity was not meaningfully constrained, but the \logg\ value with the marginally better fit was chosen from the two available. The best-fit parameters are given in Tables~\ref{results1b} and \ref{results2b}. For \four, the broad band colours computed from the models are $V-I$=1.89, $J-K$=0.94 and $V-K$=3.71.

\subsection{Tsuji08 model atmospheres (T)}
\label{sect:T}
The T analysis was provided by T. Tsuji. Modelling was restricted to \four, as the recent focus of the contributing author has been on the analysis of near infrared molecular lines \citep{2008A&A...489.1271T}. With the given photometric data (see Table~\ref{star34}) it was not easy to infer fundamental parameters of this star. 
According to \citet{1998AaA...331..619P} $J-K = 1.21$ corresponds to an M7 giant (see their Table~3), which would have a typical $ T_{\rm eff}$ of 3087$\pm94$\,K (see their Table~5).
$T_{\rm eff}$ = 3100\,K was, thus, chosen for \four.  In the absence of any information on mass, radius, etc. it was simply assumed that $M = 2 {\rm M}_\odot$ and $R = 200 {\rm R}_\odot$. These model parameters were adopted, together with a microturbulence $v_{\rm mic}$ = 3km\,s$^{-1}$, from a grid of model photospheres for red giant stars presented in \citet{2008A&A...489.1271T}.  The abundance pattern used for the modelling is described in Table 4, case (b),
in \citet{2008A&A...489.1271T} for CNO and in Table 1 in \citet{2002ApJ...575..264T} for other elements.
Line data for CO 3--0 lines are listed in Table~\ref{tab:tsuji}. Line positions and gf values are referenced 
in  \citep{2008A&A...489.1271T}. In particular, log $A_{\rm C}/A_{\rm H}$ = $-$3.92, log $A_{\rm N}/A_{\rm H}$ = $-$3.55, log $A_{\rm O}/A_{\rm H}$ = $-$3.31, and  [Fe/H] = 0.0, were chosen.  
 
\begin{figure}
\resizebox{\hsize}{!}{\includegraphics{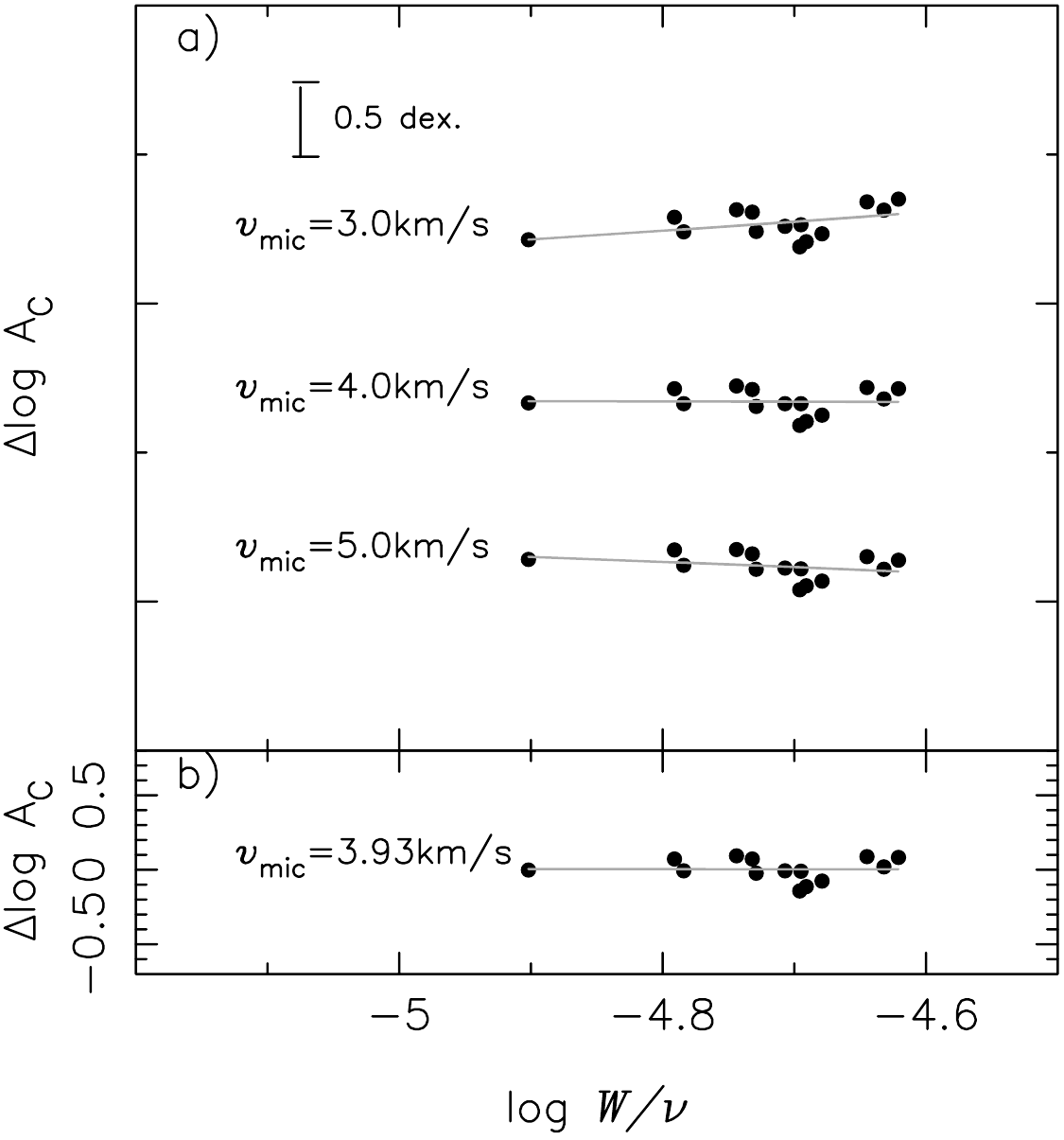}} 
\caption{T analysis: a) Line-by-line analysis of CO lines in \four, i.e. logarithmic abundance corrections to the assumed carbon abundance derived from the observed log\,$W/\nu$ values are plotted against the log\,$W/\nu$ values for $v_{\rm mic}$ = 3, 4, and 5 km\,s$^{-1}$.  b) Confirmation of the null abundance corrections for $  {\rm log} A_{\rm C}/A_{\rm H} =   -3.56  $ and $v_{\rm mic} = 3.93$ km\,s$^{-1}$.}
\label{fig:tsuji1}
\end{figure}

For the abundance analysis, 14 CO lines relatively free from blending in the given spectrum of \four\ were measured. The resulting equivalent width (EW) data are given in Table~\ref{tab:tsuji}. A measurement of OH and CN lines was also attempted but it was difficult to find a sufficient number of lines free from disturbance.  Line-by-line (LL) analysis as described in detail in \citet{1986A&A...156....8T, 2008A&A...489.1271T}  was applied to measure the selected CO lines. Results are shown in Fig.~\ref{fig:tsuji1}a, in which logarithmic abundance corrections to the assumed carbon abundance for different lines  are  given for  $v_{\rm mic}$ = 3, 4, and 5 km\,s$^{-1}$. For the microturbulence value given in Table~\ref{results2b}),
the abundance corrections were found to be independent of EWs, and the mean logarithmic abundance correction was found to be $+0.36 \pm 0.13$. This resulted in $  {\rm log} A_{\rm C}/A_{\rm H} = -3.92 + 0.36  =  -3.56 \pm 0.13$. Fig.~\ref{fig:tsuji1}b shows that the logarithmic abundance corrections are on average zero for these parameters.

\begin{table}
\center
\caption{Measured CO 3--0 lines in \four\ from the T analysis (Sect.\,\ref{sect:T}). For each line, the wavelength $\lambda$ (air), the $gf$ value, the lower level
excitation potential $E_{\rm low}$, the equivalent width over wavenumber
log\,$W/\nu$, and the identification 
are given.}
\label{tab:tsuji}
\begin{tabular}{rrrrr}
\hline\hline
$\lambda$ &   log($gf$) & $E_{\rm low}$ &  log\,$W/\nu$  &  Identification\\
(nm) & & [cm$^{-1}$] & & \\
\hline
            1560.082 &   $-$7.797 &  886.903  &   $-$4.696    & $^{12}$C$^{16}$O  3--0  R   21\\
            1560.447 &   $-$7.823 &  806.383  &   $-$4.632    & $^{12}$C$^{16}$O  3--0  R   20\\
            1560.839 &   $-$7.850 &  729.678  &   $-$4.691    & $^{12}$C$^{16}$O  3--0  R   19\\
            1561.257 &   $-$7.877 &  656.789  &   $-$4.679    & $^{12}$C$^{16}$O  3--0  R   18\\
            1561.533 &   $-$7.249 & 5055.604  &   $-$4.732    & $^{12}$C$^{16}$O  3--0  R   51\\
            1561.702 &   $-$7.907 &  587.721  &   $-$4.621    & $^{12}$C$^{16}$O  3--0  R   17\\
            1562.002 &   $-$7.234 & 5252.117  &   $-$4.784    & $^{12}$C$^{16}$O  3--0  R   52\\
            1562.499 &   $-$7.220 & 5452.275  &   $-$4.744    & $^{12}$C$^{16}$O  3--0  R   53\\
            1563.748 &   $-$8.038 &  349.698  &   $-$4.645    & $^{12}$C$^{16}$O  3--0  R   13\\
            1564.157 &   $-$7.179 & 6074.532  &   $-$4.791    & $^{12}$C$^{16}$O  3--0  R   56\\
            1564.326 &   $-$8.076 &  299.766  &   $-$4.695    & $^{12}$C$^{16}$O  3--0  R   12\\
            1564.930 &   $-$8.116 &  253.667  &   $-$4.707    & $^{12}$C$^{16}$O  3--0  R   11\\
            1565.561 &   $-$8.159 &  211.404  &   $-$4.729    & $^{12}$C$^{16}$O  3--0  R   10\\
            1566.763 &   $-$7.125 & 6954.739  &   $-$4.902    & $^{12}$C$^{16}$O  3--0  R   60\\
\hline
\end{tabular}
\end{table}
 
\begin{figure}
\resizebox{\hsize}{!}{\includegraphics{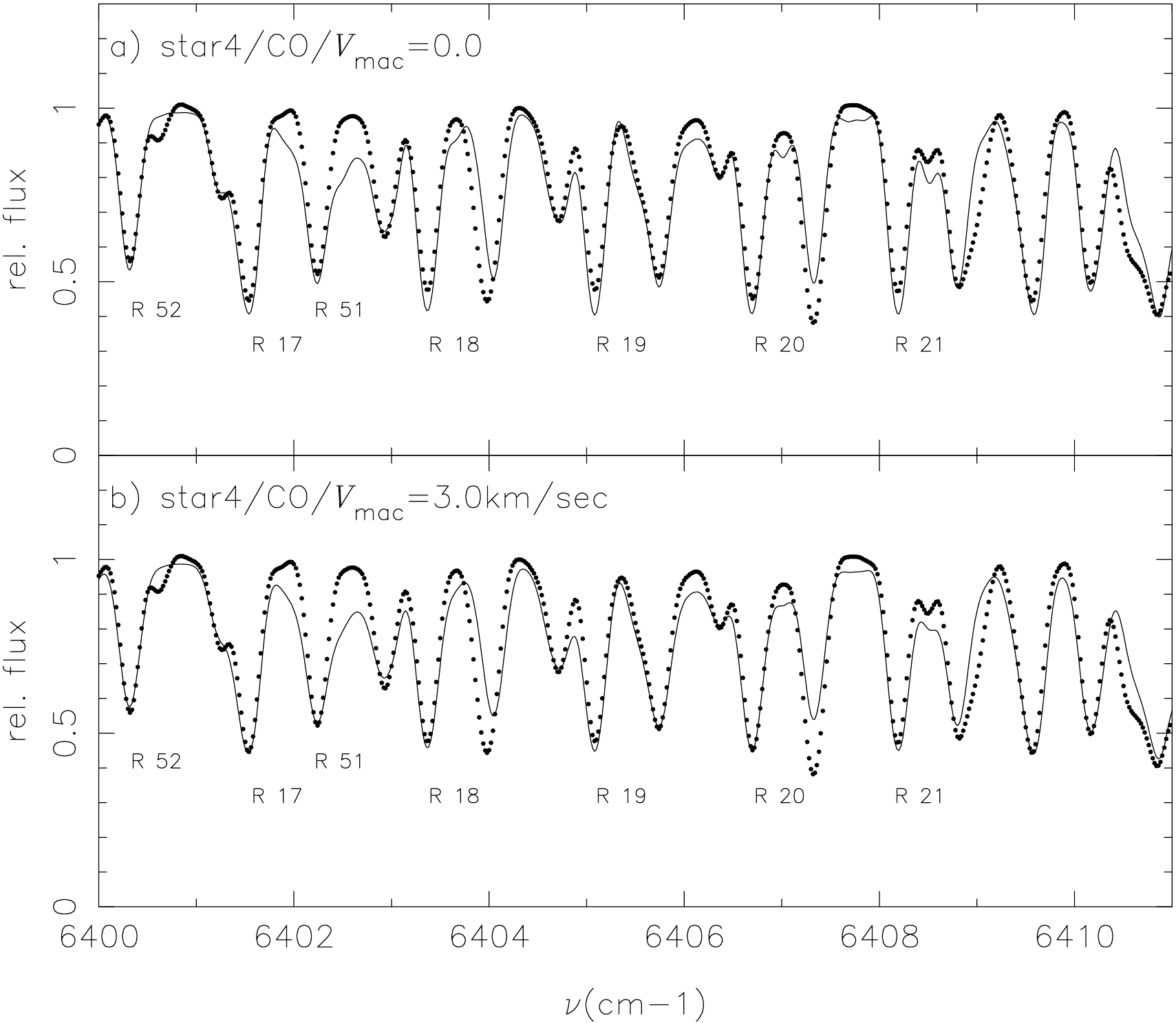}} 
\caption{T analysis: a) Observed spectrum (dots, relative flux vs. wave number) and model spectrum (solid line) with a macroturbulence of zero, for the CO lines between 1.5601 and 1.5625~$\mu$m of \four.  b) The same as a) but with a macroturbulence of 3.0\,km\,s$^{-1}$. }
\label{fig:tsuji2}
\end{figure}

For the resulting microturbulence and carbon abundance, a synthetic spectrum for the region of the CO 3--0 band was computed, and the result is compared with the observed one in Fig.~\ref{fig:tsuji2}a.  The CO lines used in the LL analysis are indicated by the rotational identifications for the CO 3--0 lines.  Only these CO lines should be considered for estimating the quality of the fit, but fits are not so good even for these lines.  
To improve the fits in Fig.~\ref{fig:tsuji2}b, some additional broadening may be required, and
several values of macroturbulent velocities were tried.  It was found that the fits are somewhat
improved for the selected CO lines with an assumed macroturbulent velocity  of 3.0 km s${-1}$
(which could have been determined empirically if a sufficient number of weak lines could be
measured, as in Fig. 3 in \citet{1986A&A...156....8T}).

Because the stellar parameters ($ T_{\rm eff}, M, R$ etc.) cannot be 
pre-defined well and because the number of lines is too small for the LL analysis,
only the basic procedure to analyze the stellar spectrum has been shown and
a final converged solution was not pursued. However, 
we may conclude that the abundance determination is possible
even for very cool stars in this way, if dozens of weak lines could be
measured.

For the T analysis, the classical theory of
line formation assuming the presence of microturbulence was applied.
However, the applicability of the classical model of line formation
may be limited to relatively weak lines with log\,$W/\nu < -4.75$
\citep{2008A&A...489.1271T}. Although the sample of CO lines of \four\ analyzed here
includes some lines slightly stronger than this limit, the
modelling was quite successful because most of the lines used are
weak and below or near the limit mentioned \citep[cf. a similar 
previous study of the CO second overtone lines]{1991A&A...245..203T}.
The strongest lines with log\,$W/\nu > -4.5$ clearly deserve an
extra contributions from outside the photosphere \citep{1988A&A...197..185T}.
Thus, if a large number of the intermediate strength lines
with $-4.75 < {\rm log}\,W/\nu < -4.5$  are included as in a
previous study of the CO first overtone lines \citep{1986A&A...156....8T}, the
results cannot be correct as was recognized 20 years later \citep{2008A&A...489.1271T}.
Such a problem will appear not only in very cool stars 
but also in earlier type stars such as K giants \citep{2009A&A...504..543T}.

\section{Results} 
\label{sect:results}
As described above, two experiments were conducted. Experiment~1 had the aim to find the best fitting model to the four sample spectra provided, using the preferred method of each participating researcher or team. It was to be completed before the beginning of the workshop.
A time limit of several weeks was set for the groups to deliver their
results. This limit was introduced not only for practical reasons but also
to simulate the typical amount of time used for parameter determination
from a stellar spectrum in the current age of multi-object spectrographs and
large surveys. Naturally, this time limit set some constraints on the
level of elaboration of the resulting fits.
Experiment~2 resulted from the discussions at the workshop and was aimed at a direct comparison of synthetic spectra from different model codes for a given set of stellar parameters.

\subsection{Experiment 1 -- stellar parameter determination}
\label{sect:exp1results}
We start with an overview of the main aspects of the different modelling approaches (for details see Sections~\ref{sect:MARCS} to \ref{sect:T}).
About half of the groups (six of fourteen) based their analysis on MARCS atmospheric models calculated in spherically-symmetric geometry. Five groups used the plane-parallel ATLAS suite of codes \citep{1993KurCD..13.....K,1993KurCD..18.....K,Lester2008}. The three software packages PHOENIX \citep{hauschildt_b99}, CODEX \citep{Ireland08} and Tsuji08 \citep{2008A&A...489.1271T} were employed by one group each (all of them spherical). The first character of the group identifier indicates this aspect.
The stellar radius, a key parameter, is defined as the layer where the Rosseland optical depth is equal to one in the case of the MARCS 
models, the model of Tsuji, and the CODEX models. 
In the case of the PHOENIX models, it is assumed that the stellar mass $M$ is one solar mass,
and the radius $R$ is computed accordingly from the specified surface gravity ($R=(GM/g)^{1/2}$).
For the plane-parallel ATLAS models the radius is undefined.

To improve the fit of the line widths  and to take into account various additional line broadening mechanisms 
like stellar rotation
a macroturbulence has been added by most of the groups, i.e. the synthetic spectra have been convoled with a Gaussian profile. 

The set-up did not require all participants to complete the experiment for all four spectra. Indeed, due to methodological constraints, some participants restricted their attempts to either the visual (M2, M3, M4, M6, A2, A4, A5, P) or the near infrared spectra (M5, A3, T). Naturally, the team that prepared the artificial observations for \three\ and \four\ (M1) did not participate in the fitting of these spectra.
Two groups analysed both wavelength domains -- A1 (\one\ and \three) and C (\two\ and \four).
In summary, \one\ was analysed by eleven groups, of which five also analysed \two. 
\three\ and/or \four\ were analysed by five groups.



\begin{figure*}
\resizebox{\hsize}{!}{\includegraphics{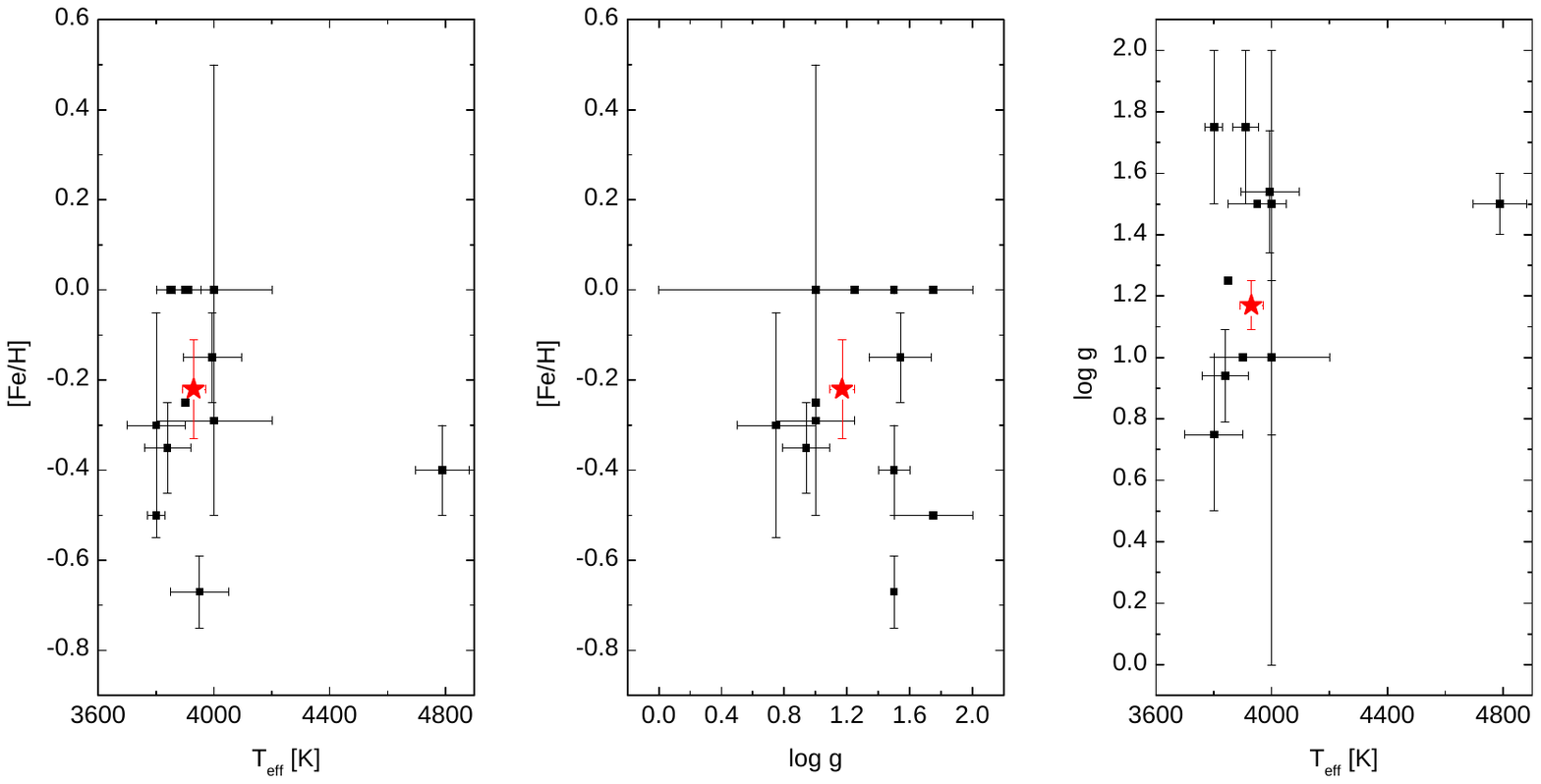}} 
\caption{Results of the various model fits in three different two-parameter planes for \one\ ($\alpha$~Tau).  The red star marks the values of the direct parameters given in Table~\ref{narvaldata}.}
\label{fig:alphataucomp}
\end{figure*}

\subsection{$\alpha$ Tau (\one) and $\alpha$ Cet (\two)}
\label{sect:results1}
The best fits to these two spectra as provided by the participating teams are summarized in Tables~\ref{results1a} and
\ref{results1b}. For $\alpha$~Tau we have the more extensive set of model fits. The results are presented in Fig.\,\ref{fig:alphataucomp} for the three possible combinations of the three main parameters derived: $T_{\rm eff}$, \logg, and [Fe/H].
The (unweighted) mean values of all results are: $T_{\rm eff}$\,=\,3980$\pm$250~K, \logg\,=\,1.3$\pm$0.3, and [Fe/H]\,=\,$-$0.2$\pm$0.2~dex, 
close to the literature values discussed in Section~\ref{sect:alftau} and Table \ref{narvaldata}. 
No systematic differences between the model families (MARCS, ATLAS, and others) are apparent.
Within the error bars given by the different groups, the derived $T_{\rm eff}$ values of almost all groups agree with each other.
The only exception is the A2 analysis, which deviates by 800~K from the mean value.
Discarding the A2 value, the mean \Teff\ of $\alpha$~Tau as a result of this experiment becomes 3910$\pm$80~K.
Note that in the case of our experiment, contrary to a series of repeated measurements, the mean value does not necessarily result in a more precise value than the individual determinations.

We have no obvious explanation for the deviating result of A2, but the analysis approach differs in various aspects from the bulk of the other analyses. First, A2 uses a large model atmosphere grid including \Teff\ values up to 6000~K, which makes it possible to probe higher temperatures than other analyses. However, the same can be said for the M3a and M3b analyses, which do not find a high-\Teff\ solution using a grid up to 8000~K.  
Second, A2 did not include molecular lines in the analysis, but the same is true for M6, A1, and A5.
A final aspect is that A2 did an equivalent width analysis, and not a synthetic spectrum fit, which seems problematic when dealing with crowded spectra. The A5 analysis also used equivalent widths, and obtained a \Teff\ close to the mean value. However, there are important differences between the two analyses, which might explain the different results.
A2 measured equivalent widths for selected lines of several different species from the observed \emph{and} computed spectra, while A5 computed equivalent widths line-by-line, adjusting the abundances for lines of \ion{Fe}{I} and \ion{Fe}{II} only. Also, A2 derived all parameters simultaneously from a fit to all lines, while A5 used the classical approach of excitation and ionization equilibrium for Fe lines. Note that there are only three lines in common in the line lists of A2 and A5.

Concerning gravity, the results for $\alpha$~Tau group around two values, slightly lower (M1, M3b, M6, A4, A5) and slightly higher (M2, M3a, A1, A2, P) than the published 
values given in Table~\ref{narvaldata}, with the two M4 results falling in both groups.
The majority of the modelers found $\alpha$~Tau to exhibit a lower metallicity than the Sun, the lowest values being suggested by A1. For the cases where [Ca/Fe] values were also determined, most groups find a mild underabundance of Ca.

\begin{figure}
\resizebox{\hsize}{!}{\includegraphics[viewport=70 50 770 554]{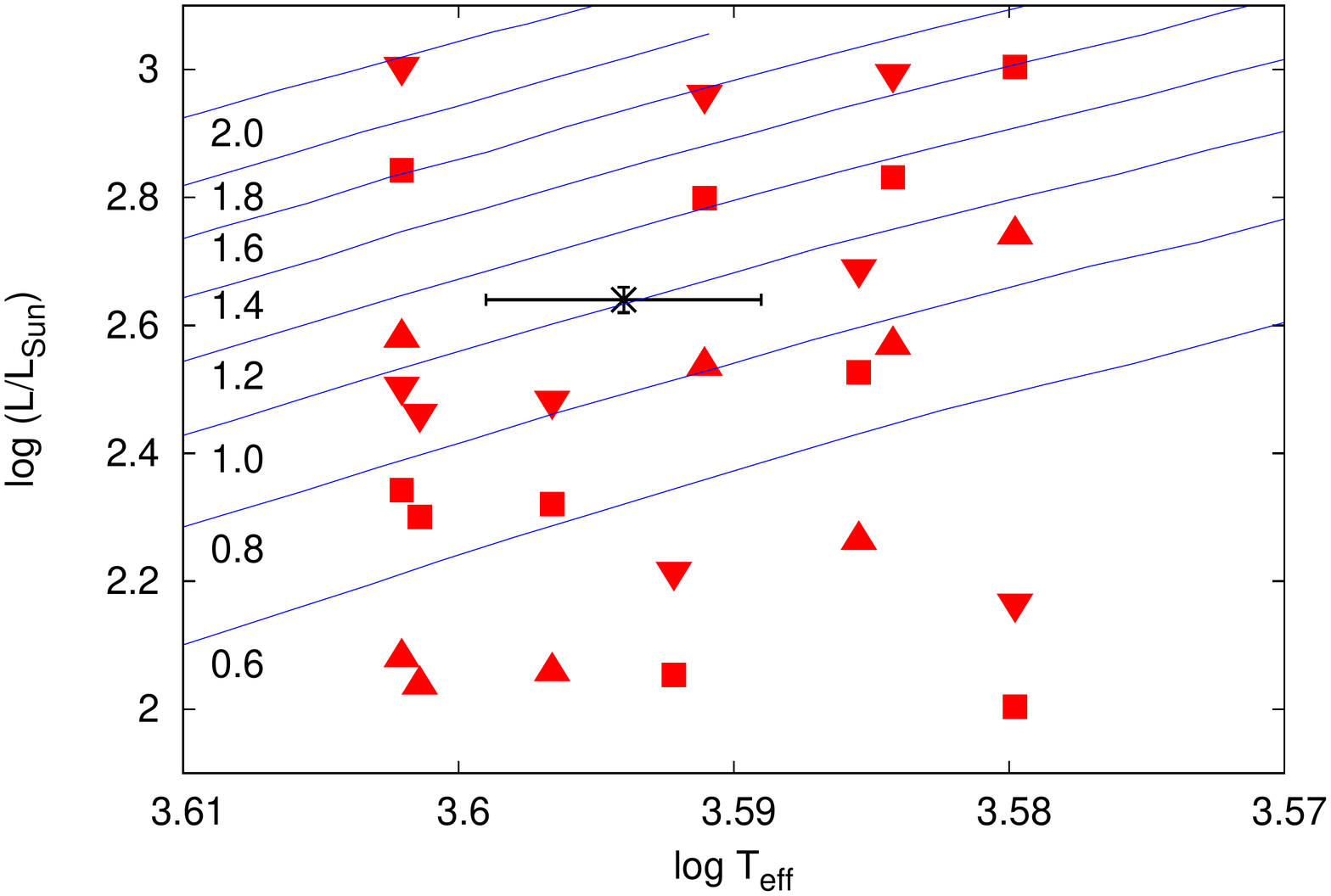}} 
\caption{Best-fit \Teff\ and \logg\ for \one\ ($\alpha$~Tau) converted to luminosity, using three different mass 
values (triangles up -- 0.6~$M_\odot$, squares -- 1.1~$M_\odot$, triangles down -- 1.6~$M_\odot$), and 
direct \Teff\ and \logg\ values from Table~\ref{narvaldata} (black cross with error bars). Solid lines are evolutionary tracks for 
$Z$=0.008 and $Y$=0.26 by the Padova group \citep{2008A&A...484..815B,2009A&A...508..355B}, labelled to the left with 
mass in solar masses.}
\label{fig:hrd}
\end{figure}

Figure~\ref{fig:hrd} shows the best-fit \Teff\ and \logg\ for \one\ ($\alpha$~Tau) from all participants except A2 
converted to luminosity, using three different mass values, together with 
the direct \Teff\ and \logg\ values from Table~\ref{narvaldata}. We also show evolutionary tracks for metal-poor models by the Padova group for several different mass values. It is obvious that some of the derived pairs of parameters are inconsistent with the chosen stellar evolution models.

Fewer model fits are available for $\alpha$~Cet, and thus we have 
omitted a graphical presentation.  The (unweighted) mean values and standard deviations of all results are: $T_{\rm eff}$\,=\,3830$\pm$220~K, \logg\,=\,1.0$\pm$0.8, and [Fe/H]\,=\,$-$0.2$\pm$0.2~dex, which is close to the published values discussed in Section~\ref{sect:alfcet}.  As for $\alpha$~Tau we find  rather good agreement in the derived temperature, with the exception of A2 (deviating by 500~K).
Discarding the A2 value, the mean \Teff\ of $\alpha$~Cet as a result of this experiment becomes 3760$\pm$70~K.
Concerning the surface gravity, the scatter is somewhat larger than for $\alpha$~Tau, and the majority of the values are larger than the published value.
For [Fe/H] the model fits to the spectrum of $\alpha$~Cet suggest metallicities from solar to subsolar. 
In general, one would expect a correlation between \Teff\ and \FeH\ for the differing results, since line strength increases towards lower temperatures and higher \FeH. This can be seen to some extent in Fig.~\ref{fig:alphataucomp}, and the $\alpha$~Cet results show a similar trend. On the other hand, there is no obvious correlation between \logg ~and \Teff, or \logg\ and \FeH. 

\begin{table*}
\caption{\label{results1a} Experiment 1: parameters determined 
for $\alpha$~Tau (\one) from the best fits to the optical spectra by the various participating teams.}
\centering
\begin{tabular}{l|ccc|cccc|ccc}
\hline\hline
 & \multicolumn{3}{c}{starting parameters from photometry} & \multicolumn{4}{c}{best fit parameters from spectroscopy} & \multicolumn{3}{c}{additional quantities}\\
Code & $T_{\rm eff}$ & log $g$ & [Fe/H] & $T_{\rm eff}$ & log $g$ & [Fe/H] & [Ca/Fe] & RV  &  $v_{{\rm mic}}$  & $v_{{\rm mac}}$  \\
& (K) & ($g$ (cm\,s$^{-2}$)) & & (K) & ($g$ (cm\,s$^{-2}$)) & & & (km\,s$^{-1}$) & (km\,s$^{-1}$) & (km\,s$^{-1}$) \\
\hline
M1 &  &  & $-$0.3\tablefootmark{a}       & 3800$\pm$100 & 0.75$\pm$0.25 & $-$0.3$\pm$0.25   &    & 54 & 2.5\tablefootmark{f} & 3\tablefootmark{f}    \\
M2 & 3850/4000\tablefootmark{b} & & & 3850         & 1.25          &    0.0            & 0.0$\pm$0.1 & 54.5 & 1--2\tablefootmark{f} & 4.5\\
M3a & & & & 3994$\pm$100 & 1.54$\pm$0.20 & $-$0.15$\pm$0.10 & & 53.9$\pm$0.9 & 2\tablefootmark{f} & \tablefootmark{g}\\
M3b & & & & 3839$\pm$80  & 0.94$\pm$0.15 & $-$0.35$\pm$0.10 & $-$0.2$\pm$0.1\tablefootmark{c} & 53.9$\pm$0.9 & 2\tablefootmark{f} & \tablefootmark{g}\\
M4 & & & & 4000\tablefootmark{d} & 1.5\tablefootmark{d} &    0.0\tablefootmark{d}  & & 55 & 2\tablefootmark{f} & 0\tablefootmark{f}\\
   & & & & 3900\tablefootmark{d} & 1.0\tablefootmark{d} & $-$0.25\tablefootmark{d} & & 55 & 2\tablefootmark{f} & 0\tablefootmark{f}\\
M6 & 3750--4250\tablefootmark{e} & 1.0--1.5\tablefootmark{e} & & 4000$\pm$200 & 1.0$\pm$1.0   &    0.0$\pm$0.5   & $-$0.3$\pm$0.1 & 54.3$\pm$0.5 & 2\tablefootmark{f} & \\
A1 & 3900 & 0.0 & 1.0 & 3950$\pm$100 & 1.5           & $-$0.67$\pm$0.08 & $-$0.23$\pm$0.17 & & 2 & --\tablefootmark{i}\\
A2 & & & & 4788$\pm$92  & 1.5$\pm$0.1   & $-$0.4$\pm$0.1   & & 54.1$\pm$0.4 & 2\tablefootmark{f} & \\
A4 & & & & 3900         & 1.0           &    0.0           & & 54.3 & 1.3 & 1.5 \\
A5 & & & & 4000$\pm$200 & 1.0$\pm$0.25  & $-$0.29          & & & 1.6$\pm$0.5 &  \\
P  & & & & 3910$\pm$45\tablefootmark{d} & 1.75$\pm$0.25\tablefootmark{d} & 0.0\tablefootmark{d} & & 54.71 & 3\tablefootmark{h} & 0\\
   & & & & 3800$\pm$30\tablefootmark{d} & 1.75$\pm$0.25\tablefootmark{d} & $-$0.5\tablefootmark{d} & & 54.71 & 3\tablefootmark{h} & 0\\
\hline
\end{tabular}
\tablefoot{
\tablefoottext{a}{Photometric colours used to resolve a \Teff\-[Fe/H] degeneracy after spectral fitting.}
\tablefoottext{b}{\Teff($V-K$)/\Teff($J-K$).}
\tablefoottext{c}{[$\alpha$/Fe].} 
\tablefoottext{d}{Degeneracy in \Teff\ and [Fe/H].}
\tablefoottext{e}{Ranges.}
\tablefoottext{f}{Fixed value(s).}
\tablefoottext{g}{Not taken into account.}
\tablefoottext{h}{Microturbulence increased from 2.0 to 4.0~km\,s$^{-1}$ as \logg\ decreased from 2.5 to 1.0.}
\tablefoottext{i}{Not determined.}\\
{\it Methods for photometric parameter determination:} M1, M2: MARCS models. 
Theoretical colour ($V-K$ and $J-K$) -- \Teff\ calibrations \citep{1998A&A...333..231B,2011ASPC..445...71V}.
M6: \citet{2000AJ....119.1448H}. A1: \citet{2005AaA...442..281K}.
}
\end{table*}

\begin{table*}
\caption{\label{results1b} Experiment 1: parameters determined 
for $\alpha$~Cet from the best fits to the optical spectra by the various participating teams.}
\centering
\begin{tabular}{l|ccc|cccc|ccc}
\hline\hline
 & \multicolumn{3}{c}{starting parameters from photometry} & \multicolumn{4}{c}{best fit parameters from spectroscopy} & \multicolumn{3}{c}{additional quantities}\\
Code & $T_{\rm eff}$ & log $g$ & [Fe/H] & $T_{\rm eff}$ & log $g$ & [Fe/H] & [Ca/Fe] & RV  &  $v_{{\rm mic}}$  & $v_{{\rm mac}}$  \\
& (K) & ($g$ (cm\,s$^{-2}$)) & & (K) & ($g$ (cm\,s$^{-2}$)) & & & (km\,s$^{-1}$) & (km\,s$^{-1}$) & (km\,s$^{-1}$) \\
\hline
M1 & & & $-$0.5\tablefootmark{a} & 3675$\pm$50 & 0.5$\pm$0.25 & $-$0.5$\pm$0.2 & & & 2.5\tablefootmark{c} & 3\tablefootmark{c} \\
M2 & $\lesssim$3700 & & & 3700        & 1.0          & 0.0            & & $-$25.9 & 1--2\tablefootmark{c} & 5.0\\
M3a & & & & 3867$\pm$100 & 1.15$\pm$0.20 & $-$0.22$\pm$0.10 & & $-$27.6$\pm$1.3 & 2\tablefootmark{c} & \tablefootmark{d}\\
M3b & & & & 3718$\pm$80  & 1.31$\pm$0.15 & $-$0.02$\pm$0.10 & $-$0.32$\pm$0.10\tablefootmark{b} &  $-$27.6$\pm$1.3 & 2\tablefootmark{c} & \tablefootmark{d}\\
A1 & 3500 & 1.0 & $-$2.0 & & & & & & & \\
A2 & & & & 4310$\pm$136 & 1.5$\pm$0.1 & $-$0.6$\pm$0.1 & & $-$26.3$\pm$0.4 & 2\tablefootmark{c} & \\
P  & & & & 3750$\pm$30 & 1.75$\pm$0.25 & $-$0.25$\pm$0.25 & & $-$26.22 & 3\tablefootmark{e} & 0\\
C  &               &               &                  & 3820$\pm$50 & $-$0.5 & 0.0$\pm$0.2 & & & 2.8\tablefootmark{c} & 2.3\\
\hline
\end{tabular}
\tablefoot{
\tablefoottext{a}{Photometric colours used to constrain metallicity after spectral fitting.}
\tablefoottext{b} [$\alpha$/Fe]. 
\tablefoottext{c}{Fixed value(s).}
\tablefoottext{d}{Not taken into account.}
\tablefoottext{e}{Microturbulence increased from 2.0 to 4.0~km\,s$^{-1}$ as \logg\ decreased from 2.5 to 1.0.}\\
{\it Methods for photometric parameter determination:} M1, M2: MARCS models. 
Theoretical colour ($V-K$ and $J-K$) -- \Teff\ calibrations \citep{1998A&A...333..231B,2011ASPC..445...71V}.
A1: \citet{2005AaA...442..281K}.
}
\end{table*}

\subsection{Star 3 and Star 4}
The fitting results for the two artificial stars are summarized in Tables~\ref{results2a} and \ref{results2b}. In this part of the experiment we asked the participants to derive also a C/O ratio. 

The fitting of spectra in the optical range is still more widely used than the application of the same technique to spectra in the $H$ and $K$ band. This may be due to the longer and wider availability of optical high-resolution spectrographs and the better knowledge of the resident features. It is therefore not surprising that also the number of teams providing a fit to one of the near-infrared spectra is smaller than for the case of the optical ones. Furthermore, the lower temperature, especially of \four, requires an extensive inclusion of molecular line data, which is not needed for the usual work of some of our participating teams and, therefore, not included in their codes.

In comparing the fit results with each other we have to remember that the fitted star' was a MARCS model. Therefore, one could expect that systematic differences between the underlying codes of the model fits should show up more clearly. For \three\ the result is a bit surprising, because the best fits of all three groups gave a solar or supersolar metallicity, while the \three\ model is clearly subsolar.  Also, the derived C/O ratio differs significantly from the model input value. Temperature values, on the other hand, nicely agree with each other and are in reasonable agreement with the \three\ value. 
For \four, all model fits are close to the real values or at least point into the correct direction (in particular to a higher C/O ratio).

A conclusion may be that the spectra of cool stars can be understood
reasonably well if the attention is confined to relatively weak
lines. However, we are still far from understanding complete
spectra of cool stars including the intermediate strength lines
and the strong lines, and the status of the modelling of these strong lines
is still not at a satisfying stage. It appears to be no problem in analyzing the 
CO lines in \four, but this may be simply because the provided
spectrum of \four\ does not include strong lines, and we should
not be fully satisfied with the present result.

\begin{table*}
\caption{\label{results2a} Experiment 1: parameters determined 
for the artificial spectrum of \three\ from the best fits of the various participating teams.}
\centering
\begin{tabular}{l|ccc|cccccc}
\hline\hline
 & \multicolumn{3}{c}{starting parameters from photometry} & \multicolumn{5}{c}{best fit parameters from spectroscopy} \\
Code & $T_{\rm eff}$ & log $g$ & [Fe/H] & $T_{\rm eff}$ & log $g$ & [Fe/H] & C/O &  $v_{{\rm mic}}$ &  $v_{{\rm mac}}$\\
& (K) & ($g$ (cm\,s$^{-2}$)) & & (K) & ($g$ (cm\,s$^{-2}$)) & & & (km\,s$^{-1}$) & (km\,s$^{-1}$)\\
\hline
M2 & 4225 & & & & & & & & \\
M5 & 4250 & 2.0 & 0.0 & 4250 & 2.0 & 0.0 & 0.46 & 1.7\tablefootmark{a} & 4.5 \\
A1 & 4300 & 2.0 & $-$2.5 & 4350$\pm$100 & 2.0 & 0.32$\pm$0.04 & 0.63\tablefootmark{b} & 2 &  --\tablefootmark{c}\\
A3 & 4200 & 3.0 & 0.0\tablefootmark{a} & 4400 & 3.5 & 0.6 & 0.63 & 2.0\tablefootmark{a} \\
\hline
\end{tabular}
\tablefoot{
\tablefoottext{a}{Fixed values.}
\tablefoottext{b}{[C/H]=0.31, [N/H]=0.02, [O/H]=0.21.}
\tablefoottext{c}{Not determined.}\\
{\it Methods for photometric parameter determination:} M2: MARCS models. 
Theoretical colour ($V-K$ and $J-K$) -- \Teff\ calibrations \citep{1998A&A...333..231B,2011ASPC..445...71V}.
M5: \citet{Worthey11}. A1: \citet{2005AaA...442..281K}.
}
\end{table*}

\begin{table*}
\caption{\label{results2b} Experiment 1: parameters determined 
for the artificial spectrum of \four\ from the best fits of the various participating teams.}
\centering
\begin{tabular}{l|ccc|cccccc}
\hline\hline
 & \multicolumn{3}{c}{starting parameters from photometry} & \multicolumn{5}{c}{best fit parameters from spectroscopy} \\
Code & $T_{\rm eff}$ & log $g$ & [Fe/H] & $T_{\rm eff}$ & log $g$ & [Fe/H] & C/O &  $v_{{\rm mic}}$ &  $v_{{\rm mac}}$ \\
& (K) & ($g$ (cm\,s$^{-2}$)) & & (K) & ($g$ (cm\,s$^{-2}$)) & & & (km\,s$^{-1}$) & (km\,s$^{-1}$)\\
\hline
M2 & $\gtrsim$3300~K/3500~K\tablefootmark{a} & & & & & & & & \\
M5 & 3250 & 0.0 & 0.0 & 3500 & 0.0 & 0.0 & 0.69\tablefootmark{b} & 3.0\tablefootmark{c} & 5.0 \\
A1 & 3800 & 1.0 & $-$1.5 & & & & & & \\
A3 & 3000 & 1.0 & 0.0\tablefootmark{c} & 3200 & 0.25 & 0 & 0.79 & 2.0\tablefootmark{c} & 0\\
C & & & & 3244$\pm$50 & 0.0 & $-$0.1$\pm$0.2 & 0.48 & 2.8\tablefootmark{c} &1.2\\
T & 3087$\pm$94 & & & 3100 & 0.14 & 0.0 & 0.56 & 3.9$\pm$0.4 & 3.0\\
\hline
\end{tabular}
\tablefoot{
\tablefoottext{a}{\Teff($V-K$)/\Teff($J-K$).}
\tablefoottext{b}{C/O not consistent with model structure.}
\tablefoottext{c}{Fixed values.}\\
{\it Methods for photometric parameter determination:} M2: MARCS models. 
Theoretical colour ($V-K$ and $J-K$) -- \Teff\ calibrations \citep{1998A&A...333..231B,2011ASPC..445...71V}.
M5: \citet{Worthey11}. A1: \citet{2005AaA...442..281K}. T: \citet{1998AaA...331..619P}.
}
\end{table*}

\subsection{Experiment 2 -- synthetic spectra comparison}
For Experiment 2 the participants were asked to calculate a synthetic spectrum using the stellar parameters
listed in Sect.\,\ref{exp2descr}. Spectra should be provided for the wavelength ranges 4900--5400 \AA,
6100--6800 \AA, and 8400--8900 \AA, respectively.
Eight high-resolution spectra ($R$=300\,000 to 500\,000) were produced for the comparison, namely P, A1, A2, A3, M1, M2, M3, and M4. Note that M3 is at that stage (no parameter determination) very similar to M2 since the spectra are based on the same line-lists and model atmospheres but due to interpolation
and the selected parameters the spectra are slightly different. 
A3 provided both a plane-parallel and a spherical solution, which are very similar. The plane-parallel version is used in the following as it had the higher spectral resolution. A1 and A2 both used ATLAS9+SYNTHE, but under two different operating systems. They differ only at very few wavelength points, so that we show only A1 in the figures below. An overview of the modelling details can be found in Table~\ref{tab:spectraTable}.

\begin{table*}
\center
\caption{Overview of model inputs used for Experiment 2.
The listed codes were also used for Experiment 1.}
\label{tab:spectraTable}
\begin{tabular}{llrl}
\hline\hline
Label & Codes & Spectral & Comments\\
 & & resolution & \\
\hline
M1 & MARCS + coma08 & 300000 & spherical\\
M2 & MARCS  & 500000 & mass = 1 M$_\odot$; spherical\\
       & + TURBOSPECTRUM &  & \\
M3 & MARCS  & 500000 & mass = 1 M$_\odot$, \logg=1.5, [Fe/H]=$-$0.25\\
       & + TURBOSPECTRUM &  & spherical\\
M4 & MARCS  + BSYN & 300000  at 6000\,\AA & mass = 2 M$_\odot$, $v_{{\rm mic}}$=2\,km\,s$^{-1}$; spherical\\
       &  &  & spectrum convolved to a resolution of 80\,000 \\
A1 & ATLAS9 + SYNTHE & 500000 & model atmospheres interpolated within Kurucz grid\\
A2 & ATLAS9 + SYNTHE & 300000 & model atmospheres and spectra calculated\\
  & Linux vers. (Sbordone) & & for [Fe/H]=0 and $-$0.5, interpolated to [Fe/H]=$-$0.2\\
A3 & ATLAS12 & 500000 & plane-parallel\\
P & Phoenix V.15 & 300000 & \\     
\hline
\end{tabular}
\end{table*}

To compare these synthetic spectra we first made a linear interpolation of the normalized flux values to a common wavelength grid with a step size of 0.01\,{\AA}. Then, for each wavelength point the mean flux of all six spectra (P, A1, A3, M1, M2, M4) was calculated, and the root-mean-square (rms) difference of all normalized fluxes from the mean normalized flux was determined.  Next, the rms differences were averaged over 2\,{\AA}-wide wavelength intervals. In addition, the difference of each individual normalized spectrum from the mean normalized flux was calculated.

\begin{figure*}  
\resizebox{\hsize}{!}{\includegraphics[trim=55pt 40pt 25pt 55pt,clip]{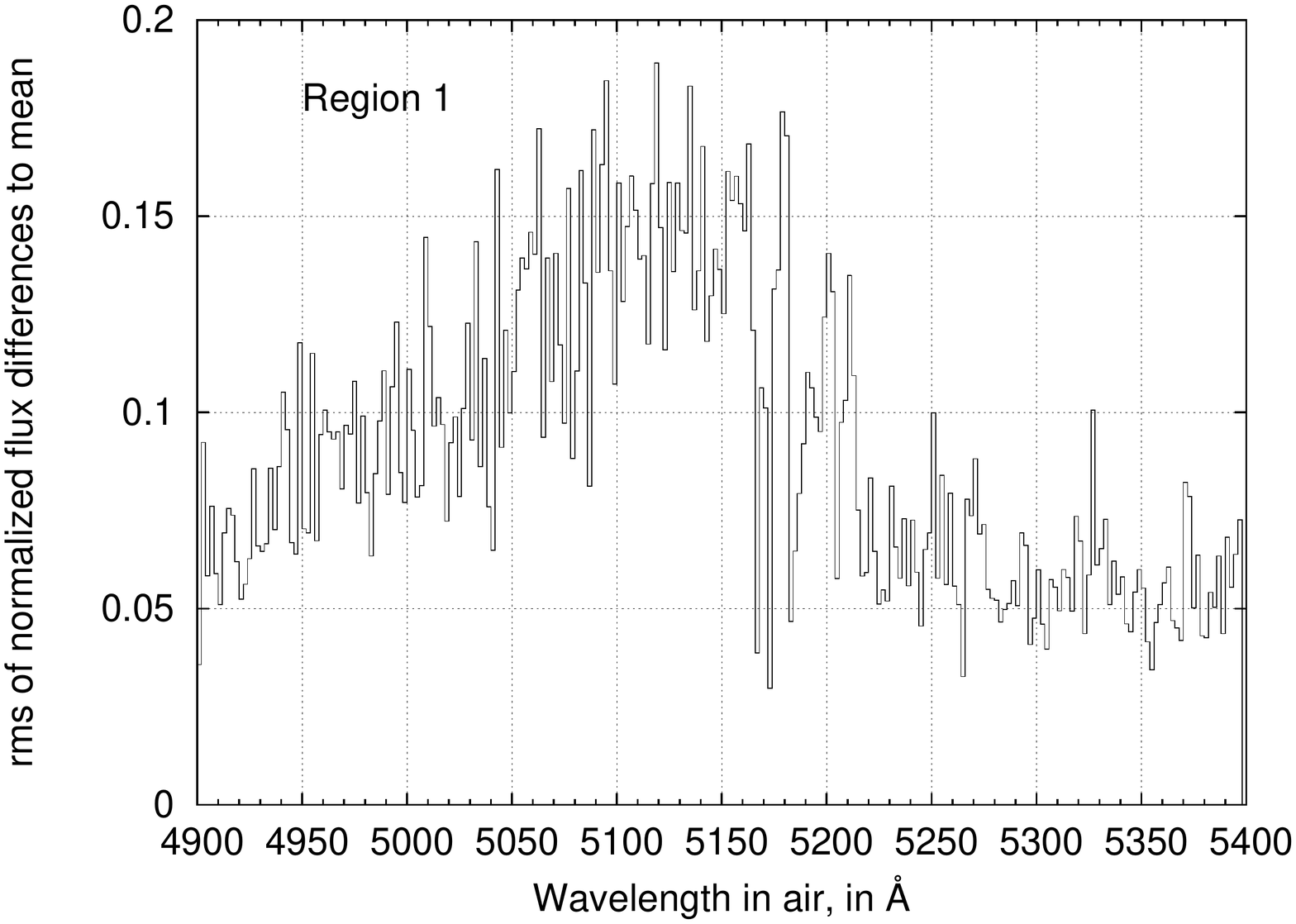}\includegraphics[trim=55pt 40pt 25pt 55pt,clip]{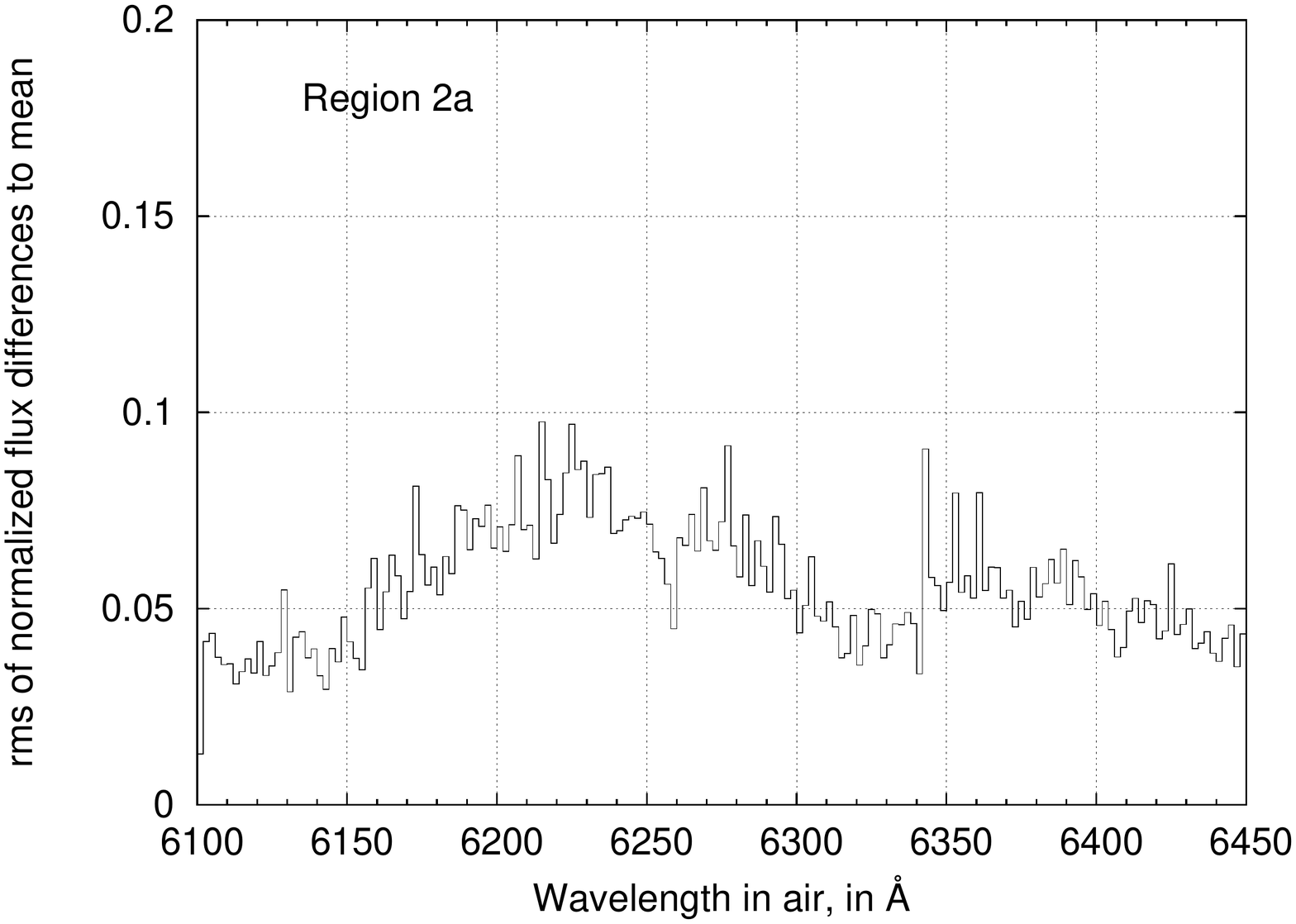}}
\resizebox{\hsize}{!}{\includegraphics[trim=55pt 40pt 25pt 55pt,clip]{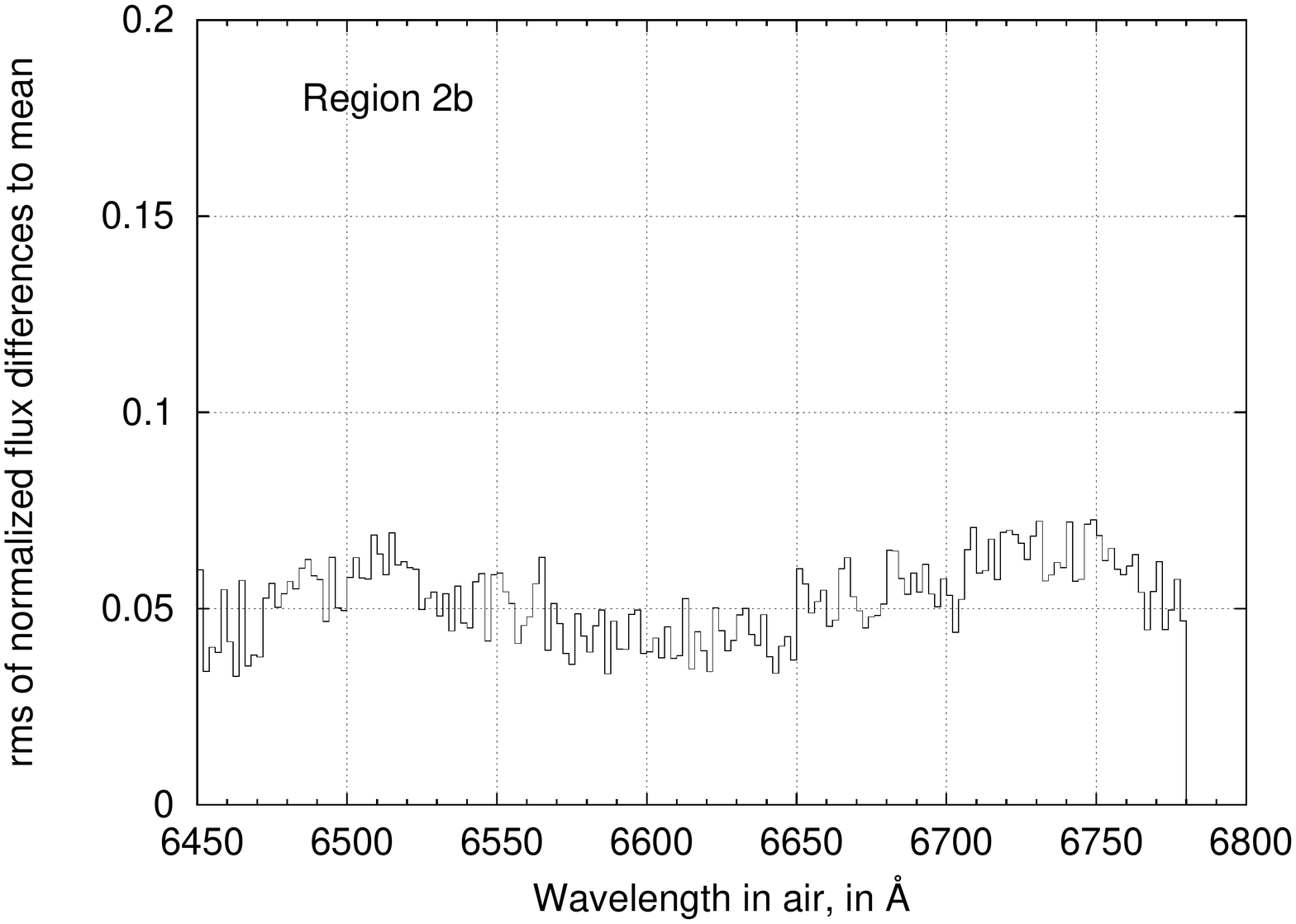}\includegraphics[trim=55pt 40pt 25pt 55pt,clip]{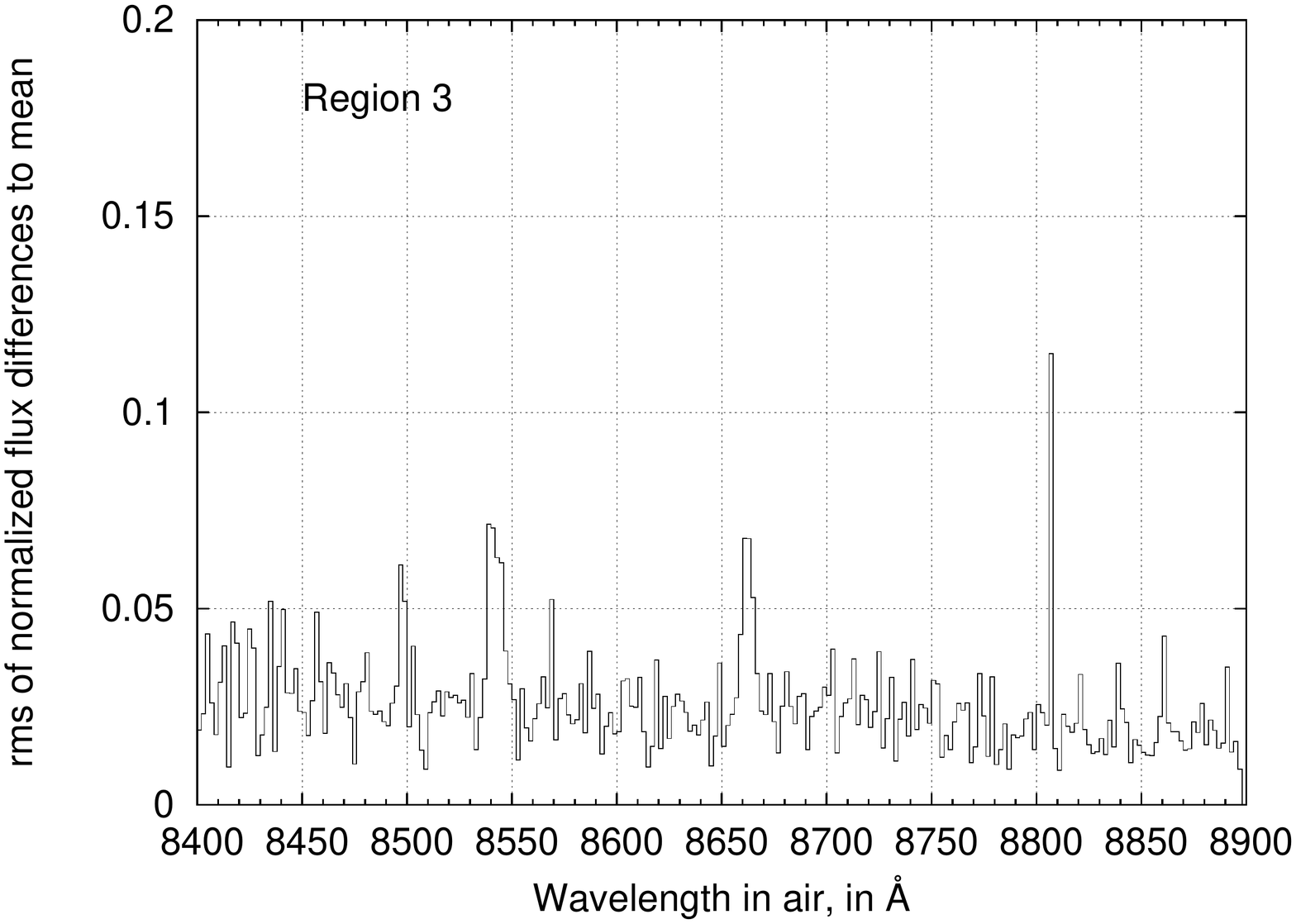}}
\caption{Root-mean-square (rms) 
of the differences between the normalized flux of six model spectra (P, A1/A2, A3, M1, M2/M3, M4) and the mean of these six spectra, averaged over 2~{\AA}-wide wavelength intervals. The different panels show four different wavelength regions, referred to in the text.}
\label{fig:rms}
\end{figure*}

\begin{figure*}  
   \resizebox{\hsize}{!}{\includegraphics[trim=55pt 40pt 55pt 55pt,clip]{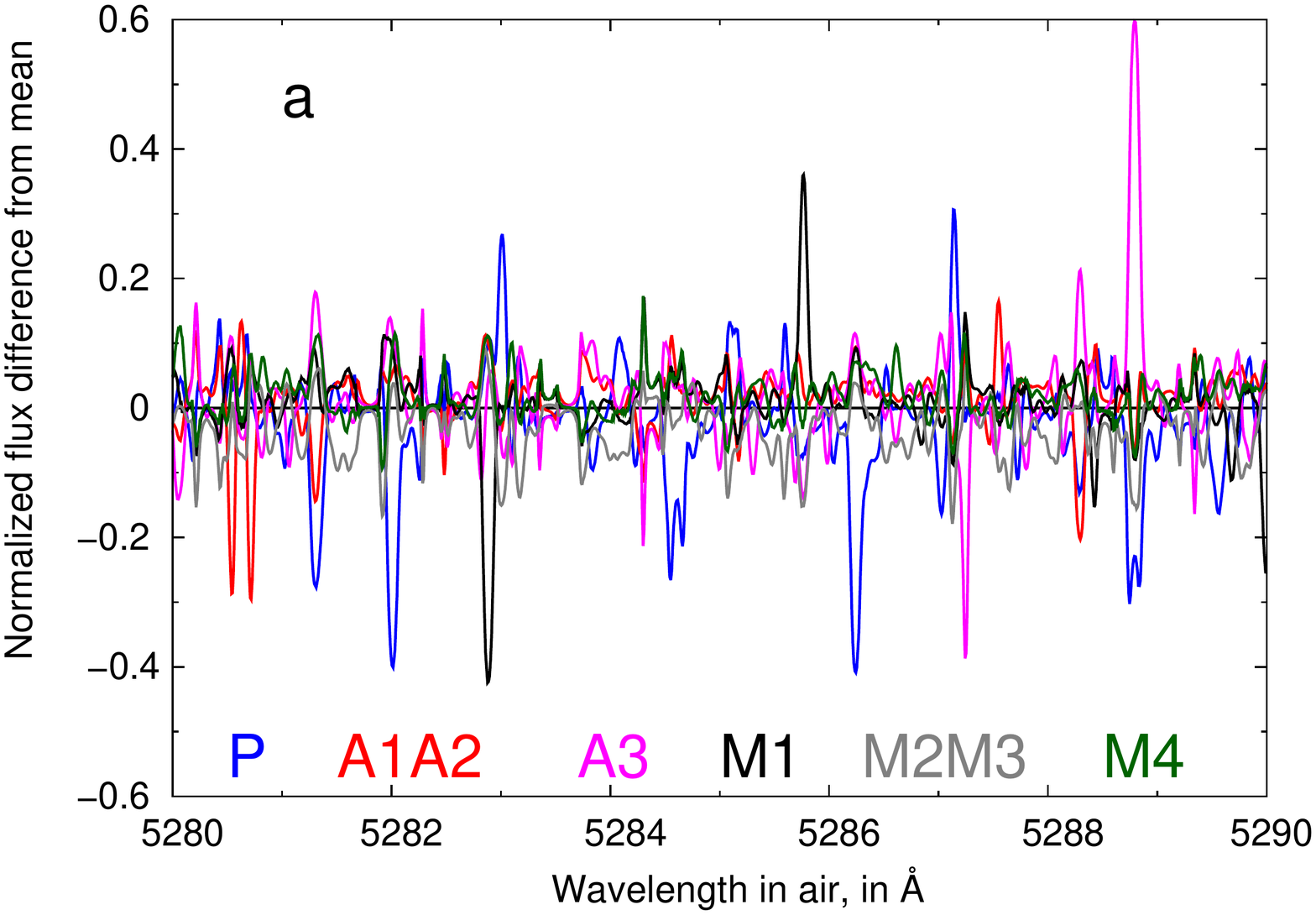}\includegraphics[trim=55pt 40pt 55pt 55pt,clip]{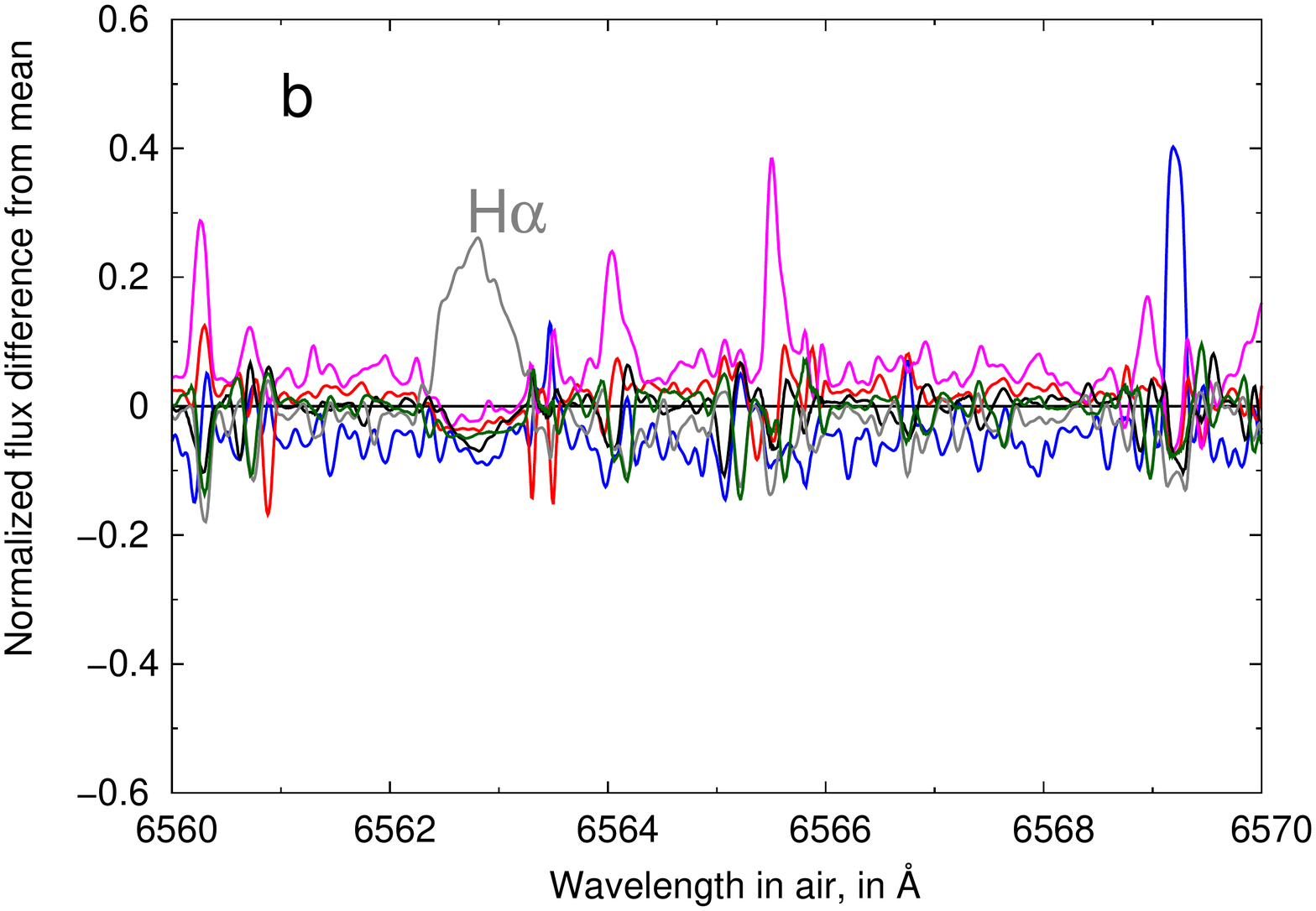}}
   \resizebox{\hsize}{!}{\includegraphics[trim=55pt 40pt 55pt 55pt,clip]{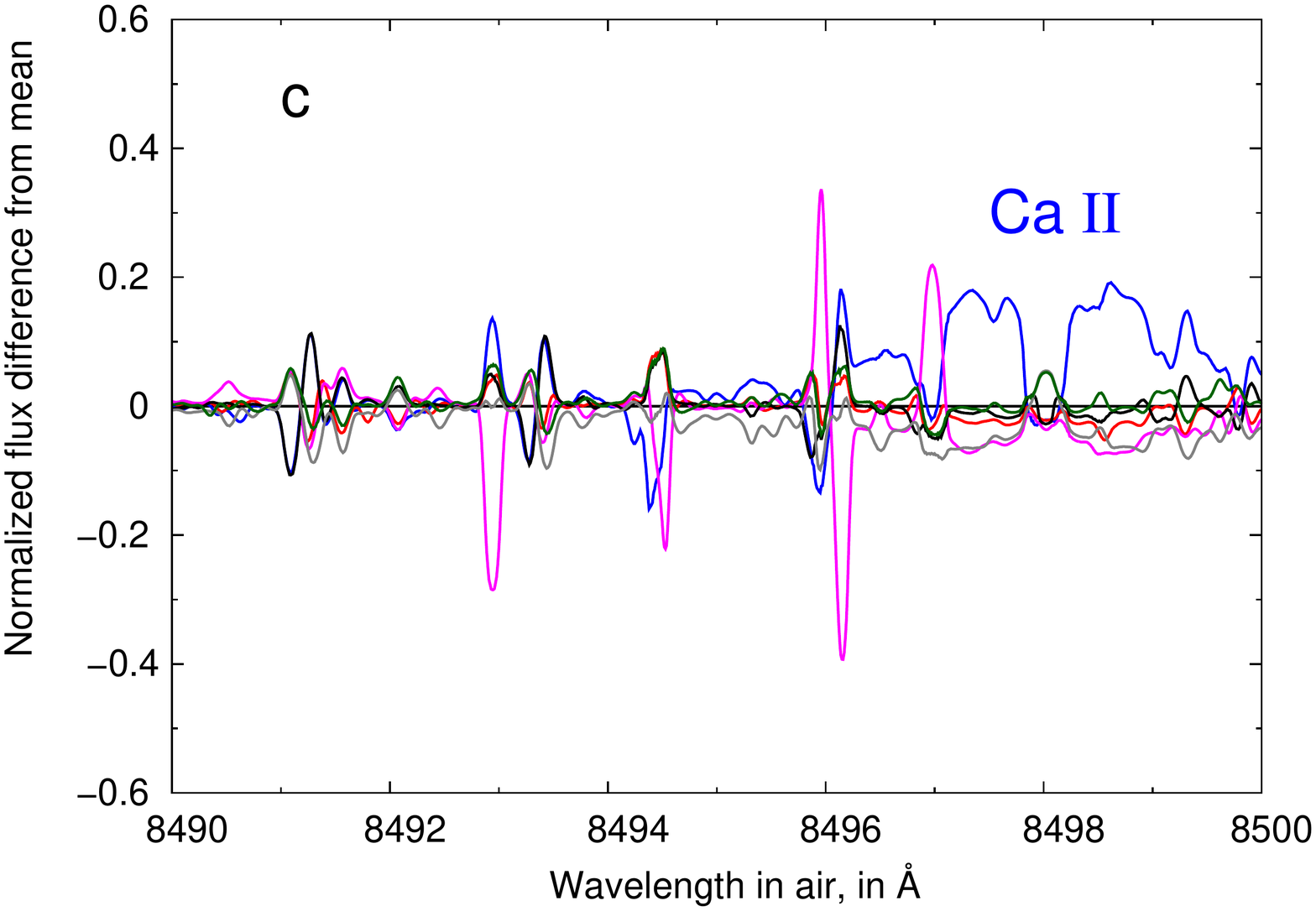}\includegraphics[trim=55pt 40pt 55pt 55pt,clip]{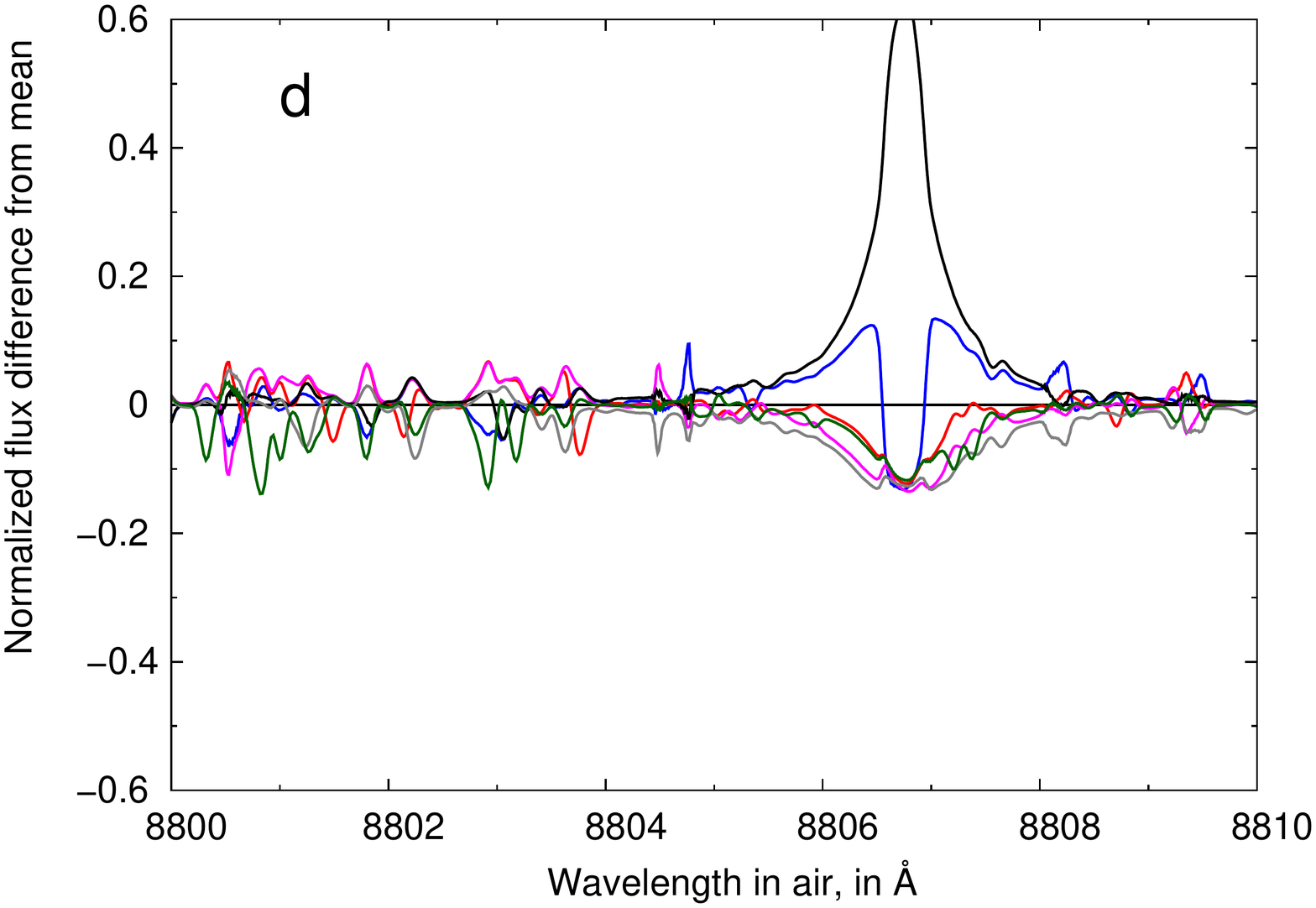}}
\caption{Normalized flux differences from the mean for six different models, for four different 
wavelength regions. The six model codes appear across the bottom of panel (a) in the colours used in the plots. (a) blue region with relatively low rms difference; (b) region containing H$\alpha$; (c) region containing the IR \ion{Ca}{II} triplet 8498~\AA\ line; (d) region with high rms difference due to differences in a strong \ion{Mg}{I} line at 8807~\AA.}
\label{fig:diff}
\end{figure*}

Figure~\ref{fig:rms} shows the rms flux differences.
We note that a large rms value can be caused by one or more of the following: a large difference for an individual line, small differences for several lines, or a small continuum difference.
To explore the features seen in Figure~\ref{fig:rms}, we looked at the individual spectra in more detail.
In Fig.~\ref{fig:diff} we show the differences of the six spectra from the mean, for selected 10~{\AA}-wide wavelength intervals.
Starting with Region~1 in Fig.~\ref{fig:rms}, we found that 
MgH is the dominant molecular absorber in the region from 5000 to 5210~\AA, where we see high rms differences. 
Comparison with a spectrum calculated with MgH lines only demonstrated that the differences in the spectra produced by different investigators 
are mainly due to differences in the MgH line data that they used. 
The broad rms peaks around 6220~\AA\ in Region~2a and around 6740~\AA\ in Region~2b coincide with regions where TiO absorption dominates, while the smaller peak around 6510~\AA\ in Region~2b coincides with dominant CN absorption.

The three sharp peaks in Region~2a close to 6350~\AA\ are due to three atomic lines. Two lines at 6343 and 6362~\AA\ are present in four out of the six spectra, and are much broader in the A3 model than in the others. These lines are not present in the observed spectrum of $\alpha$~Tau. One line at 6354~\AA\ is present in all models, but with varying line depths (0.3, 0.4, 0.6, 0.8, 0.9) -- a comparison with the observed spectrum of $\alpha$~Tau points towards intermediate values. A careful check of the line lists revealed that these three lines can likely be
associated with wrong data for some \ion{Ca}{i} transitions. A3 included theoretical iron line lists from Kurucz, which are based on
predicted energy levels. These lines appear to cause the strong and broad
features seen in A3's synthetic spectra at the mentioned wavelengths.
A minor sharp peak can be seen in Region~2b close to the H$\alpha$ line, which was not included in the line list of the M2/M3 model (see Fig.~\ref{fig:diff}b).
We note that none of the models reproduce the H$\alpha$ line in the observed spectrum of $\alpha$~Tau. The observed line depth is more than twice as large as any of the calculated ones.
In Region~3 in Fig.~\ref{fig:rms}, the most obvious features are centered on the IR \ion{Ca}{II} triplet lines and are due to different modelling of the broadening of these lines. A3 and M2/M3 show the broadest lines, A1/A2, M1, and M4 are narrower, and P is narrowest. An example is shown in Fig.~\ref{fig:diff}c.
Finally, the very sharp peak seen in Region~3 in Fig.~\ref{fig:rms} close to 8810~\AA\ is due to deviations for a strong \ion{Mg}{I} line in two of the models. The line is absent in M1 and narrower than in the others in P (see Fig.~\ref{fig:diff}d). A comparison with the observations supports the models with larger broadening.

In general, the outcome of the comparison can be summarized in the following way:
\begin{itemize}
   \item Good agreement is seen for strong atomic lines, with a few exceptions (e.g. the wings of \ion{Mg}{I} 8807~\AA).
   \item Medium-weak atomic lines are in partial agreement. 
   \item Large differences are seen for the IR \ion{Ca}{II} triplet lines.
   \item Molecular lines differ to varying degrees.
\end{itemize}

\section{Discussion} 
\label{sect:discussion}
As pointed out above, there seems to be no obvious difference between the results from different `code-families' (MARCS, ATLAS). Experiment~2, 
the `fixed parameter comparison', revealed quite significant differences in the synthetic line strengths, 
in particular for some medium to weak lines, molecular lines, and for the \ion{Ca}{II} IR triplet.
The main reason for these differences seems to be the different line parameters used by the different groups. However, 
the spectrum modelling differences are only one aspect leading to the resulting scatter of the derived parameters of, e.g., \one.
In the following we briefly discuss various aspects that may affect the result.

\subsection{Method of parameter determination}
In an ideal world, observations with their uncertainties would be combined with models and their uncertainties via a Bayesian 
approach to estimate stellar parameters, including correlations between parameters. This is especially difficult in the real world, 
because uncertainties are often difficult to estimate. Some of the observing parameters (e.g., spectral point-spread function, telluric 
correction) require complex models to determine uncertainties, and, to our knowledge, no modelling code currently propagates uncertainties 
in, e.g., oscillator strengths and pressure broadening coefficients through to wavelength-dependent uncertainties on the output spectra. 
Therefore, different groups tried different approaches to fitting what they regard as the most reliable parts of the spectra with differing 
weighting schemes.

The participating teams have chosen a remarkable variety of approaches for deriving the stellar parameters. 
The groups can be divided into those which measured equivalent widths from the provided spectra which they compared to calculated equivalent widths (A2, A5, T), and those which employed spectrum synthesis.
The latter groups synthesized either continuous wavelength intervals (M2, M3a, M4, M5, A3, P, C), or intervals around selected spectral lines (M1, M3b, M6, A1, A4; see Fig.~\ref{fig:wavelengths}).
The main difference between the two M3 analyses was in the characteristics of the spectra that were fed through an automated pipeline into the parameter determination code MATISSE. M3a was tailored to low resolution ($R\sim$6\,500) spectra covering almost continuous wavelengths within a region of $\sim$300\AA. M3b was applied to spectra at a higher resolution ($R\sim$15\,000), using discontinuous wavelength regions totaling $\sim$435\AA\ spread over $\sim$1800\AA.

The M2, M5, M6, A1, A3, and T, analysis started by determining approximate stellar parameters (or at least one parameter) from a comparison of observed photometry and synthetic colours (see Tables~\ref{results1a} to \ref{results2b}, and Section~\ref{sect:A3}). M1 used the colour information at a later step to partly lift the degeneracy between \Teff / \logg\ and metallicity.
The remaining groups did not use the provided colour information.

The experiment participants used different methods to calculate equivalent widths or synthetic spectra.
A2 applied automatic measurement of equivalent widths from synthetic spectra calculated with SYNTHE, using fixed-width integration around the center of each spectral line. A5 used the plane-parallel code MOOG \citep[original version described in][]{1973PhDT.......180S} to calculate the equivalent widths, while T employed the method described in \citet{1986A&A...156....8T}. 
A1, A3, and A4 worked with the plane-parallel line-formation code SYNTHE \citep{1993KurCD..18.....K} for spectrum synthesis.
M1 used a spherically-symmetric spectrum synthesis code developed for the study of line profile variability in 
AGB stars, described in \citet[][their Section 2.4]{2010A&A...514A..35N}.
M2, M3, and M5 used the spectral line formation code TURBOSPECTRUM \citep{Alvarez1998}, while M4 used the code BSYN developed in Uppsala. Both TURBOSPECTRUM and BSYN provide the option to compute the radiative transfer in spherically-symmetric geometry, and both have a number of subroutines in common with the MARCS code.
M6 used the MOOG code in spectrum-synthesis mode. The P and C analyses employed the spectrum synthesis integrated in the PHOENIX and CODEX packages, respectively.

The methods used for parameter optimization with the provided spectra are briefly summarized as follows.
M1, M2, M4, M5, A2, A3, P, and C based their fit on minimizing a weighted squared difference ($\chi^{2}$) between the observed and the synthetic spectrum, either for the whole wavelength range available or for selected regions containing parameter-sensitive features. Some of the groups complemented the automatic fit with an inspection by eye (M1, M2, M5).
M3 used the MATISSE algorithm, which has been developed for fast automatic parameter determination 
of low- to high-resolution stellar spectra, and is based on a local multi-linear regression method (for details see Section~\ref{sect:M3}).
A1 verified by eye that the observed spectrum was reproduced by the model spectrum with the photometric \Teff\ values, and adjusted the \logg\ and \FeH\ values manually in order to fit selected features sensitive to these parameters.  
M6, A4, and A5 derived the stellar parameters by removing line abundance trends. This ``classical'' method effectively combines the information from spectrum points within spectral lines, and uses knowledge on spectral line behaviour to infer the parameters. The difference was that M6 and A4 used spectrum fits for each line, and A4 also fitted the wings of strong lines, while A5 worked with equivalent widths.
In the T analysis, \Teff\ was derived from photometry, a fixed value was chosen for \logg, and the abundance analysis was performed as described in \citet{2008A&A...489.1271T}.


A prime issue that has to be 
discussed is how these different strategies affect the result. There are good reasons to start with photometry data to constrain the stellar
parameters.  Photometric colours have the advantage of providing information over a wide baseline in the spectrum. Some of them -- of course 
depending on the spectral types of the target stars -- are nearly independent of metallicity over a wide range of temperatures (for cool stars 
$V-K$ can be used very well), while others involve a dependence on surface gravity. A clear disadvantage, however, that can introduce considerable 
error for this approach, is interstellar reddening, although this will not play a significant role in the context of our experiment.  
Narrow-band photometry -- e.g. using filters to study the strength of individual molecular bands \citep[see e.g.][]{wing} -- may be a 
way to circumvent this weakness.

Furthermore, photometric colours are affected by the actual chemical composition of the star, especially for cool 
objects with altered C/O ratios (e.g. \four\ in Experiment~1). In such cases, the uncertainty in the derived \Teff\ is expected to increase significantly.
To make use of the photometry, a calibration relating colours and parameters is required, which can be either purely theoretical or 
semi-empirical. A considerable number of such calibrations have been published throughout the years. Thus, it is not surprising that each 
of the groups using the colour information in Experiment~1 applied a different calibration (see Tables~\ref{results1a} to \ref{results2b}). The resulting 
estimates for \Teff\ agree rather well, except for \four, as expected from the discussion above (see also discussion in Section~\ref{sect:M5}). 
For \logg\ and \FeH, the values disagree in the few applicable cases.

All of the methodologies used here for the parameter determination suffer from several sources of uncertainty.
First, in the squared-difference minimization method, similar $\chi^{2}$ values can be obtained for a certain range of models, 
and so the ``best parameters'' must be constrained by additional considerations.
An interesting aspect is that the groups using this method defined the difference measure ($\chi^2$) in different ways, 
in particular regarding the weight factor. It is not clear how this might affect the outcome of the analysis. Note 
that we did not compare individual $\chi^2$ values to assess the results from the different groups, as those will also depend on 
the length and location of the wavelength sections.
Second, none of the methods provides a straight-forward way for estimating the uncertainties in the derived parameters, except for 
MATISSE (M3) which automatically provides error estimates
based on the signal-to-noise ratio. 
Third, separating the effects of different line-broadening parameters such as micro- and macro-turbulence or rotation may be difficult in the 
case of the spectral synthesis method. On the other hand, the full spectrum fit can -- in principle -- better handle spectra with lots of 
blended lines, and it includes a large number of spectral indicators at once. 

Since M2 and M3 used identical model spectra for their analysis, the differences in the results are entirely due to the different methods 
applied. In addition, the influence of different methods can be seen by comparing the M3a and M3b results.
M2 and M3b both used the optical wavelength region (4900 -- 6800~\AA) for the fit, although slightly different subintervals were selected. 
Furthermore, M3b worked at decreased spectral resolution, while M2 used the full resolution. The resulting \Teff\ values are comparable for 
both $\alpha$~Tau and $\alpha$~Cet, while \logg\ and \FeH\ differ. This is in line with the findings of M4 that the optical region is mostly 
sensitive to \Teff\ variations. Hence, the ``best-fit'' values of \logg\ and \FeH\ will be rather arbitrary. Using only the RVS wavelength region 
(8470-8740 {\AA})
and a low resolution, M3a gives different results for all three parameters. The \Teff\ values are higher for both benchmark stars (compared to M3b), 
while \logg\ and \FeH\ are higher for one star and lower for the other.

The P analysis for $\alpha$~Tau resulted in the highest value for \logg. This could be explained by low values of the pressure 
broadening parameters for strong atomic lines used by the P code, apparent in Fig.~\ref{fig:diff} c and d. To achieve a fit 
to broad observed line wings, the P analysis requires a higher \logg\ value than the other models.

\subsection{Limitations in accessible parameter range}
Typically, models are optimized for application to a specific range of stellar parameters. In Experiment~1 we also see this in some limits of the 
tested parameter range. For example the CODEX code (C) has an upper limit in \logg\ of 0.0, and thus the `best fit' result in the case of \two\ is 
naturally somewhat offset from the findings  of the other groups.  

As mentioned in Section~\ref{sect:results1}, most participating groups explored only a small volume in parameter space around values estimated 
from photometry or comparisons with spectra of standard stars. Only M3 and A2 based the analysis of \one\ and \two\ on a large model grid, and 
they obtained results similar to the others, except for \Teff\ in the case of A2.

\subsection{LTE and non-LTE}
\label{sect:nlte}
As one of the underlying assumptions in this comparative work, all the participants used
LTE stellar model atmospheres with LTE line formation (except for the M6 Ca abundance analysis described below).
A full non-LTE treatment of all elements in the atmospheres of cool giants is a very complex task that can not yet be achieved due to 
the lack of quantitative atomic data.
Nevertheless, non-LTE theoretical models of giant stars have been computed by \citet{2005ESASP.560..967S}, taking into account the main 
electron contributors. They showed that non-LTE effects on the structure of the atmosphere increase with decreasing metallicity, especially 
deep and high up in the photosphere. The change in the structure will change the line formation domains and the intensity of the photospheric 
lines. Fortunately, the change in the structure seems to be negligible at solar metallicity. Thus, one can use LTE model atmospheres for a 
partial non-LTE analysis by applying line synthesis of a given element in non-LTE. This has been done by many authors, e.g., 
\citet{1999ApJ...521..753T,2001ApJ...550..970S}. More recently, \citet{2011A&A...528A..87M} showed with a very complete model atom of 
\ion{Fe}{I/II} that the non-LTE corrections do not exceed 0.1~dex for solar and mildly low metallicities.

The LTE assumption may strongly affect the surface gravity determination when this astrophysical parameter is constrained by the ionization 
equilibrium to follow Saha's law.
The surface gravity deduced from the photometry and evolutionary tracks is usually used as a starting point for spectroscopic iteration between 
two ionization stages of iron.
This latter procedure relies on the assumption of LTE and tends to give LTE values of \Teff\ and \logg\ smaller than those from photometric determinations. 
Therefore, this can also impact the iron abundance analysis.
A temporary solution to this problem can be to trust the surface gravity given by the photometric estimation, and to use only \ion{Fe}{II} lines, 
which are known to be free of non-LTE effects \citep[see, e.g.][]{2011A&A...528A..87M}, to determine [Fe/H]. An investigation of non-LTE effects 
on Fe line formation is not within the scope of the present paper. However, the M6 team determined the abundance of Ca in $\alpha$~Tau using 
non-LTE radiative transfer in an LTE model atmosphere, as described in the following paragraph.         

\paragraph{\ion{Ca}{I}/\ion{Ca}{II} analysis of $\alpha$~Tau:}
The M6 team aimed to find the same abundance for \ion{Ca}{I} and \ion{Ca}{II}. It was decided to compare the abundance 
determined using the IR triplet (8498, 8542, 8662)~\AA\ with the abundance determined using the optical triplet (6102, 6122, 6162)~\AA. 
A non-LTE analysis was performed, using a \ion{Ca}{I} atom with 153 energy levels (taking into account fine structure), 2120 bound-bound 
radiative transitions and 11476 bound-bound collisional transitions with electrons, and 81 photoionization tables from TopBase. 
They used a \ion{Ca}{II} model atom with 74 energy levels, 422 bound-bound radiative transitions, 2628 bound-bound collisional transitions with 
electrons, and 40 photoionization tables from TopBase. The contribution of inelastic collisions with neutral hydrogen was neglected due to the 
lack of accurate quantum-mechanical calculations to determine their importance, and the Stark effect in the line-broadening parameters was also 
neglected (see \citealt{2011arXiv1107.6015M}). Spherical MARCS model atmospheres with standard composition were adopted. The code MULTI2.2, 
slightly modified for the collisional transition part, solves the radiative transfer and statistical equilibrium equations consistently and 
computes line profiles, equivalent widths and contribution functions for all the radiative transitions. They performed several computations for 
the model atmosphere 4000/1/0 (\Teff/\logg/[Fe/H]) with different calcium abundances for \ion{Ca}{I} and \ion{Ca}{II} and decided by eye the best 
metallicity fitting both the \ion{Ca}{I} lines and the \ion{Ca}{II} triplet lines. This turns out to be 
[Ca/Fe] = $-0.3 \pm 0.1$ with an adopted [Fe/H]=0.0 (see Table~\ref{results1a}). 
We note that the core of the \ion{Ca}{II} IR lines is still poorly reproduced, because of the existence of a 
chromosphere where the line cores form 
\citep[see][]{2011A&A...535A..59B}.

\subsection{Geometry}
In addition to \Teff, \logg\ and abundances, stellar
spectra are in principle also affected by the extension
of the atmosphere, which can be parameterized by 
(\Teff/5770)$(R/{\rm R}_\odot)(M/{\rm M}_\odot)^{-1}$. If the atmospheric
extension could be modelled from spectra, then this would uniquely
constrain $M$, $L$ and distance independently of evolutionary models.
None of the models submitted in the study of stars 1 through 4 came up
with a meaningful constraint on extension, but this by itself does not
mean it is insignificant.

Especially extended atmospheres can show very strong effects of
extension such as molecular shells, where it is not even obvious if a
strong line in a cool extended shell will appear in absorption or
emission, \citet[e.g.][]{2004A&A...421.1149O}. Extreme effects 
observed in highly evolved stars are difficult to
model completely in studies such as this paper, because they
necessarily require departures from hydrostatic equilibrium (e.g.
pulsation and shocks) in order to elevate the material. The effects in
stars with only low-amplitude oscillations such as Star 2 are much
smaller, but are significant enough that many modelling teams chose to
use spherical codes. This mirrors the experience of the various groups 
that even in a purely hydrostatic world the  sphericity effects on spectra increase
towards cooler giants.

The extension of an atmospheric model is typically parameterised by a
single statement: whether the model is spherical (with a reasonable
value of extension) or plane-parallel. For $T_{\rm eff}$ greater than
4000\,K, \citet{2006A&A...452.1039H} show typical abundance differences
between spherical and plane-parallel codes of 0.1\,dex. This is
comparable to or smaller than the metallicity errors estimated from
models in this paper, but suggests that if sphericity is not taken
into account, the atmospheric parameters will be mis-estimated. We suggest
that further work on this topic is required, informed by detailed
observations of a range of stars with well-known parameters and a
range of atmospheric extensions.

\subsection{Atomic and molecular line data}
\label{sect:linedata}
Accuracy and completeness of atomic and molecular data play a key role in fitting observed spectra with synthetic ones. In the experiments presented in this article we find various approaches to the compilation of input data. 
All MARCS-based analyses relied on atomic data extracted from the VALD database (in 2008--2009), while all ATLAS-based analyses used atomic data from the Kurucz web site.
A few groups calibrated the $gf$-values on high-resolution spectra of standard stars -- M3a and M5 on the Sun and Arcturus, and A4 on several standard stars, including the Sun, Arcturus, and $\mu$~Leo. A2 verified the atomic line data for the cases of the Sun and Procyon.
Molecular data were included in all calculations except for M6 and A5.
Typically, line lists for 8--11 different species were included. In the case of M2, M3, and M4, the data stem from the same sources \citep[described in][]{2008A&A...486..951G}. For M1, the data sources are given in \citet{2009A&A...494..403L}, and for M5 they are specified in Section~\ref{sect:M5}.
A1, A2, A3, A4, P, and C obtained molecular line data from the same
    sources (via the Kurucz website), see respective subsections of Section 3
    for details. Data sources for T are given in Tsuji (2008).

The VALD database for atomic lines was therefore the prime source of atomic data, although the line lists have been extracted from different versions of the database.
Sources for molecular line lists are less homogeneous, and also the number of molecules included varies 
from group to group. All line lists, both for atoms and for molecules, still require significant improvement. 
A way around missing line data is offered by the use of astrophysical oscillator strengths. 

However, the derivation of astrophysical $gf$-values requires a reference object, for which atmospheric parameters 
\emph{and element abundances} are known precisely. Currently, this is only the case for the Sun (even in that case 
some uncertainties remain). As the nature and strength of spectral lines appearing in the object under study (e.g. a 
cool giant) and the Sun will differ significantly, the astrophysical $gf$-values will be in most cases of limited 
use. 

A possible approach would be to work differentially using a ``chain'' of reference stars, starting from parameters close to solar and leading towards the required area in parameter space. We assume that for each neighbouring pair of stars it will be possible to define a common set of spectral lines. The line data calibrated on the star ``closer'' to the Sun will then define the element abundances of the ``new'' star. 
However, astrophysical $gf$-values will always remain an unsatisfying solution. Good experimental data are certainly highly desired. 

We note that oscillator strengths are only one element of the line data set required for each transition. Various damping constants used to parametrize collisional processes are equally important and equally uncertain. Differential work can remedy these uncertainties to a certain extent, in a similar way as for $gf$-values. For certain transitions, specialised theories are available, as for example for the hydrogen lines\footnote{http://www.astro.uu.se/{$<$tilde$>$}barklem/hlinop.html}, but these are not implemented in all codes.

\subsection{Convection}
\label{sect:convection}
An important physical process to be included in the calculation of model stellar atmospheres is convection.
The main effects of convection are 1) the contribution of the convective flux to the energy transport, leading to 
a modified temperature distribution compared to the purely radiative case; and 2) the generation of horizontal and vertical 
velocity fields, which modify the positions and shapes of spectral lines.
The efficiency of convective energy transport and its effect on the temperature structure in the line-forming region depends 
on all three atmospheric parameters (\Teff, \logg, \MonH) in a complex way (the contribution of the convective flux to the 
total flux is shown e.g. in Fig.~4 in \citealt{2002A&A...392..619H}). At \Teff$\approx$4000~K, the efficiency of convection 
reaches a maximum and seems to be independent of gravity or metallicity. 

In all of the atmospheric models used for the experiments described in the present paper, the first effect of convection is 
included in an approximative way, based on the mixing length theory \citep[MLT,][]{1958ZA.....46..108B}. The practical implementation 
of the MLT varies somewhat between the different codes.
In the ATLAS models, MLT convection is implemented as described in \citet{1970SAOSR.309.....K} and \citet{1996ASPC..108...85C}, 
similar to the formulation according to \citet{1978stat.book.....M}. A variant of the latter is used in the PHOENIX models 
\citep{2002A&A...395...99L}.
In the MARCS models, the MLT formulation of \citet{1965ApJ...142..841H} is used.
%
This variation in the implementation results in different numbers of free parameters, and different meanings of these parameters,
for each model, hampering a direct 
comparison (see Section 5.3.1 in \citealt{2002A&A...395...99L} for a discussion on the importance of the MLT formulation).

The free parameter dominating any discussion of MLT convection is the ``mixing length,'' i.e. the distance which convective elements can 
travel before they dissolve. It is usually specified in terms of the local pressure scale height, and typical values are between 0.5 and 3, 
depending on stellar parameters and whether spectral features or global stellar parameters are modelled.
However, a number of additional parameters have to be set.
The size of convective elements is usually assumed to be a single value and equal to the mixing length. The need for 
this ``one-eddy'' approximation is 
overcome in the alternative ``full-spectrum-turbulence'' convection models \citep[e.g.][]{1991ApJ...370..295C}, where the distribution of 
kinetic energy generated by convection over different spatial scales is computed from a turbulence model. This treatment of convection has 
been implemented in ATLAS models by \citet{2002A&A...392..619H} and was compared to MLT models for \Teff$\ge$4000~K, \logg$\ge$2 and a range 
of metallicities, but has not been used by any of the groups in the present paper.
Additional free parameters are needed to specify the geometry of the convective elements and the temperature distribution within them. This 
determines the energy lost by radiation and thus the efficiency of convection.
Furthermore, viscous energy dissipation is parametrized by a coefficient in the equation for the convective flux. 
The default values for these parameters are different in the ATLAS and MARCS models. However, since all free parameters basically are scaling 
factors in a single equation, this may be compensated by using different values for the mixing length.

An additional degree of freedom is provided in the MLT implementation of ATLAS models through the ``overshooting option.'' A smoothing procedure 
allows for a positive convective flux in stable layers next to the convection zone. 
\citet{1997A&A...318..841C} investigated the effect of this option for \Teff$\ge$4000~K, \logg$\ge$2.5 and a range of metallicities. Comparison 
with observation indicated that except for the Sun, the atmospheric parameters derived with different methods were more consistent when the 
overshooting option was switched off.

The second effect of convection, the three-dimensional velocity field, is not treated at all by MLT. Additional Doppler broadening of spectral 
lines caused by convective motions is taken into account in the one-dimensional models by specifying the ad-hoc parameters micro- and macroturbulence 
(in units of velocity, see modelling descriptions in Section~\ref{sect:modelling} and Tables~\ref{results1a} to \ref{results2b}).

Realistic modelling of the effects of surface convection on stellar spectra can only be achieved with time-dependent, three-dimensional, 
radiation-hydrodynamic numerical simulations (hereafter referred to as ``3D models'').
Such simulations have been done for the Sun for several decades, and have become increasingly accurate. This has been demonstrated by 
comparison to a variety of observations, not the least the profiles of numerous individual resolved spectral lines. 
Without applying any micro- or macroturbulence parameters, the wavelength shifts and asymmetries of the solar line profiles are in general 
very well reproduced \citep[see, e.g., the detailed review by][]{2009LRSP....6....2N}.

3D surface convection models and spectra for stars other than the Sun are not yet widely available, due to the large amount of computing power 
and human post-processing time required. However, individual regions in the parameter space have been investigated thoroughly. The most relevant 
study in the context of this paper is that by \citet{2007A&A...469..687C}, who examined the effects of convection on spectral lines for red giant 
stars with \logg\,=\,2.2 at two different effective temperatures and \FeH\,=\,$0, -1, -2, -3$. The largest effects when compared to 1D model atmospheres 
where found at the lowest metallicities, while the effects were minor at solar metallicity (abundance differences of up to 0.8~dex and within 
$\pm$0.1~dex, respectively). However, these models were significantly hotter and at lower \logg\ than the stars investigated in the present paper.

The situation may improve rapidly in the near future, as at least two groups are computing larger grids of 3D models, using two different 
simulation codes.
The ``CIFIST grid'' \citep{2009MmSAI..80..711L} and the ``StaggerGrid'' \citep{2011JPhCS.328a2003C} both will include on the order of 100 models 
for main-sequence and red giant stars (25--30 \Teff/\logg\ combinations and four metallicities). Within the latter project, average temperature 
structures calculated from the full 3D models will be made publicly available, as well as corresponding values for micro- and macroturbulence for 
implementation with classical line-formation codes. This will be the first step towards a more realistic spectrum analysis of a variety of stars.
We note that the parameters of the two benchmark stars $\alpha$~Tau and $\alpha$~Cet lie just outside of the parameter ranges investigated for 
convection effects so far, both in grids of 1D models and in sets of 3D models.
However, both of the future 3D grids include a solar-metallicity model at (\Teff,~\logg) = (4000~K,~1.5), which should be close enough for 
establishing the importance of convection for spectrum modelling for these stars.

Finally, we point out that the spatial extension of hydrodynamic simulations can be set up in two different ways. In the surface gravity domain 
of red giants, the ``box-in-a-star'' approach is applicable (cf the two model grids mentioned above). In this approach, the simulation is 
performed for a box inside the stellar atmosphere, covering several convective cells. The horizontal boundaries of the box are periodic, 
the vertical boundaries are open, and the extensions of the box are small compared to the stellar radius. However, the characteristic size of 
the convective cells scales with pressure scale height and thus increases with decreasing gravity. In the extreme case of red supergiants such 
as Betelgeuse, the atmosphere contains only a few giant convective cells, and hydrodynamic simulations of the ``star-in-a-box'' type must be 
employed \citep{2002AN....323..213F}. 
$\alpha$~Tau and $\alpha$~Cet lie in the transition region between these two approaches \citep[close to \logg\,=\,1,][]{2011JPhCS.328a2012C}. Since the 
influence of convection on \emph{spectrum modelling} has not been investigated in this region so far, these two stars are very 
interesting test objects for future studies (see, however, Section 3.5 in \citealt{2005AaA...442..281K} for a 3D hydrodynamical model and 
convection-related effects on \emph{colours} of a late-type giant).

\subsection{Other issues}

An often neglected fact is that model atmosphere and synthetic spectrum codes rely on a wealth of physical input data besides 
the spectral line data discussed in Section~\ref{sect:linedata}. Apart from parameters describing convective processes, an example 
for these are data used for chemical equilibrium calculations (establishing the equation of state), namely partition functions and 
molecular dissociation energies. Another important ingredient are data describing bound-free and free-free transitions providing the 
continuous opacity.
Although widely used standard references exist for much of these data (e.g. the ``Kurucz routines'' for continuous opacities, or 
\citealt{1981ApJS...45..621I} for partition functions), these are evolving slowly and often are not described in detail. These 
``hidden data'' leave room for differences of unknown degree between codes from different authors. Furthermore, they can lead to 
inconsistencies between model atmospheres and spectral synthesis. In this regard, the first step towards consistent spectrum modelling is 
to use spectrum synthesis and model atmosphere codes developed by the same author or group.

For example, the MARCS model atmosphere code and the BSYN synthetic spectrum code share many subroutines (e.g. for the chemical equilibrium) 
and the computation of the continuous opacities is consistent in both codes. The same is true for MARCS and TURBOSPECTRUM, and the ``Kurucz suite'' 
consisting of ATLAS9, ATLAS12, and SYNTHE.

\section{Conclusions}
Estimating stellar parameters by comparison of observed and synthetic spectra is always affected by inaccuracies 
in line data and assumptions of model stellar atmospheres for cool stars 
(cf. the extensive discussion of this issue in the context of cool, 
evolved giants by \citealt{gustafsson06}). Due to these limitations even a `perfect' fit of a spectrum cannot be seen 
as proof for a perfect model. We stress that it is not the intention of this paper to rank the various existing 
models of stellar atmospheres, but to learn from the effects caused by their differences. 

Experiment~1 clearly illustrates the need to be cautious when comparing or combining stellar parameters that were 
derived using different model atmospheres and fitting strategies. We tried to illuminate various possible influences, 
but a clear trend in terms of a systematically higher or lower value for some parameter for a given assumption could 
not be derived. This is a very complex problem with various aspects that seem to partly compensate for each other. 

It would be desirable to repeat the experiment using different modelling codes, but exactly the same method (e.g. a 
chi-square fit to preselected wavelength regions). Also, differences in atomic and molecular line lists should be 
sorted out. It seems to be extremely difficult to implement the use of equal input line data in the various codes, 
simply because of different requirements for the data format. This annoying obstacle could possibly be mitigated in 
the future by the use of a common infrastructure for atomic and molecular line data. An attempt to create such an 
infrastructure is currently underway in the VAMDC EU project\footnote{http://www.vamdc.eu/} 
\citep{2010JQSRT.111.2151D}. Finally, repeating the experiment using exactly the same tools
may give a clearer idea about the reasons for different results between different codes.

One could ask whether the data given for Experiment~1 are sufficient to derive a unique set of parameters. No doubt additional data would 
set further constraints and resolve one or another degeneracy.  A more extensive analysis of the existing data may produce some 
progress as well, but in our view it will be always limited by the accuracy of the line data and inaccuracies in our models, which 
definitely require further attention in the future.

However, we think that Experiment~1 represents quite a typical situation in observational astrophysics: a piece of the spectrum and some photometric 
data are given, and some fixed strategy to analyze the data is adopted. 
With our limited experiment we want to point out that the atmospheric model and the analysis method used can have a significant effect on the absolute 
values of the derived stellar parameters.
Bringing together such a variety of groups fitting stellar spectra by the use of various methods, this paper provides also an unprecedented snapshot of 
the current status of this field.

For future analysis projects, we recommend working differentially as far as possible. For stellar samples with parameters significantly different 
from the Sun, a set of suitable benchmark stars may serve as stepping stones for a piecewise differential analysis.

\acknowledgements
The workshop on which this paper is based was kindly supported by the ESF through the GREAT initiative, by the Robert F. Wing Support Fund at 
Ohio State University, and the Department of Astronomy at the University of Vienna.
The work of TL was funded by the Austrian Science Fund FWF projects P20046, P21988, and P23737.
UH acknowledges support from the Swedish National Space Board.
CA acknowledges partial support by the Spanish grant AYA2008-08013-C03-03.
HN acknowledges financial support from the Alexander von Humboldt foundation.
KE gratefully acknowledges support from The Swedish Research Council.
CW acknowledges the financial support of CNES, OCA and ESO. GK acknowledges the financial
support of CNES and CNRS.
UH, KE, TM, and FT acknowledge the role of the SAM collaboration
   (http://www.anst.uu.se/ulhei450/GaiaSAM/) in stimulating this research through
   regular workshops.
BA thanks for support by the Austrian Science Fund FWF under project number P23006.
We thank all participants in the workshop for contributing to the fruitful discussions there.
We thank B. Edvardsson and B. Gustafsson for comments on a draft version of the paper.
Based on data obtained within the Gaia DPAC (Data Processing and Analysis Consortium) and coordinated by the GBOG (Ground-Based Observations for Gaia) working group.

\bibliographystyle{aa}
\bibliography{19142}

\end{document}